\documentclass[twocolumn,showpacs]{revtex4}
\usepackage{times,xspace}
\usepackage{amsbsy,amssymb,amsmath,bm}
\usepackage{graphicx,color,epsfig,rotate}
\usepackage{fancyheadings}

\def\fa{\mathfrak{a}}

\def\fe{\mathfrak{e}}
\def\fb{\mathfrak{b}}
\def\fc{\mathfrak{c}}

\def\fg{\mathfrak{g}}

\def\fmm{\mathfrak{M}}

\def\one{{\mathchoice {\rm 1\mskip-4mu l} {\rm 1\mskip-4mu l} {\rm
1\mskip-4.5mu l} {\rm 1\mskip-5mu l}}}
\def\bbbc{{\mathchoice {\setbox0=\hbox{$\displaystyle\rm C$}\hbox{\hbox
to0pt{\kern0.4\wd0\vrule height0.9\ht0\hss}\box0}}
{\setbox0=\hbox{$\textstyle\rm C$}\hbox{\hbox
to0pt{\kern0.4\wd0\vrule height0.9\ht0\hss}\box0}}
{\setbox0=\hbox{$\scriptstyle\rm C$}\hbox{\hbox
to0pt{\kern0.4\wd0\vrule height0.9\ht0\hss}\box0}}
{\setbox0=\hbox{$\scriptscriptstyle\rm C$}\hbox{\hbox
to0pt{\kern0.4\wd0\vrule height0.9\ht0\hss}\box0}}}}
 
\newcommand{\Integers}{\ensuremath{\mathbb{Z}}\xspace}

\newcommand{\ket}[1]{|{#1}\rangle}

\newcommand{\bi}{{\bf i}}
\newcommand{\bj}{{\bf j}}
\newcommand{\bk}{{\bf k}}
\newcommand{\bl}{{\bf l}}
\newcommand{\bmm}{{\bf m}}
\newcommand{\bn}{{\bf n}}

\newcommand{\np}{\mbox{$\in \hspace*{-0.28cm}/$}}

\newcommand{\ignore}[1]{}
\newcommand{\cComment}[1]{}
\newcommand{\gComment}[1]{}
\renewcommand{\cComment}[1]{\textcolor{blue}{Cristian: #1}}
\renewcommand{\gComment}[1]{\textcolor{red}{Gerardo: #1}}
\begin{document}
\title{Algebraic Approach to Interacting Quantum Systems}
\author{C.D. Batista and G. Ortiz}
\affiliation{Theoretical Division, 
Los Alamos National Laboratory, Los Alamos, NM 87545}
\date{Received \today }

\begin{abstract}
We present an algebraic framework for interacting extended quantum
systems to study complex phenomena characterized by the coexistence and
competition of different states of matter. We start by showing how to
connect different (spin-particle-gauge) {\it languages} by means of
exact mappings (isomorphisms) that we name {\it dictionaries} and prove
a fundamental theorem establishing when two arbitrary languages can be
connected. These mappings serve to unravel symmetries which are hidden
in one representation but become manifest in another. In addition, we
establish a formal link between seemingly unrelated physical phenomena
by changing the language of our model description. This link leads to
the idea of {\it universality} or equivalence. Moreover, we introduce
the novel concept of {\it emergent symmetry} as another symmetry
guiding principle. By introducing the notion of {\it hierarchical
languages}, we determine the quantum phase diagram of lattice models
(previously unsolved) and unveil hidden order parameters to explore new
states of matter. Hierarchical languages also constitute an essential
tool to provide a unified description of phases which compete and
coexist. Overall, our framework provides a simple and systematic
methodology to predict and discover new kinds of orders. Another aspect
exploited by the present formalism is the relation between condensed
matter and lattice gauge theories through quantum link models. We
conclude discussing applications of these dictionaries to the area of
quantum information and computation with emphasis in building new
models of computation and quantum programming languages.
\end{abstract}


\maketitle

\section{Introduction}
\label{sec1}

Unveiling the fundamental principles behind complex behavior in matter
is a problem at the frontiers of condensed matter physics and embraces
cases like the high-temperature superconductors, heavy fermions, and
low-dimensional electron liquids. Generically, describing the structure
and behavior of matter involves studying systems of interacting quantum
constituents (bosons, fermions, spins, gauge fields) whose fields
satisfy the basic laws of quantum mechanics. Nevertheless, the plethora
of complex phenomena exhibited by nature exceeds our ability to explain
them, in part, because {\it the whole is not necessarily the sum of its
parts} \cite{anderson0} and thus typical perturbation-like-theory
arguments or standard mathematical techniques are not appropriate to
disentangle its mysteries. In this paper we present a unifying
algebraic framework for interacting extended quantum systems that
enables one to study complex phenomena characterized by the coexistence
and competition of various states of matter.

The emergence of such complex phenomena may be the result of very 
simple, undiscovered, principles that conspire against any 
straightforward explanation. Nonetheless, we expect that two pillars of
modern science, symmetry and topology, are key guiding principles
behind those fundamental laws of emergence. Group theory and geometry
have been fundamental to the physics of the twentieth century and we
count on them to continue playing such a role. Indeed, the notion of
symmetry and its breakings have shaped our current conception of
nature. Understanding the idea of invariance and its corresponding
conservation laws turns out to be as fundamental as determining the
causes that prevent such harmony and leads to more complex behavior.
Other kinds of order, not described by broken symmetries, are possible.
For example, quantum orders of topological nature distinguish the
internal structure of the state without breaking any {\it local}
symmetry, and its study is beyond the scope of the present paper
\cite{wen}. Another example is provided by the notion of {\it emergent}
symmetries where, as the name indicates, the number of symmetries of
the state increases as the temperature is lowered, contrary to the
broken symmetry case. These concepts are at the heart of the field of
quantum phase transitions that studies the changes that may occur in
the macroscopic properties of matter at zero temperature (i.e., $T=0$)
due to changes in the parameters characterizing the system. 

In this regard, the development of exact algebraic methods is one of
the most elegant and promising tools toward the complete understanding
of quantum phases of matter and their corresponding phase transitions.
Typically, there are no distinct length or time scales which separates
the different competing orders and often these systems are near quantum
criticality which makes their study extremely complicated, if not
impossible, by the traditional techniques, such as standard mean-field
or perturbation theories. The reason which prevents the effective use
of these theories is precisely the key for the successful application
of the algebraic methods, namely, the absence of a small parameter for
the various complex quantum orderings. In other words, one cannot
systematically apply renormalization group ideas and easily integrate
out irrelevant degrees of freedom. On the other hand, those competing
orders are frequently related by symmetry  principles characterizing
the critical behavior caused by the competing interactions, thus
increasing the symmetry group of the effective low-energy physics.

In the present manuscript we will be concerned with quantum lattice
systems. A quantum lattice is a composite system identified with
$\Integers^{N_s}$ where $N_s$ is the total number of lattice sites, and
associated to each lattice site (or mode) ${\bf i} \in \Integers^{N_s}$
is a Hilbert space ${\cal H}_{\bf i}$ of finite dimension $D$
describing the {\it local} modes. The total Hilbert space, i.e., state
space of the lattice, is the tensor product $\otimes$ of the local
modes state spaces, ${\cal H} = \bigotimes_{\bf i} {\cal H}_{\bf i}$,
in the case of distinguishable subsystems. (Notice that ${\cal H}$ may
support  inequivalent tensor product decompositions.)
Indistinguishability places an additional constraint on the space of
admissible states which manifests itself in their symmetry properties
(e.g., antisymmetric states in the case of fermions) and, consequently,
the physical Hilbert space ${\cal H}$ is a subspace of $\bigotimes_{\bf
i} {\cal H}_{\bf i}$.  A pure state of the system (at $T=0$) is simply
a vector $| \Psi \rangle$ in ${\cal H}$, and an observable is a
selfadjoint operator $\hat{\cal O}: {\cal H} \rightarrow {\cal H}$. The
dynamical evolution of the system is determined by its Hamiltonian $H$.
The topology of the lattice, determined by the connectivity of the
graph induced by the interactions in $H$, is an important element in
establishing complexity. Unless specified, we will always consider
regular lattices of coordination ${\sf z}$. In the case of quantum
continuous systems we can still use the present formalism after
discretizing, for example, the space coordinates. Going beyond this
approach is deferred for a possible later publication.

Every given physical system is represented by a {\it language} which
consists of a set of operators that acts irreducibly on the Hilbert
space ${\cal H}$. For example, if our system consists of a collection
of interacting electrons a natural language is a set of spin 1/2
fermion field operators. Defining a language amounts to establishing
the state space ${\cal H}$ used to describe the physical system, while
a Hamiltonian written in terms of that language comes to setting up the
quantum dynamics expected to depict the phenomenon of interest. Can we
connect the different (spin-particle-gauge) languages of nature within
a single algebraic framework? The answer to this question is {\it yes}
and the key outcome is a set of {\it dictionaries} relating the
languages representing the different physical systems. More precisely,
by dictionary we mean an isomorphic (one-to-one) mapping connecting two
languages. The existence of dictionaries provides not only a tool to
explore complexity but leads naturally to the fundamental concept of
{\it universality}, or equivalence, meaning that different physical
systems display the same behavior. The concepts of language and
dictionary are introduced in section \ref{sec2}. Previous to our work
there were two seemingly unrelated examples of these types of mappings:
The Jordan-Wigner (1928) \cite{jordan} and Matsubara-Matsuda (1956)
\cite{matsubara} transformations. We have not only generalized these
($SU(2)$) transformations to any spin, spatial dimension and particle
statistics but have proved a fundamental theorem that permits the
connection of the generators of the different languages which can be
used to describe a given physical problem.

Section \ref{sec2} starts defining the concept of a {\it bosonic
language}. Then we prove our fundamental theorem and establish the
necessary conditions for application of a given bosonic language to
describe the physical system under consideration. The main result of
this section is a proof of the existence of mappings between the
generators of any pair of languages which belong to the same class,
i.e., that can be used to describe the same physical system. The
necessary and sufficient condition for any pair of languages to be
connected is that the dimension of their local Hilbert spaces $D$ be
the same. In addition we prove that for each class of languages, there
is at least one that is realized by the generators of a Lie algebra. To
complete this picture, we introduce the notion of {\it hierarchical
language} and prove that there is at least one hierarchical language in
each class. Those languages are, in general, the most convenient ones
to characterize the different quantum phases of the system under
consideration.

In section \ref{sec3}, we extend these conclusions to non-bosonic
systems by introducing operators which transmute the modes statistics.
These operators have a local and a non-local component. The non-local
part is a trivial generalization of the transformation introduced by
Jordan and Wigner \cite{jordan} to map spins $S$=1/2 into spinless
fermions. The local component transmutes the statistics associated to
the interchange of particles which are sharing the same lattice site
$\bi$. Adding the transmutation of statistics to the fundamental
theorem completes the characterization of each class of equivalent
languages. In other words, the classes of bosonic languages defined in
section \ref{sec2} are expanded in section \ref{sec3} to include
fermionic, anyonic, or hybrid (para-) languages. 

The most natural choice of local Hilbert space ${\cal H}_\bi$ is the
one spanned by a single site (or mode) $\bi$ basis of dimension $D$.
The possibility to decompose the total Hilbert space of the problem
$\cal H$ into subspaces or subsystems which are not necessarily single
sites opens up the possibility to generate other dictionaries with
unforeseen applications. (Remember that $\cal H$ may support different
tensor product decompositions.) Section \ref{sec3n} expands on this
concept and show the simple case of a dictionary that uses as local
Hilbert space a bond $(\bi,\bj)$ state space ${\cal H}_\bi \bigotimes
{\cal H}_\bj={\cal H}_{\bi\bj}$ of dimension $D^2$ (${\cal
H}=\bigotimes_{(\bi,\bj)}{\cal H}_{\bi\bj}$), and which is mapped onto
a site Hilbert space of the same dimension.

Section \ref{sec4} is devoted to show explicit connections between
equivalent languages. The main purpose of this section is to illustrate
through examples the application of the dictionaries developed in the
previous two sections. Out of the many possible transformations
predicted by our fundamental theorem and the transmutation of
statistics, we selected a few of them which are useful for the
applications described in the later sections. However, it is important
to note that the procedure described to find the mapping connecting
two languages in the same class is very straightforward. In addition,
we show that the fractional exclusion statistics algebras emerge
naturally from the present formalism. 

Given a model Hamiltonian operator representing a quantum system, we
can use our dictionaries to {\it translate} it into another equivalent
Hamiltonian written in a different language. In other words, we can
write down the same Hamiltonian operator in many distinct ways
according to the different languages that are included in the
corresponding class. Since some of these languages are naturally
associated to specific physical entities (like spins, particles, gauge
fields, etc.), the corresponding translation provides a rigorous
connection between seemingly unrelated physical systems. This is the
main subject of section \ref{sec5}, where we show examples of
completely different quantum lattice systems described by the same
model. Another interesting aspect of these mappings is the potential to
unveil hidden symmetries of the Hamiltonian, possibly leading to exact
or quasi-exact solutions \cite{ushveridze}. Sometimes particular
languages allow us to recognize invariant subspaces of our Hamiltonian.
When the action of our Hamiltonian is restricted to an invariant
subspace, the corresponding operator can always be written in a
language which is more {\it elementary} than the original one (i.e., it
realizes a reduced number of degrees of freedom). In some cases, this
procedure enables one to recognize the hidden symmetries which lead to
the quasi-exact solution of the model considered (i.e., the model is 
integrable within the invariant subspace). The quasi-exact solution of
the one-dimensional $t$-$J_z$ \cite{ours4} model belongs to this class
of problems. It is important to remark that this quasi-exact solution
leads to the exact determination of the quantum phase diagram and the
charge excitations of the $t$-$J_z$ model. 

A new notion beyond Landau's concept of broken symmetry
\cite{Landau,andersonrev} is the complementary idea of {\it emergent
symmetry}, i.e., the fact that new symmetries not realized in the
Hamiltonian describing the system can emerge at low energies. There are
many instances in which the high-symmetry, low-energy, effective theory
can be derived in an exact way, i.e., without appealing to approximate
schemes like the renormalization group. For instance, the exact
effective theory (ground state plus any charge excitations) of the
one-dimensional  $t$-$J_z$ model is the $S$=1/2 $XXZ$ model
\cite{ours4}. It is well known that the $XXZ$ model has an infinite
number of symmetries which make it exactly solvable by the Bethe
ansatz. Another example is provided by the family of spin Hamiltonians
for which the ground state is a product of spin singlets. This family
includes the Majumdar-Ghosh \cite{Majumdar} model and many other
generalizations \cite{Majumdar,Shastry1,Shastry2,Shastry3,Lin}. These
and other simple examples introduced in section \ref{sec5-6} illustrate
the fundamental concept of emergent symmetry which, as we will show,
can be used as a guiding principle to find new states of matter. In a
sense that will become clear later, this concept not only formalizes
but  also provides a systematics to the principle of adiabatic
continuity that P.W. Anderson \cite{andersonrev} has advocated as one
of the two most important principles of condensed matter physics
(together with the concept of broken symmetry). States like the Fermi
liquid or the band insulator which do not correspond to any broken
symmetry can be characterized by their corresponding emergent
symmetries. In this way one can establish a formal connection between
different phases. For instance, the emergent symmetry of the band
insulator is analog to the emergent symmetry of the singlet dimer
magnetic states. In most cases, the notion of emergent symmetry is
approximate; however, still provides a guiding principle to identify
the relevant degrees of freedom, and the nature of the ground state and
low-energy  excitations. 

Coexistence and competition of different quantum orderings associated
to broken continuous symmetries is one of the main subjects of section
\ref{sec6}. There we show the fundamental role played by the
hierarchical languages for the classification of the possible order
parameters and the calculation of the quantum phase diagram of a given
model. We note that the hierarchical languages are the most
natural ones to provide a unified description of the order parameters
characterizing each phase. To illustrate the procedure we consider the
example of the Heisenberg $SU(N)$ Hamiltonians written in different
languages. We then take the local $SU(N)$ order parameter and reduce
its components according to the different subgroups of $SU(N)$ that can
be used to generate an equivalent language (for instance $SU(2)$). As
an illustration, we show that the local order parameter for an $S$=1  
$SU(2)$-spin Hamiltonian can be either the usual magnetization or a
spin-nematic order parameter. To conclude this section, we show how to
obtain the quantum phase diagram of the bilinear-biquadratic $S=1$
Heisenberg model (for spatial dimensions $d>1$) just by writing the
Hamiltonian in the $SU(3)$ hierarchical language. It is important to
remark that this zero temperature phase diagram was only known for
semi-classical spins \cite{Papanicolau}. The fact that there is no
calculation involved in the elaboration of this quantum phase diagram
permits the reader to appreciate the power and potential applications
of this algebraic framework. 

The formal connection between lattice models in condensed matter
physics and lattice gauge theories of high-energy physics is described
in section \ref{sec7}. There we take advantage of the existing quantum
link models and the connection between spins and other degrees of
freedom, such as gauge fields, which emerges from our algebraic
approach. 
  
In addition to the fascinating field of quantum phase transitions and
statistical mechanics, our algebraic approach can be applied to the
field of quantum information \cite{qcbook}. This new paradigm of
information processing takes advantage of the fundamental laws of
quantum mechanics to perform operations which can be done at least as 
efficiently (i.e., with polynomial complexity) as with classical
devices. The device that performs the manipulation of information is
named quantum computer and the basic unit of information is the {\it
qubit} (i.e., a two-level system). The quantum computer (a quantum
many-body system in disguise) consists of an isolated set of quantum
degrees of freedom which can be controlled and manipulated at the
quantum level to perform the required operations. The different
languages associated with the description of these quantum degrees of
freedom are possible programming languages and indeed may realize
different models of quantum computation. The development of
dictionaries connecting these languages is relevant not only to improve
the efficiency of a quantum computer, but also to model and simulate
different physical phenomena. The applications of our algebraic
framework to the fields of quantum information and quantum computing
are discussed in section \ref{sec8}. There we show how each physical
realization of a quantum computer has a class of languages associated
to the realization of the quantum operations. For instance, if we
consider the typical case of interacting $S$=1/2 spins the most natural
language is the one generated by the Pauli matrices. Hence we can use
the dictionary provided by the Jordan-Wigner (JW) \cite{jordan}
transformation to simulate a fermionic system \cite{sfer}. In this
regard, the purpose of this paper is to generalize this idea to other
possible realizations of quantum computers. We also discuss the
potential applications of this algebraic framework to the description
of recent experiments on bosonic systems in optical lattices.

Finally, in section \ref{sec9} we summarize the main concepts
introduced in this paper.  

 
\section{General Lie Algebraic Framework}
\label{sec2}

The notions of Hilbert space and linear maps or operators are central
to the formulation of quantum mechanics \cite{gQM}. Pure states (wave
functions) are elements of a Hilbert space and physical observables
are associated to Hermitian (self-adjoint) operators which act on that
space, and whose eigenvalues are selected by a measuring apparatus. The
role of linear operators in quantum mechanics is not restricted to the
representation of physical observables. Non-Hermitian operators are
very often used. That is the case, for instance, of the creation and 
annihilation operators in second quantization. Moreover, the unitary
operator that describes the time evolution of the system is in general
non-Hermitian. These observations simply tell us that linear operators
in general (Hermitian and non-Hermitian) play a more general role in
quantum mechanics since they  provide the mathematical {\it language}
which is required to describe the dynamical behavior of our quantum
system.

What do we mean by the term mathematical {\it language}? In order to
provide a rigorous answer for this question it is important to first
determine what is the mathematical structure associated to the set of
quantum operators. Quantum operators form a complex linear (vector)
space under the sum and the multiplication by a scalar over the field
of complex numbers. If we augment this vector space with a bilinear
operation (product $\Box$ between two operators), the set form an  {\it
algebra}. Quantum mechanics also requires this operation being 
non-commutative and associative. This additional structure makes the
set of quantum operators form an {\it associative algebra}. In
principle, any operator in this algebra can play a role in the 
description of our quantum system. However, one can always select a
subset (which forms a basis) in such a way that any quantum operator
can be expressed as a function of the selected ones. For instance, we
can use the  Pauli matrices $\{ \sigma^{x}, \sigma^{y}, \sigma^{z} \}$
\begin{equation}
\sigma^x=\begin{pmatrix} 0&1 \cr 1&0\end{pmatrix} , \
\sigma^y= \begin{pmatrix} 0&-i \cr i&0\end{pmatrix} , \
\sigma^z=\begin{pmatrix} 1&0 \cr 0&-1\end{pmatrix} , 
\end{equation}
to express any quantum operator associated with a spin $1/2$. The
particular subset of operators that we choose to express any quantum
operator is the mathematical language that we will use for the 
quantum description of our system. The elements of this subset will be
called {\it generators} of the language.

What are the conditions a given set of quantum operators must satisfy
to become a language? How many different languages can be used to
describe a quantum system? What is the connection between the different
languages? What is the most appropriate language to describe a
particular system? What is the  relation between language and symmetry
generators? How can languages help us understand the phenomenon of
universality? (It turns out that the notions of language and
universality are closely related  whenever a common language can be
used to describe seemingly unrelated physical phenomena.) A great part
of this paper is dedicated to answer these questions. In particular, in
this section we introduce the notion of {\it bosonic language}, we
establish a formal connection or {\it dictionary} between the different
bosonic languages associated with a given bosonic system, and we also
establish a formal relation between bosonic languages and {\it Lie
algebras}.

One of the fundamental steps toward a deeper understanding of classical
systems was the recognition that dynamical variables which define the
phase space are generators of continuous transformations. The set of
continuous transformations forms a {\it group} and the infinitesimal
generators provide a basis for a Lie algebra which is related to
the group. A group is a non-empty set which is closed under
an associative product $\Box$, it contains an identity, ${\bf 1}$,  and
all of its elements are invertible. When the transformations in the
group  leave the equations of motion invariant the set becomes a
symmetry group for the considered system. A real (complex) Lie
algebra $\cal L$ is a linear space over the field $F$ of real
(complex) numbers  which is closed under a non-associative Lie product
$[\;,\;]$ that satisfies ($\fa,\fb,\fc \in {\cal L}$ and $\alpha,\beta
\in F$):
\begin{eqnarray}
{[} \alpha \fa + \beta \fb , \fc {]}&=& \alpha {[}\fa,\fc{]} + \beta
{[}\fb,\fc{]} \nonumber \\
{[} \fa , \fb {]}&=&-{[} \fb , \fa {]} \nonumber \\
0&=&{[} \fa , {[} \fb , \fc {]} {]}+ {[} \fb ,{[} \fc , \fa {]}{]} + {[}
\fc , {[} \fa , \fb {]}{]}
\end{eqnarray}
Whenever a continuous transformation is a symmetry of our physical
system, the infinitesimal generator, which is an element of the
corresponding  Lie algebra, becomes a conserved quantity. As a
consequence, it is convenient to choose that quantity as one of the
coordinates for the phase space of our system.

In the same way, it may be appropriate to choose the quantum operators
(generators of the language) for the description of a quantum
system  in such a way that they simultaneously provide a basis for a 
Lie algebra. In other words, the use of symmetry generators  for
the description of an interacting quantum system incorporates symmetry
as a guiding principle to find out the possible solutions. As we
demonstrate below, it is always possible to find a language whose
generators form a basis of a Lie algebra.

We start by considering the bosonic languages of quantum
mechanics. To introduce the definition of a bosonic language, we need
to define first the concept of {\it monoid}. A monoid is a triple
$(\fmm,\Box,{\bf 1})$ in which $\fmm$ is a non-empty set, $\Box$ is an
associative product in $\fmm$, and ${\bf 1}$ is an element of $\fmm$
such that $\Box({\bf 1,\fa})=\fa=\Box({\bf \fa,1})$ for all $\fa\in
\fmm$. In this way we see that the the concept of monoid generalizes
the notion of group; a group is a monoid all of whose elements
are invertible \cite{Noten0}. A {\it bosonic language} is a set of
operators which can be grouped in subsets $S_{\bf i}$ (associated to
each mode) and satisfy the following conditions: 

\begin{itemize}

\item  Each element $\fb^\mu_{\bf i}$ of $S_{\bf i}$ ($\mu \in
[1,N_\fg]$) belongs to the algebra of endomorphisms for the vector
space ${\cal H}_{\bf i}$ over the field of complex numbers $\bbbc$,
$\fb^\mu_{\bf i}:{\cal H}_{\bf i}\rightarrow {\cal H}_{\bf i}$, and
these elements are linearly independent.

\item  The elements of $S_{\bf i}$ generate a monoid of linear
transformations under the associative product in the algebra which
acts irreducibly on ${\cal H}_{\bf i}$ in the sense that the only
subspaces stabilized by $S_{\bf i}$ are ${\cal H}_{\bf i}$ and
${\bf 0}$ (${\bf 0}$ is the null vector).

\item If $\fb^\mu_{\bf i}$ and $\fb^\nu_{\bf j}$ are elements of
different subsets $S_{\bf i}$ and $S_{\bf j}$, then $\fb^\mu_{\bf
i}\fb^\nu_{\bf j}=\Box(\fb^\mu_{\bf i},\fb^\nu_{\bf
j})=\Box(\fb^\nu_{\bf j},\fb^\mu_{\bf i})$.

\end{itemize}
The elements of the sets $S_{\bf i}$ are called generators of the
bosonic language. Combining the associative product  and the additive
operation in the algebra of endomorphisms of ${\cal H}_{\bf i}$, ${\rm
End}_\bbbc {\cal H}_{\bf i}$, we can define the non-associative Lie
product $[\;,\;]$, which is called commutator: $[\fb^\mu_{\bf
i},\fb^\nu_{\bf j}]=\fb^\mu_{\bf i}\fb^\nu_{\bf j}-\fb^\nu_{\bf
j}\fb^\mu_{\bf i}$ \cite{Noten1}. Then, the last condition can be
reformulated by establishing that the commutator of elements in
different subsets is zero
\begin{equation} 
{[} \fb^\mu_{\bf i} , \fb^\nu_{\bf j} {]}=0 \ ,\;\; \mbox{ if}\;\; {\bf
i}\neq {\bf j}.
\label{bosl}
\end{equation}
It is also important to notice that the set $S_{\bf i}$ is not
necessarily closed under the regular product (composition) or the Lie
product (commutator). If the set $S_{\bf i}$ is closed under the Lie
product, the elements of $S_{\bf i}$ generate a Lie algebra. We will
denote the Lie algebra by ${\cal S}_{\bf i}$. In addition, since each
generator is represented by an endomorphism of ${\cal H}_{\bf i}$ there
is a particular representation $\Gamma_{\cal S}$ of ${\cal S}_{\bf i}$
associated to the bosonic language. The second condition for a bosonic
language implies that $\Gamma_{\cal S}$ is irreducible.  The third
condition implies that the global Lie algebra associated to the
complete set of generators is the direct sum of local algebras ${\cal
S}_{\bf i}$, ${\cal S}= \bigoplus_{{\bf i}} {\cal S}_{\bf i}$.
Therefore, if the set $S_{\bf i}$ is closed under the Lie product, we
can represent the bosonic language by the conjunction of the Lie
algebra ${\cal S}$, and the irreducible representation $\Gamma_{\cal
S}$: ${\cal S}\wedge \Gamma_{\cal S}$. The dimension of $\Gamma_{\cal
S}$ is equal to the dimension of the local Hilbert space ${\cal H}_{\bf
i}$: ${\rm dim}\Gamma_{\cal S}=D$. The algebra of endomorphisms for the
vector  space ${\cal H}_{\bf i}$ is the enveloping algebra of ${\cal
S}_{\bf i}$.

The following fundamental theorem shows that two languages are {\it equivalent} if 
they have in common the dimension $D$ of their local Hilbert space
${\cal H}_{\bf i}$, and $D$ is finite. In other words, there is always a
dictionary connecting both languages, and a given physical
phenomena can be described with either one. A corollary of the theorem
is that given a particular language one can always find an equivalent
one which can be expressed as ${\cal S}\wedge \Gamma_{\cal S}$, i.e.,
the operators are generators of a Lie algebra ${\cal S}_{\bf i}$ in the 
irreducible representation $\Gamma_{\cal S}$. In general there are
various complex Lie algebras which can be used to build equivalent
languages. The only condition is that they have to admit an irreducible
representation of dimension $D$. Clearly, the number of Lie algebras
satisfying this condition increases with $D$. 

\subsection{Equivalent Classes of Bosonic Languages}

The demonstration of the fundamental theorem of this section is a
direct consequence of the classical theorem of Burnside
\cite{jacobson}, which plays an important role in the theory of rings.
Given the above definitions we can enunciate the theorem in the
following way:

{\it Burnside's theorem}.
Let $G$ be a monoid of linear transformations in a finite dimensional 
vector space $V$ over an algebraically closed field $F$, that acts
irreducibly on $V$ in the sense that the only subspaces stabilized by
$G$ are $V$ and ${\bf 0}$. Then $G$ contains a basis for End$_{F}V$
over $F$. 

End$_{F}V$ is an abbreviated notation for the ring of endomorphisms of
$V$ over the field $F$. In quantum mechanical systems, the field $F$
corresponds to the complex numbers $\bbbc$. The demonstration of the 
Burnside's theorem can be found, for instance, in Ref. \cite{jacobson}.
An immediate consequence of this theorem is the following one which is
the basis for connecting different bosonic languages:

{\it Fundamental theorem} (``On the equivalence of languages''). Given
two bosonic languages having the same finite dimension $D$ of their
local Hilbert spaces, ${\cal H}_{\bf i}$, the generators of one of them
can be written as a polynomial function of the generators of the other
language and vice versa. 

{\it Proof}:
The proof is a trivial application of Burnside's theorem. We need to
notice first that since the dimension $D$ is the same for both
languages, the spaces ${\cal H}_{\bf i}$ are also the same (vector
space of dimension $D$ over the field of complex numbers $\bbbc$). Let
us consider the monoid of transformations $G_1$ generated by
multiplying the generators of the first language until the enlarged set
becomes closed under the product. The second condition in the definition
of a bosonic language states that $G_1$ acts irreducibly on ${\cal
H}_{\bf i}$.  Since the dimension $D$ of ${\cal H}_{\bf i}$ is finite,
Burnside's theorem guarantees that $G_1$ admits a basis for
End$_{\bbbc}{\cal H}_{\bf i}$. Therefore, any endomorphism in ${\cal
H}_{\bf i}$ can be written as a linear combination of endomorphisms in
$G_1$. In particular, the generators of the second language can be
written in this way because they belong to End$_{\bbbc}{\cal H}_{\bf
i}$. Since each element of $G_1$ is a product of generators of the
first language, this concludes the demonstration. 

This theorem establishes an isomorphism between the algebras of
endomorphisms associated to each of the two languages. Motivated by
this observation we can introduce the notion of classes of equivalent
bosonic languages. We will say that two bosonic languages belong to the
same class if they have the same dimension $D$ of their local Hilbert
spaces ${\cal H}_{\bf i}$. The fundamental theorem establishes the
existence of dictionaries connecting languages within the same class.
As a consequence, we can use any bosonic language in the class to
describe a given physical phenomena. The natural question which emerges
from this result is: What is the most appropriate language in a given
class for describing our particular problem? There is no generic answer
to this question. Nonetheless, the following two corollaries give an
important hint for problems which are invariant under particular
transformations, because they relate the notion of language to the
generators of symmetry groups.

{\it Corollary I}: In each class of bosonic languages there is at least
one which is the conjunction of a Lie algebra ${\cal S}$ and an
irreducible representation ${\Gamma_{\cal S}}$ (${\cal S}\wedge
\Gamma_{\cal S}$), i.e., the generators of the bosonic language are
generators of the Lie algebra ${\cal S}_{\bf i}$ in the representation
${\Gamma_{\cal S}}$.

{\it Proof}:
First, we need to notice that each class is characterized by the
dimension $D$ of the local Hilbert space ${\cal H}_{\bf i}$. Let us
consider the group $U(1) \otimes SU(2)$. The Lie algebra associated to
this group is ${\cal L}_{\bf i}=u(1) \bigoplus su(2)$. The generators
of ${\cal L}_{\bf i}$ are $\{{\bf 1}^{\;}_\bi,S^x_\bi,S^y_\bi,S^z_\bi
\}$, 
\begin{equation}
\left [ S_\bi^\mu, S_\bj^\nu \right ] = i \delta_{\bi\bj} \epsilon_{\mu
\nu \lambda} S_\bi^\lambda \ , \;\;\; \mu,\nu,\lambda=x,y,z
\end{equation}
($S_\bi^\pm=S_\bi^x\pm i S_\bi^y$ and $\epsilon$ is the totally
antisymmetric Levi-Civita symbol), and there is one irreducible
representation $\Gamma^D_{\cal L}$ of dimension $D$ for each possible
value of $D$ \cite{Noten6}. Therefore, the set ${\cal S}_{\bf
i}=\{{\bf 1}^{\;}_\bi,S^x_\bi,S^y_\bi,S^z_\bi \}$ in the representation
$\Gamma^D_{\cal L}$ fulfills the three requirements for a bosonic
language and operates in a local Hilbert space of dimension $D$. Since
${\cal S}_{\bf i}$ is the set of generators of the Lie algebra ${\cal
L}_{\bf i}$, making ${\cal S}_{\bf i}={\cal L}_{\bf i}$ and
${\Gamma_{\cal S}}=\Gamma^D_{\cal L}$, we have proved the corollary I. 

The proof of corollary I shows that any bosonic problem with a
local Hilbert space of dimension $D$ can be described with
$SU(2)$-spins of magnitude (representation) $S=(D-1)/2$. The
Matsubara-Matsuda (MM) \cite{matsubara} transformation is the simplest
application of this corollary to $D=2$. (Indeed, one can construct
generalized MM transformations for any $D$ \cite{ours1}.) 

We introduce now another definition which is motivated by the next
corollary. A given bosonic language will be called {\it hierarchical}
if any local physical operator $\hat{\cal O}_\bi$ can be written as a
linear combination of the generators of the language, i.e., 
\begin{equation}
\hat{\cal O}_\bi=\sum_{\mu=1}^{N_\fg} \lambda_\mu \fb^\mu_\bi \ ,
\end{equation} 
where $\lambda_\mu \in \bbbc$, and it is the conjunction of a  Lie
algebra ${\cal S}$ and an irreducible representation ${\Gamma_{\cal
S}}$.

{\it Corollary II}: In each class of bosonic languages there is 
one which is hierarchical and its generators are the identity and
the generators of $su(N=D)$ in the fundamental representation.

{\it Proof}: 
For each class, with dimension $D$ of the local Hilbert space ${\cal
H}_{\bf i}$, we consider the group $U(1) \otimes SU(N)$ with $N=D$.
The generators of the Lie algebra associated to this group, ${\cal
L}_{\bf i}=u(1) \bigoplus su(N)$, are the identity plus the generators
of $su(N)$. Since the fundamental representation $\Gamma^F_{\cal L}$ of
${\cal L}_{\bf i}$ has dimension $D=N$, the conjunction of ${\cal L}=
\bigoplus_\bi {\cal L}_{\bf i}$ and this representation is one possible
language for the class considered. Since the dimension of  ${\cal
L}_{\bf i}$ is $D^2$, which is the dimension of the vector space
End$_{\bbbc}{\cal H}_{\bf i}$, then the generators of ${\cal L}_{\bf
i}$ also form a basis for End$_{\bbbc}{\cal H}_{\bf i}$.

The first consequence of corollary II is that the generators of any
language can be expressed as a linear combination of generators of a
hierarchical language in the same class. Again, the most trivial
example is given by the class of bosonic languages containing the spin
$S$=1/2 lattice. The generators of any language (like hard-core bosons
or any two level system) in the same class can be written as a linear
combination of the identity and the Pauli matrices. We will see later
that corollary II is the key to get a hierarchical classification of
the possible broken symmetries of a given physical Hamiltonian.

We consider now two additional examples that illustrate in detail the
contents of  the fundamental theorem and the subsequent corollaries.
The first example corresponds to hard-core bosons with $N_f$ different
flavors $\alpha$. Since they are hard core only single occupancy is
allowed, i.e., the eigenvalues of ${\bar n}_{\bf i}=\sum_\alpha 
\bar{n}_{{\bf i}\alpha}$ are either $0$ or $1$
($\bar{b}^\dagger_{\bi\alpha} \bar{b}^\dagger_{\bi\beta}=0$, and $\bar{n}_{{\bf
i}\alpha}=\bar{b}^{\dagger}_{\bi\alpha} \bar{b}^{\;}_{\bi\alpha}$ is
the number operator for the flavor $\alpha$ at the site ${\bf i}$). To
distinguish between site and flavor indices we will adopt the following
convention: Sites are denoted by latin indices while the different
flavors (or orbitals within the same site) are labeled by greek
indices. The minimal set ${S}_{\bf i}$ of operators that we can use to
generate a bosonic language which is appropriate for hard-core bosons
is: 
${ S}_\bi=\{{\bf 1}^{\;}_\bi,\bar{b}^\dagger_{\bi\alpha},
\bar{b}^{\;}_{\bi\alpha} \}$ with $1 \leq \alpha \leq N_f$. It can be
shown that this set satisfies the three requirements for a bosonic 
language. The dimension of the local Hilbert space for these
endomorphisms is $D=N_f+1$. Then by varying the total number of flavors
we can generate all possible values of $D$. Since each class of bosonic
languages is characterized by the value of $D$, these hard-core bosons
provide an example of a bosonic language  in each class. It is clear
that the set ${ S}_\bi$ is not closed under the Lie product.
Therefore, we cannot associate a Lie algebra to this minimal bosonic
language. However, if we increase the number of generators in the set
${S}_\bi$ by including bilinear forms of the type
$\bar{b}^\dagger_{\bi\alpha}\bar{b}^{\;}_{\bi\beta}$, then the new set
${\cal S}_\bi=\{{\bf 1}^{\;}_\bi,\bar{b}^\dagger_{\bi\alpha},
\bar{b}^{\;}_{\bi\alpha}, \bar{b}^\dagger_{\bi\alpha}
\bar{b}^{\;}_{\bi\beta} \}$, with $1 \leq \alpha,\; \beta \leq N_f$,
becomes closed under the Lie product
\begin{eqnarray}
\begin{cases}
{[} \bar{b}^{\;}_{\bi\alpha}, \bar{b}^{\;}_{\bj\beta} {]} =
{[} \bar{b}^\dagger_{\bi\alpha}, \bar{b}^\dagger_{\bj\beta} {]} = 0  \ , \\
{[} \bar{b}^{\;}_{\bi\beta}, \bar{b}^\dagger_{\bj\alpha} {]}
= \delta_{\bi\bj} 
(\delta_{\alpha \beta}- {\bar n}_\bi \delta_{\alpha
\beta}-\bar{b}^\dagger_{\bi\alpha} \bar{b}^{\;}_{\bi\beta}) \ ,
\\
{[}\bar{b}^\dagger_{\bi\alpha}\bar{b}^{\;}_{\bi\beta}, 
\bar{b}^\dagger_{\bj\gamma} {]}=\delta_{\bi\bj}
\delta_{\beta \gamma} \bar{b}^\dagger_{\bi\alpha}  \ . 
\end{cases}
\label{conm}
\end{eqnarray}
This means that the extended set ${\cal S}_\bi$ is now a set of
generators for a Lie algebra in a particular representation. From the
commutation relations (Eq.~(\ref{conm})) we can conclude that ${\cal
S_\bi}$ is the direct sum of an $u(1)$ algebra, generated by the
identity ${\bf 1}^{\;}_\bi$, and an $su(N)$ ($N=D=N_f+1$) algebra
generated by  $\{\bar{b}^\dagger_{\bi\alpha},\bar{b}^{\;}_{\bi\alpha},
\bar{b}^\dagger_{\bi\alpha} \bar{b}^{\;}_{\bi\beta} \}$: ${\cal
S}_\bi=u(1)\bigoplus su(N)$. The representation $\Gamma_{\cal S}$ is
the fundamental representation of $su(N)$ (${\rm dim}\Gamma_{\cal
S}=N$). Therefore, the new language is a hierarchical one. In section
\ref{sec4a} we give a more detailed description of this particular
language. Here we only want to emphasize the practical consequences of
the fundamental theorem and its corollaries. The first non-trivial
observation is that for each system of interacting multiflavored
hard-core bosons, there is an equivalent system of interacting
$SU(N)$-spins in the fundamental representation (with the minimal
non-zero magnitude). For $N_f=1$ we recover the well-known MM
transformation  \cite{matsubara}. We can see now that this is the
generalization to $su(N)$ of the MM transformation. With this example
we can envision the broad set of applications derived from the
fundamental theorem. Another consequence of the theorem is that any
physical theory for a bosonic system can be formulated in terms of
multiflavored hard-core bosons if the dimension of the local Hilbert
space is finite. The usefulness of this formulation will depend on the
particular system as it is illustrated in the next sections. Since the
second language is hierarchical we can write down any local physical
operator (endomorphism in ${\cal H}_{\bf i}$) as a linear combination
of its generators. Therefore, each of these generators will appear in
the Hamiltonian under consideration with a power not larger than one.
In addition, if the Hamiltonian has a global symmetry generated by a
direct sum of local transformations: $\bigoplus_\bi \hat{T}_\bi$, the
symmetry will become explicit by writing the Hamiltonian in terms of a
hierarchical language. The most basic example is the case of MM
hard-core bosons ($N_f=1$) in a lattice, described by a Hamiltonian
with a kinetic energy term and a nearest-neighbor density-density
interaction. The expression for the Hamiltonian in terms of the first
language defined by the set $S_\bi$ is
\begin{equation}
H_{\sf xxz}= t \sum_{\langle \bi,\bj \rangle} (\bar{b}^\dagger_\bi
\bar{b}^{\;}_\bj + \bar{b}^\dagger_\bj \bar{b}^{\;}_\bi) 
+ V \sum_{\langle \bi,\bj \rangle} (\bar{n}_\bi
-\frac{1}{2}) (\bar{n}_\bj-\frac{1}{2}) \ .
\label{version1}
\end{equation}
where $\langle \bi,\bj \rangle$, refers to nearest-neighbors in an
otherwise regular $d$-dimensional lattice.  Since $\bar{b}^\dagger_\bi$
and $\bar{b}^{\;}_\bi$ are not generators of a Lie algebra, the
eventual global symmetries of $H_{\sf xxz}$ remain hidden in this
particular language. However, if we {\it translate} $H_{\sf xxz}$ to
the second $SU(2)$-spin language using the dictionary provided by
Matsubara and Matsuda \cite{matsubara}
\begin{eqnarray}
S^{z}_{\bi}&=& {\bar n}_\bi- \frac{1}{2}\ ,
\nonumber \\
S^{+}_{\bi}&=& \bar{b}^\dagger_\bi \ ,
\nonumber \\
S^{-}_{\bi}&=& \bar{b}^{\;}_\bi  \ , 
\label{set2m}
\end{eqnarray}
we can immediately unveil the hidden symmetries of $H_{\sf xxz}$.
The well-known expression for $H_{\sf xxz}$ in terms of the $su(2)$
generators (i.e., the equivalent spin Hamiltonian) is
\begin{equation}
H_{\sf xxz}= \sum_{\langle \bi,\bj \rangle} J_z S^{z}_{\bi}S^{z}_{\bj}+ 
\frac{J_{\perp}}{2} 
(S^{+}_{\bi}S^{-}_{\bj}+S^{-}_{\bi}S^{+}_{\bj}).
\label{version2}
\end{equation} 
The magnetic couplings, $J_z$ and $J_{\perp}$, are related to the
original parameters, $t$ and $V$, by the relations: $J_z=V$ and
$J_{\perp}=2t$. It is clear from the last version of $H_{\sf xxz}$
(Eq.~(\ref{version2})) that the original model has a global $SU(2)$
invariance if $V=2t$, i.e., it is in the isotropic Heisenberg point.
The existence of this $SU(2)$-symmetric point has a very important
consequence for the phase diagram of the bosonic model of
Eq.~(\ref{version1}): If there is a charge density wave (CDW)
instability at that point, the $SU(2)$ invariance implies that {\it
there is also} a Bose-Einstein condensation and vice versa. The order
parameters of both phases are different components of a unique order
parameter in the spin language, i.e., the staggered magnetization of
the antiferromagnetic phase ($t>0$). The $z$-component of the staggered
magnetization is mapped onto the CDW order parameter for the bosonic
gas, while the transverse component is equivalent to the order
parameter for the Bose-Einstein condensation. Only one of these two
phases, which are coexisting at the $SU(2)$ invariant point, is stable
when we depart from the symmetric point in any of both directions in
parameter space  (Bose-Einstein condensation if $V<2t$ and CDW if
$V>2t$). In this very simple example  we can see the advantages of
using a hierarchical language ($su(2)$ in this case). In the first
place, we can immediately recognize the high symmetry points. Secondly,
we can describe an eventual broken symmetry state at those points in
terms of a unified order parameter \cite{commzhang}. If we were to use a
non-hierarchical language to describe the same problem, we would find
coexistence of more than one phase at the high symmetry points. The
order parameters of each of those phases are different components of
the unified order parameter that we simply found with the hierarchical
language. These ideas are developed in more detail in section
\ref{sec6}. Here the aim is to give a flavor of the potential
applications of our fundamental theorem. 

Using the concept of transmutation of statistics, to be introduced
in the next section, we will extend the notion of classes of bosonic
languages to more general classes containing non-bosonic languages
(i.e., languages for which Eq.~(\ref{bosl}) is replaced by a different
algebraic condition). The most well-known examples of non-bosonic
languages are the fermionic ones for which the commutator of
Eq.~(\ref{bosl}) is replaced by an anticommutator. We will see that
there are non-local transformations which map non-bosonic languages
into bosonic ones. In this way, the simultaneous application of these
transformations and the fundamental theorem provides the natural path
toward a unification of the different languages used to describe
quantum systems.


\section{Transmutation of Statistics}
\label{sec3}

In the previous section we have shown that two different bosonic
languages can be connected (i.e., there is a dictionary connecting the
two languages) if they have the same dimension of their local Hilbert
spaces $D$. However, we know that the bosonic languages do not exhaust
all the possible languages of quantum mechanics. The best known example
is the fermionic language, whose creation and annihilation operators
obey a closed set of {\it anticommutation} relations. There are many
other examples of non-bosonic languages that we address below. 

Is it possible to connect these non-bosonic languages to the bosonic
ones introduced in the previous section ? A positive answer to this
question is given in this section by introducing operators which
transform commutators into anticommutators and vice versa. These
operators have a local (related to the generalized Pauli exclusion
principle (see section \ref{sec4d})) and a non-local (related to the
exchange statistics) components. The local component is derived at the
beginning of this section, while the non-local one is introduced in the
second part. We will see that an Abelian {\it anyonic} statistics can
also be achieved by introducing a continuous parameter in these
transformations. (Non-Abelian statistics is beyond the scope of the
present manuscript.) The transmutation of statistics together with the
fundamental theorem establish the framework needed to classify the
classes of equivalent languages. This means that to describe a given
physical problem one can use the original language or any other
belonging to the same class. 

Before proceeding, we would like to mention that it seems as if there
are two unrelated notions of anyonic statistics in the literature. One
tied to quantum mechanics in first quantization in the coordinate
representation, and another derived within the framework of quantum
field theory. In both cases the original motivation to introduce such
particles was basically as an inherent possibility in the kinematics of
(2+1)-dimensional quantum mechanics and clearly the concepts, if
correctly implemented, should be equivalent but are not indeed
\cite{oursanyon}. In this manuscript, our anyon notion is consistent
with the one developed in quantum field theory where the exclusion
properties are preserved under {\it statistical transmutation}. In this
way, fermions can be kinematically transformed into hard-core bosons
but not into canonical ones.

\subsection{Fermionic Languages}

In addition to bosons, the other type of fundamental particles found in
Nature are the fermions. We have seen that the notion of bosonic
languages is closely related to the concept of Lie algebras, since for
each class of bosonic languages there is at least one language whose
elements are generators of a Lie algebra. However, the same cannot be
done for fermions. This can be easily seen by noticing that the third
condition for a bosonic language (see Eq.~(\ref{bosl})) is not valid in
general for fermions. In addition, the main consequence of the third
condition for a bosonic language is that the global Lie algebra is the
direct sum of the local ones associated to each subset $S_\bi$. The
generalization of these concepts to {\it fermionic languages} can be done by
introducing the notion of Lie superalgebras (see for instance Ref.
\cite{cornwell}). The fermionic languages are associated to Lie
superalgebras in the same way the bosonic languages are associated to
Lie algebras. Therefore, to give a general definition of a fermionic
language we should first introduce and explain the notions of Grassman
algebras, which are associative superalgebras, and Lie superalgebras.
Since this is beyond the scope of the present paper, we will only
consider the fermionic language generated by the canonical creation and
annihilation operators
\begin{eqnarray}
\begin{cases}
\{c^{\;}_{\bi\alpha},c^{\;}_{\bj\beta}\}=
\{c^\dagger_{\bi\alpha},c^\dagger_{\bj\beta}\}=0  
 \ ,\\
\{ c^{\;}_{\bi\alpha},c^\dagger_{\bj\beta}\}=\delta_{\bi\bj} 
\delta_{\alpha \beta}  \ ,
\end{cases}
\label{canoc}
\end{eqnarray}
and other languages obtained by imposing particular local constraints
on the canonical fermions. These generators, together with the
identity, generate the Heisenberg Lie superalgebra. In analogy to the
bosonic languages (see Eq.~(\ref{bosl})), the Lie product (in the
superalgebra) of two elements acting in different sites (different
subsets ${\cal S}_\bi$,${\cal S}_\bj$ ) is zero. Thus, instead of
having a direct sum of local Lie algebras like in the bosonic case, we
have a direct sum of local Lie superalgebras. In the case of canonical
fermions the local Lie superalgebras are Heisenberg superalgebras. 

\subsection{Local Transmutation}
\label{sec3b}

We will start by considering multiflavor fermions  $c^\dagger_{\bi
\alpha}$ ($\alpha \in [1,N_f]$) which satisfy the canonical 
anticommutation relations (Eq.~(\ref{canoc})). Other types of fermions,
usually considered in physics, can be derived from the
canonical ones by imposing particular constraints. For this reason, the
transformations derived for canonical fermions can be extended to these
other fermionic algebras by incorporating those constraints. This
procedure is illustrated with different examples in the next section. 

The canonical fermions can be transformed into bosons
$\tilde{b}^\dagger_{\bi\alpha}$ which are hard-core in each flavor (the
eigenvalues of $\tilde{n}_{\bi \alpha}= 
\tilde{b}^{\dagger}_{\bi\alpha} \tilde{b}^{\;}_{\bi\alpha}$ are either
$0$ or $1$), i.e., two or more bosons of the same flavor are not
allowed on the same site, but two of different flavors are permitted.
In the previous section we have shown that a physical theory for
objects obeying commutation relations (Lie brackets) can be formulated
in terms of a bosonic language. By the present connection we will be
able to extend this statement to fermions (or anyons, in general)
through a transmutation of statistics. To this end, we  need to define
a different type of multiflavored hard-core boson
$\tilde{b}^\dagger_{\bi\alpha}$ satisfying the following commutation
relations  
\begin{eqnarray}
\begin{cases}
[\tilde{b}^{\;}_{\bi\alpha},\tilde{b}^{\;}_{\bj\beta} ]=
[\tilde{b}^\dagger_{\bi\alpha},\tilde{b}^\dagger_{\bj\beta}]=0  \ , \\
  
[ \tilde{b}^{\;}_{\bi\alpha},\tilde{b}^\dagger_{\bj\beta} ]= 
\delta_{\bi\bj} \delta_{\alpha\beta}(1-2\tilde{n}_{\bi\alpha})
\; , \;
[\tilde{n}_{\bi\alpha}, \tilde{b}^\dagger_{\bj\beta} ]= \delta_{\bi\bj}
\delta_{\alpha\beta} \tilde{b}^\dagger_{\bi\alpha}  \ , 
\end{cases}
\label{bef}
\end{eqnarray}
which implies $\{\tilde{b}^{\;}_{\bi\alpha},
\tilde{b}^\dagger_{\bi\alpha} \}=1$. The Lie algebra generated by these
bosonic operators is ${\cal L}=\bigoplus_{\alpha,\bi} su(2)$, i.e. each
set $\{\tilde{b}^\dagger_{\bi\alpha},\tilde{b}^{\;}_{\bi\alpha},
\tilde{n}_{\bi\alpha}-1/2\}$ generates an $su(2)$ algebra. 

To show that these bosons (hard-core in each flavor) can be
expressed as a function of canonical fermions, one defines
\begin{equation}
c^\dagger_{\bj\alpha}=\tilde{b}^\dagger_{\bj\alpha} \hat{\cal
T}^{\dagger}_{\bj\alpha} \ ,
\label{rotloc}
\end{equation}
where $\displaystyle \hat{\cal T}^{\;}_{\bj\alpha}=\exp[i \pi
\sum_{\beta < \alpha} \tilde{n}_{\bj\beta}]$ is the {\it local
transmutator}, and we are assuming a particular ordering for the flavor
index $\alpha$. From the expression for $\hat{\cal T}^{\;}_{\bj\alpha}$
it is clear that
\begin{eqnarray}
{\hat{\cal T}^{2}_{\bj\alpha}}=I,\;\;\;
\hat{\cal T}^{\dagger}_{\bj\alpha}=\hat{\cal T}^{\;}_{\bj\alpha} \ .
\nonumber \\
\end{eqnarray}
It is straightforward to verify that the $c_\bj$-operators satisfy
local canonical anticommutation relations (Eq.~(\ref{canoc}) when
$\bi=\bj$). 

In this way we have established a mapping between fermions and bosons
which are operating locally (on a given orbital or mode $\bj$). In
other words, we have related the subset ${\cal S}_\bj=\{
\tilde{b}^\dagger_{\bj\alpha},\tilde{b}^{\;}_{\bj\alpha},
\tilde{n}_{\bj\alpha}-1/2 \}$ of local generators of a bosonic
language to the subset ${\hat{\cal S}}_\bj=\{{c}^\dagger_{\bj\alpha},
{c}^{\;}_{\bj\alpha},\hat{n}_{\bj\alpha}-1/2\}$ of local generators  of
canonical fermions ($\hat{n}_{\bj\alpha}={c}^\dagger_{\bj\alpha}
{c}^{\;}_{\bj\alpha}$).

\subsection{Non-Local Transmutation}

So far, we have only transmuted the commutation relations between
generators which belong to the same site or subset ${\cal S}_\bi$.  For
commutation relations of two generators of different sites we need to
introduce a non-local operator $K_{\bf j}$. Jordan and Wigner
\cite{jordan} were the first to introduce such an operator in
connection to their one-dimensional ($1d$ or $d=1$) transformation
between spins $S$=1/2 and spinless fermions. The so-called kink
operator that they introduced is
\begin{equation}
K^{1d}_{\bj} = \exp[i \pi \sum_{\substack{\bl<\bj}} 
\ \bar{n}_\bl] \ ,
\label{1d} 
\end{equation}
where $\bar{n}_\bl$ is the number operator for spinless fermions at the
site $\bl$. It is clear that for multiflavor canonical fermions we only
need to replace $\bar{n}_\bl$ by $\hat{n}_\bl=\sum_{\alpha}
\hat{n}_{\bl\alpha}$, where $\alpha$ denotes the flavor. Therefore,
Eq.~(\ref{rotloc}) must be replaced by
\begin{equation}
c^\dagger_{{\bf j}\alpha}=
\tilde{b}^\dagger_{{\bf j}\alpha} \hat{\cal T}^{\dagger}_{{\bf j}\alpha}
K_{\bj}^\dagger=\tilde{b}^\dagger_{{\bf j}\alpha} {\cal
K}_{\bj\alpha}^{\dagger},
\label{trans}
\end{equation}
where $K_{\bj}=K^{1d}_{\bj}$ for a $1d$ lattice. Even though this is a
non-local operator, it does not introduce long-range interactions if
the model has only short-range terms \cite{Noten5}. In other
words, a given $1d$ Hamiltonian can be written in terms of bosons,
fermions or anyons, and the interactions remain short range for all the
cases. This is a special characteristic of one dimension. The only
consequence of changing the statistics of the particles is a change of
the short-range interactions in the original basis. Therefore, the
concept of particle statistics in one dimension becomes irrelevant
since any physical system  can be described with a bosonic language
without changing the short-range character of the interactions
\cite{rpmbt}. 

\begin{figure}[htb] \hspace*{0.0cm}
\includegraphics[angle=0,width=7.4cm,scale=1.0]{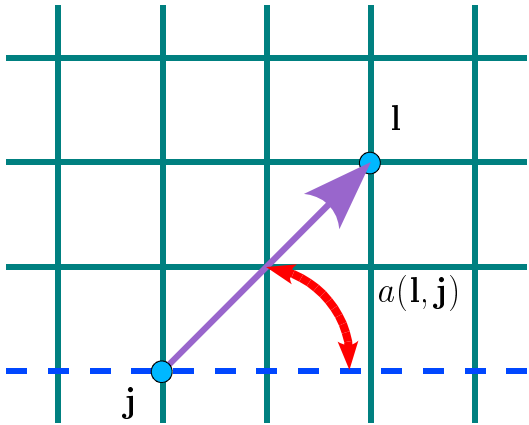}
\caption{Schematics of the geometry and function describing one of the
possible ways of defining the statistical transmutator $K^{2d}_{\bf j}$
in 2$d$. The dotted line represents a fixed direction on the lattice.}
\label{fig0}
\end{figure}

The next step is the generalization of $K_{\bj}$ to higher dimensions.
This has been done by Fradkin \cite{eduardo} and Y.R. Wang
\cite{wang2}, who considered the generalization of the traditional JW
transformation for the two-dimensional ($2d$) case, and Huerta-Zanelli
\cite{huerta} and S. Wang \cite{wang}, who did the same for higher
dimensions. We will see that, in the same way we did for the $1d$ case,
these generalizations can be extended to transform canonical fermions
into bosons and vice versa. The generalization given by Fradkin
\cite{eduardo} for the $2d$ case is
\begin{equation}
K^{2d}_{\bf j} = \exp[i \sum_{\substack{{\bf l}}} a({\bf l},{\bf j}) 
\ \bar{n}_{\bf l}] \ , 
\label{2d}
\end{equation}
Here, $a({\bf l},{\bf j})$ is the angle between the spatial vector
${\bf l}-{\bf j}$ and a fixed direction on the lattice, and $a({\bf
j},{\bf j})$ is defined to be zero (see Fig.~\ref{fig0}). Again for
the case of multiflavor canonical fermions we just need to replace
$\bar{n}_{\bf l}$ by $\hat{n}_{\bf l}$. We comment that the $1d$ kink
(or string) operators constitute a particular case of Eq.~(\ref{2d})
with $a(\bl,\bj)=\pi$ when $\bl<\bj$ and equals zero otherwise.


A general procedure for finding the string operators in $d$ dimension 
has been given by Wang \cite{wang} in the way we describe below.
Following the $1d$ and $2d$ cases, the expression proposed for the
operator $K_\bj$ is
\begin{equation}
K_{\bf j} = \exp[i \sum_{\substack{{\bf l}}} \omega({\bf l},{\bf j}) 
\ \bar{n}_{\bf l}] \ ,  
\end{equation}
where $\omega({\bf l},{\bf j})$ is a function to be determined by
imposing the transmutation of statistics. It can be shown that
this is equivalent to the antisymmetric condition 
\begin{eqnarray}
\exp[i\omega({\bf l},{\bf j})]&=&-\exp[i\omega({\bf j},{\bf l})], \ 
\mbox{if } \bl \ne \bj
\nonumber \\ \omega({\bf l},{\bf l})&=& 0.
\label{condition}
\end{eqnarray}
The simplest solution for a $1d$ lattice is
\begin{equation}
\omega(\bl,\bj)=\pi \Theta(\bj-\bl)
\end{equation}
where $\Theta(x)$ is the $1d$ step or Heaviside function. This is the
solution found by Jordan and Wigner (see Eq.~(\ref{1d})). The solution
introduced in Ref. \cite{eduardo} for the 2$d$ case corresponds to 
\begin{equation}
\omega({\bl,\bj})=a({\bf l},{\bf j}) \ .
\label{arg}
\end{equation}
However, it has been pointed out by Wang \cite{wang} that this is 
not the only possible solution since
\begin{equation}
\omega(\bl,\bj)=\pi [\Theta(j_1-l_1) (1-\delta_{l_1j_1})+\Theta(j_2-l_2)
\delta_{l_1j_1}], 
\end{equation}
with $\bl=l_1{\bf e}_1+l_2{\bf e}_2$ and $\bj=j_1{\bf e}_1+j_2{\bf
e}_2$, also satisfies Eqs.~(\ref{condition}). The advantage of this
solution is that its generalization to higher dimensions is
straightforward ($\bl=\sum_{\mu} l_\mu {\bf e}_\mu$, and $\mu \in [1,d]$). 
For instance, in 3$d$ we have
\begin{eqnarray}
\omega(\bl,\bj)&=& \pi [\Theta(j_1-l_1) (1-\delta_{l_1j_1}) 
\nonumber \\
&+& \Theta(j_2-l_2)\delta_{l_1j_1} (1-\delta_{l_2j_2})
\nonumber \\
&+& \Theta(j_3-l_3)\delta_{l_1j_1}\delta_{l_2j_2}] .
\label{theta}
\end{eqnarray}
In the next section we will see that, for $d >1$, the operator $K_{\bf
j}$ introduces non-local interactions in the new representation of the
original Hamiltonian. A phase factor of the form $\exp[i \int_{\bf
x}^{\bf y} \bold{A}\cdot d{\bold s}]$ appears in between the product
$c^{\dagger}_{\bf x} c^{\;}_{\bf y}$ of the kinetic energy-like terms.
The field operator ${\bf A (x)}$ is defined by
\begin{equation}
{\bf A (x)}=\sum_{\bf l} \nabla \omega({\bf x},\bl) 
\; \bar{n}_\bl \ ,
\label{def}
\end{equation}
where ${\nabla}$ represents the lattice gradient ($\mu,\nu\in 
[1,d]$)
\begin{equation}
{\nabla}_\mu \omega({\bf x,l})=\omega({\bf x+e_\mu,l})-
\omega({\bf x,l}).
\end{equation}
It was shown \cite{eduardo} that
${\bf A (x)}$ is the vector potential of a generalized Chern-Simons
construction for a lattice. The field strength $F_{\mu \nu}$ 
associated to this vector potential is the lattice rotor
\begin{eqnarray}
F_{\mu\nu}({\bf x})&=&\nabla_\mu A_\nu({\bf x})-\nabla_\nu A_\mu({\bf
x}) ,
\nonumber \\
F_{0\mu}({\bf x})&=&\nabla_0 A_\mu({\bf x})-\nabla_\mu A_0({\bf x}),
\end{eqnarray}
where $\nabla_0$ is the time derivative and $A_0({\bf x})$ is a scalar
field. Since $A_0$ is a Lagrange multiplier field for the Chern-Simons
Lagrangian, it can be integrated out to get the {\it Gauss law}
\begin{equation}
\bar{n}_\bl=\frac{1}{\pi} F_{12}(\bl) \ .
\label{source}
\end{equation}
This relation imposes the constraint between charge and flux  giving
rise to the vector potential ${\bf A}$. The physical interpretation is
that there is a flux attached to each particle. In this way the phase
associated to the original particle statistics is now generated by a
mechanism based on the Aharonov-Bohm effect. It can be seen that the
$\omega$ function introduced by Fradkin \cite{eduardo}
(Eq.~(\ref{arg})) generates a vector potential ${\bf A}$ which is
solution of Eq.~(\ref{source}). On the other hand, the $\omega$
function proposed by Wang \cite{wang} cannot be associated to any flux
since the total change of $\omega$ on any closed loop vanishes (or
equivalently $\oint {\bf A}.d{\bf s}=0$). This result suggests that the
vector potential can be eliminated by a gauge transformation. However,
it is impossible to find a gauge transformation which does not change
the commutation relations of the particles. The advantage of the
solution given by Wang \cite{wang} is the straightforward
generalization to higher spatial dimensions. Fradkin's approach ($2d$)
\cite{eduardo} provides an alternative formulation of the original
problem and has stimulated original methods to find relevant solutions
in quantum Hall systems. The generalization of his approach to 3$d$
\cite{huerta} is more involved since it requires an extended Hilbert
space and non-Abelian gauge transformations. 

Summarizing, we have shown that one can write down the generators of
the canonical fermionic algebra as an operator function of hard-core 
bosons and then as a function of the generators of any Lie group having
an irreducible representation of dimension given by the dimension $D$
of the {\it local} Hilbert space. If we are dealing with canonical
fermions, the dimension $D$ is equal to $2^{N_f}$. However, we can
obtain other values of $D$ by imposing different constraints on the
fermionic  occupation numbers like in the bosonic case \cite{ours1}
(see next Section). This is a common situation for strongly correlated
problems where part of the Hilbert space is eliminated when deriving
effective low-energy theories. For instance, the Hilbert space of the
effective low-energy theory for the Hubbard model ($t$-$J$ model) has
the constraint of no double-occupancy and therefore $D$ is equal to
$3$. 

\subsection{Anyons}

Similarly, one can extend this idea of transmutation of statistics to
particles satisfying general equal-time anyonic canonical commutation
relations defined by an angle $\theta$. To this end we need to
generalize the transmutators to any statistical angle
$0\leq\theta\leq\pi$
\begin{eqnarray}
\hat{\cal T}^{\theta}_{\bj\alpha}&=&\exp[i 
\theta \sum_{\beta < \alpha} {n}_{\bj\beta}], 
\nonumber \\
K^{\theta}_{\bf j}&=&\exp[i \frac{\theta}{\pi} \sum_{\substack{{\bf
l}}} \omega({\bf l},{\bf j})  \ {n}_{\bf l}]
\end{eqnarray}
We have seen that there are many types of bosonic particles. For each
type we will get a different type of anyon after statistical
transmutation. For instance, if we start from canonical bosons
($n_{\bi\alpha}=b^\dagger_{\bi\alpha}b^{\;}_{\bi\alpha}$)
\begin{eqnarray}
\begin{cases}
[ b^{\;}_{\bi\alpha},b^{\;}_{\bj\beta} ] = 
[ b^\dagger_{\bi\alpha},b^\dagger_{\bj\beta} ]=0  
 \ , \\
{[}b^{\;}_{\bi\alpha},b^\dagger_{\bj\beta}{]}= \delta_{\bi\bj} 
\delta_{\alpha\beta} \; , \;
[ n_{\bi\alpha}, b^\dagger_{\bj\beta} ]=
\delta_{\bi\bj} \delta_{\alpha\beta}b^\dagger_{\bi\alpha}  \ , 
\end{cases}
\label{canob}
\end{eqnarray}
by transmutation of the statistics
\begin{equation}
a^\dagger_{\bj\alpha}= b^\dagger_{\bj\alpha} 
(\hat{\cal T}^{\theta}_{\bj\alpha})^\dagger (K^{\theta}_{\bf j})^\dagger=
b^\dagger_{\bj\alpha}({\cal K}^{\theta}_{\bj\alpha})^\dagger,\ 
\label{atr}
\end{equation} 
one gets anyons obeying commutation relations 
\begin{eqnarray}
\begin{cases}
[ a^{\;}_{\bj\alpha},a^{\;}_{\bj\alpha} ] = 
[ a^\dagger_{\bj\alpha},a^\dagger_{\bj\alpha}]=0  
 \ , \\
{[}a^{\;}_{\bj\alpha},a^\dagger_{\bj\alpha}{]}= 1 \ ,  
\end{cases}
\end{eqnarray}
and {\it deformed} commutation relations if both anyonic operators
correspond to different sites or flavors. To write down those
commutation relations we need first to define a particular ordering for
the combined site and flavor indices \cite{ordering}. In this way, for
$(\bj,\beta) > (\bi,\alpha)$
\begin{eqnarray}
\begin{cases}
[ a^{\;}_{\bi\alpha},a^{\;}_{\bj\beta} ]_\theta = 
[ a^\dagger_{\bi\alpha},a^\dagger_{\bj\beta} ]_\theta=0  
 \ , \\
{[}a^{\dagger}_{\bj\beta},a^{\;}_{\bi\alpha}{]}_\theta=0
\ .
\end{cases}
\label{anyb}
\end{eqnarray}
The requirement of an ordering for the indices comes from the fact that
$[A,B]_\theta=-\exp[i\theta] [B,A]_{-\theta}$. So unless
$\exp[i\theta]$ is a real number (i.e., bosons ($\theta=0$) or fermions
($\theta=\pi$)), we need to define a particular index ordering. It is
in this ordering that the intrinsic many-body character of the these
particles is encoded. In this simple example we can see that the Pauli
exclusion principle and the phase associated to the exchange of two
particles with different indices (flavor or site), are two distinct and
independent concepts. The exclusion behavior is determined by the local
commutation relations between operators associated to the same site and
flavor. Since these commutation relations are not changed by the
transmutation, Eq.~(\ref{atr}), the exclusion properties are preserved
when we change the statistics of the particles. In our example we can
create any number of anyons in the same orbital and flavor because the
local commutation relations are the same as the ones for canonical
bosons. However, the deformed commutation relations of Eq.~(\ref{anyb})
indicate that the result of  exchanging two anyons with different
indices is the multiplication by a phase factor $\exp[i\theta]$. It is
this second aspect, not related to the exclusion properties, 
that decides whether the particles are bosons, fermions, or anyons.

Now we will consider another example of transmutation from bosons to
anyons. In this case, we will take the multiflavored hard-core bosons
defined by the commutation relations Eq.~(\ref{bef}). The particles 
obtained after the transmutation of the hard-core bosons will be called
{\it type I hard-core anyons}
\begin{equation}
\tilde{a}^\dagger_{\bj\alpha}=\tilde{b}^\dagger_{\bj\alpha} (\hat{\cal
T}^{\theta}_{\bj\alpha})^\dagger (K^{\theta}_{\bf j})^\dagger=
\tilde{b}^\dagger_{\bj\alpha}({\cal K}^{\theta}_{\bj\alpha})^\dagger\ ,
\end{equation} 
with
\begin{eqnarray}
\hat{\cal T}^{\theta}_{\bj\alpha}&=&\exp[i 
\theta \sum_{\beta < \alpha} \tilde{n}_{\bj\beta}], 
\nonumber \\
K^{\theta}_{\bf j}&=&\exp[i \frac{\theta}{\pi} \sum_{\substack{{\bf
l}}} \omega({\bf l},{\bf j})  \ \tilde{n}_{\bf l}] \ .
\end{eqnarray}
Like in the previous example, the local commutation relations 
are preserved ($\tilde{n}_{\bj\alpha}=\tilde{a}^{\dagger}_{\bj\alpha}
\tilde{a}^{\;}_{\bj\alpha}$, $\tilde{n}_\bj=\sum_{\alpha=1}^{N_f} 
\tilde{n}_{\bj\alpha}$)
\begin{eqnarray}
\begin{cases}
[ \tilde{a}^{\;}_{\bj\alpha},\tilde{a}^{\;}_{\bj\alpha} ] = 
[ \tilde{a}^\dagger_{\bj\alpha},\tilde{a}^\dagger_{\bj\alpha}]=0  
 \ , \\
{[}\tilde{a}^{\;}_{\bj\alpha},\tilde{a}^\dagger_{\bj\alpha}{]}=1-
2\tilde{n}_{\bj\alpha} \ .  
\end{cases}
\end{eqnarray}
In this particular case, since there is a hard-core condition
$\tilde{a}^\dagger_{\bj\alpha}\tilde{a}^\dagger_{\bj\alpha}=0$, the
operators also satisfy the following local anticommutation relations
\begin{eqnarray}
\begin{cases}
\{ \tilde{a}^{\;}_{\bj\alpha},\tilde{a}^{\;}_{\bj\alpha} \} 
= \{ \tilde{a}^\dagger_{\bj\alpha},\tilde{a}^\dagger_{\bj\alpha} \} =0  
 \ , \\
 \{ \tilde{a}^{\;}_{\bj\alpha},\tilde{a}^\dagger_{\bj\alpha} \}= 1 \ .  
\end{cases}
\label{lcf}
\end{eqnarray}
Thus, the local anticommutation relations are also preserved under 
statistical transmutation. Clearly, Eqs.~(\ref{lcf}) are the local
anticommutation relations for canonical fermions. This is not
surprising since the multiflavored hard-core bosons defined by
Eq.~(\ref{bef}) can be transmuted into canonical fermions (see
Eq.~(\ref{trans})). For the commutation relations involving  operators
with different indices, we have to define an index ordering like in the
previous example. For $(\bj,\beta) > (\bi,\alpha)$
\begin{eqnarray}
\begin{cases}
[ \tilde{a}^{\;}_{\bi\alpha},\tilde{a}^{\;}_{\bj\beta} ]_\theta = 
[ \tilde{a}^\dagger_{\bi\alpha},\tilde{a}^\dagger_{\bj\beta} ]_\theta=0  
 \ , \\
{[}\tilde{a}^{\dagger}_{\bj\beta}, \tilde{a}^{\;}_{\bi\alpha}{]}_\theta=0
\ .
\end{cases}
\label{hca}
\end{eqnarray}
$\theta=0 $ corresponds to bosons which are hard-core in each flavor
(see Eq.~(\ref{bef})), and $\theta=\pi $ to canonical fermions, i.e., 
$\tilde{a}^\dagger_{\bj\alpha}(\theta=0)={\tilde
b}^\dagger_{\bj\alpha}$ and 
$\tilde{a}^\dagger_{\bj\alpha}(\theta=\pi)=c^\dagger_{\bj\alpha}$. For
any statistical angle $\theta$ one can put up to a single particle per
mode and flavor (i.e, the particles are {\it hard-core} in each flavor and,
therefore, satisfy the Pauli exclusion principle). It is evident in
general that the transmutation of the statistics does not change the
local commutation relations and therefore the exclusion rule. In a
later section we will show how to generalize the exclusion principle
to accommodate up to $p$ particles per single quantum state.  In
particular, we will see that the para-particles, which obey this
generalized exclusion principle,  can be obtained from particles which
are partial transmutations of canonical bosons (to obtain the
parabosons) or fermions (to obtain the parafermions). 

The last example of anyonic particles corresponds to the transmutation
of the hard-core bosons defined by the commutation relations
Eq.~(\ref{conm}). This concept will be useful for the generalized 
JW transformations \cite{ours1} that we introduce in the
next section (these particles will be called {\it type II hard-core
anyons} or simply {\it JW particles})
\begin{equation}
\bar{a}^\dagger_{\bj\alpha}=\bar{b}^\dagger_{\bj\alpha} (\hat{\cal
T}^{\theta}_{\bj\alpha})^\dagger (K^{\theta}_{\bf j})^\dagger=
\bar{b}^\dagger_{\bj\alpha}({\cal K}^{\theta}_{\bj\alpha})^\dagger\ ,
\label{amon}
\end{equation} 
with
\begin{eqnarray}
\hat{\cal T}^{\theta}_{\bj\alpha}&=&\exp[i 
\theta \sum_{\beta < \alpha} \bar{n}_{\bj\beta}], 
\nonumber \\
K^{\theta}_{\bf j}&=&\exp[i \frac{\theta}{\pi} \sum_{\substack{{\bf
l}}} \omega({\bf l},{\bf j})  \ \bar{n}_{\bf l}] \ .
\end{eqnarray}
Since the local commutation relations are preserved, we have
($\bar{n}_{\bj\alpha}=\bar{a}^{\dagger}_{\bj\alpha}
\bar{a}^{\;}_{\bj\alpha}$, $\bar{n}_\bj=\sum_{\alpha=1}^{N_f} 
\bar{n}_{\bj\alpha}$)
\begin{eqnarray}
\begin{cases}
[ \bar{a}^{\;}_{\bj\alpha},\bar{a}^{\;}_{\bj\alpha} ] = 
[ \bar{a}^\dagger_{\bj\alpha},\bar{a}^\dagger_{\bj\alpha}]=0  
 \ , \\
{[}\bar{a}^{\;}_{\bj\alpha},\bar{a}^\dagger_{\bj\alpha}{]}=1-
\bar{n}_{\bj\alpha} - \bar{n}_{\bj} \ .  
\end{cases}
\label{canoa1}
\end{eqnarray}
Again, we need to define an index ordering for the deformed commutation 
relations involving operators with different indices.
For $(\bj,\beta) > (\bi,\alpha)$
\begin{eqnarray}
\begin{cases}
[ \bar{a}^{\;}_{\bi\alpha},\bar{a}^{\;}_{\bj\beta} ]_\theta = 
[ \bar{a}^\dagger_{\bi\alpha},\bar{a}^\dagger_{\bj\beta} ]_\theta=0  
 \ , \\
{[}\bar{a}^{\dagger}_{\bj\beta}, \bar{a}^{\;}_{\bi\alpha}{]}_\theta=0
\ .
\end{cases}
\label{hca2}
\end{eqnarray}

Notice that, formally, in all cases we could have considered different
angles $\theta$ instead of a single one in the expressions for the
local, $\hat{\cal T}^{\theta}_{\bj\alpha}$ (e.g., $\theta_1$), and
non-local, $K^{\theta}_{\bf j}$ (e.g., $\theta_2$), transmutators,
although we do not see how physically relevant this general situation
could be.

It is important to stress at this point when the exchange statistics 
property of the particles (i.e., the property attached to the non-local
part of the transmutator) becomes relevant in the description of a
particular physical system. It is clear that whenever the system
Hamiltonian does not permute particles on different sites $\bi$ and
$\bj$, then $K_\bj^\theta$ becomes a symmetry and the Hamiltonian is
invariant under particle exchange statistics. This is the case in many
1$d$ problems (e.g., the $XXZ$ model of Eq. (\ref{version1}) or the
Hubbard model) and in special 2$d$ problems such as the $U(1)$ gauge
magnet that we expand on in section \ref{sec7a}.

\begin{figure}[htb] \hspace*{0.0cm}
\includegraphics[angle=0,width=8.2cm,scale=1.0]{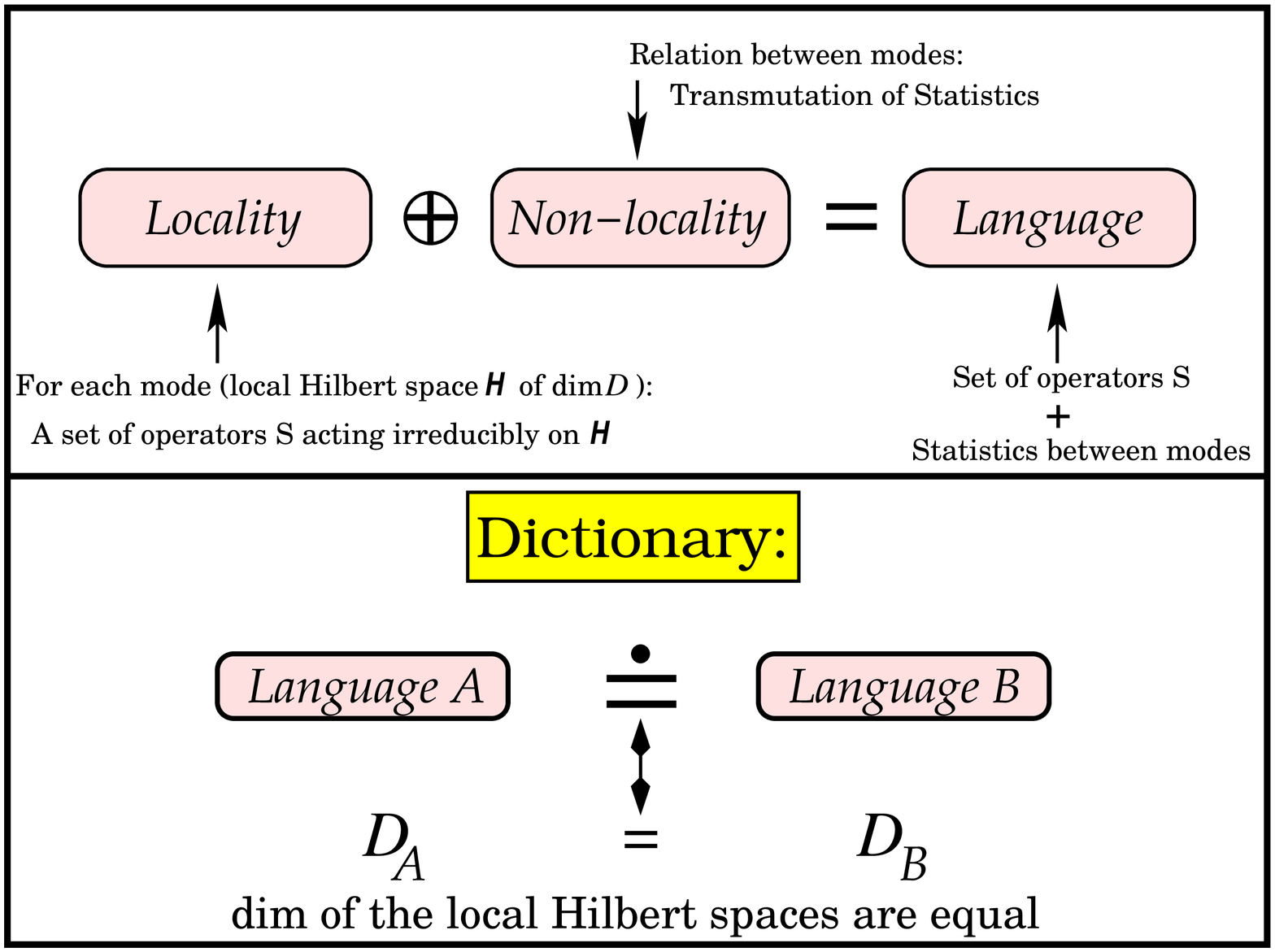}
\caption{What is a language, and when can two languages be connected? 
Summary of the main content of our fundamental theorem in
conjunction with the concept of transmutation of statistics.
}
\label{fig01}
\end{figure}

Figure \ref{fig01} summarizes the fundamental notions of languages and 
dictionaries connecting them, concepts developed in sections \ref{sec2}
and \ref{sec3}. 

\section{Hilbert Space Decomposition}
\label{sec3n}

We have seen in the previous section that the only condition to
establish an isomorphism between two different languages is that the
dimension of their local Hilbert spaces be equal (see
Fig.~\ref{fig01}). Notice, however, that the choice of local Hilbert
space ${\cal H}_\bi$ depends upon the particular decomposition of the
global Hilbert space ${\cal H}$. When we work on a lattice, the most
natural decomposition is the one where ${\cal H}_\bi$ is the Hilbert
space associated to each site (or mode). This is the decomposition that
we have adopted in our previous examples. Another possible
decomposition could be the one for which ${\cal H}_\bi$ corresponds to
the Hilbert space of a bond (i.e., two sites instead of one). In
general, there is no restriction in the way one can partition ${\cal
H}$. 

\begin{figure}[htb]
\includegraphics[angle=0,width=8.6cm,scale=1.0]{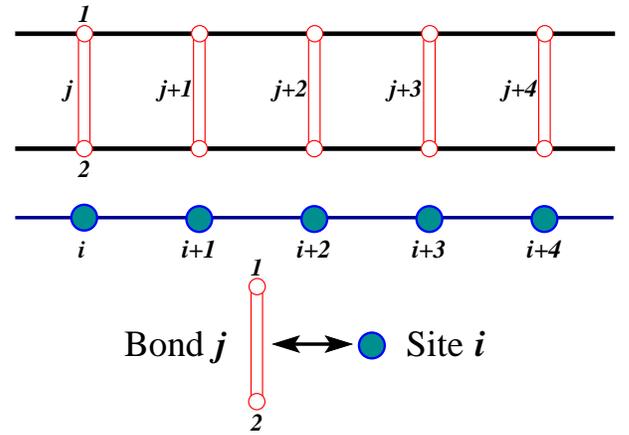}
\caption{Mapping between an $S$=1/2 ladder and a fermionic linear
chain. In this case, bonds (rungs) are mapped onto sites.}
\label{ladder}
\end{figure}

To be more concrete let us consider a $S$=1/2 ladder system (see
Fig.~\ref{ladder}). The global Hilbert space of the system can be
decompose into the direct product of the local Hilbert spaces on each
rung. By doing so, the dimension of ${\cal H}_\bi$ is enlarged from
$D=2$, for the sites decomposition, to $D=4$. According to our previous
results, the change of $D$ opens up the possibility of new mappings
between languages. For instance, $D=4$ is the dimension of the on site
Hilbert space for canonical $S$=1/2 fermions on a lattice. Therefore,
the $S$=1/2 spin ladder can be mapped onto a system of electrons on a
linear chain (see Fig.~\ref{ladder}). To be more explicit, a possible
isomorphic mapping is (see Eq. (\ref{mn}) for a different mapping)
\begin{eqnarray}
c^{\dagger}_{\bj \uparrow}&=& \sqrt{2} \ S^{z}_{\bj 2} (S^{+}_{\bj
1}+S^{+}_{\bj 2}) K^{\dagger}_{\bj} \ , \nonumber \\
c^{\dagger}_{\bj \downarrow}&=& \sqrt{2} \ (S^{-}_{\bj 2}-S^{-}_{\bj 1})
S^{z}_{\bj 2} K^{\dagger}_{\bj} \ .
\label{sdd}
\end{eqnarray}
Eq.~[\ref{sdd}] can be inverted to obtain the expression for the spin
operators as a function of the fermionic creation and annihilation
operators 
\begin{eqnarray} 
\begin{cases}
S^{+}_{\bj 1} = \frac {1}{\sqrt{2}} [(1-2 \hat{n}_{\bj \downarrow})
c^{\dagger}_{\bj \uparrow} K_{\bj} +  (1- 2 \hat{n}_{\bj \uparrow})
K^{\dagger}_{\bj}c^{\;}_{\bj \downarrow} ] \ , \nonumber \\
S^{-}_{\bj 1} = \frac {1}{\sqrt{2}} [K^{\dagger}_{\bj} c^{\;}_{\bj
\uparrow}(1-2\hat{n}_{\bj\downarrow}) +  c^{\dagger}_{\bj \downarrow}
K^{\;}_{\bj} (1-2\hat{n}_{\bj\uparrow}) ] \ , \nonumber \\
S^{z}_{\bj 1} = \frac {1}{2} [\hat{n}_{\bj \uparrow} - \hat{n}_{\bj
\downarrow} +  c^{\;}_{\bj \downarrow} c^{\;}_{\bj \uparrow} +
c^{\dagger}_{\bj \uparrow} c^{\dagger}_{\bj \downarrow} ] \ , 
\nonumber \\
\end{cases}
\end{eqnarray}
\begin{eqnarray}
\begin{cases}
S^{+}_{\bj 2} = \frac {1}{\sqrt{2}} [c^{\dagger}_{\bj \uparrow} K_{\bj}
+  K^{\dagger}_{\bj}c^{\;}_{\bj \downarrow} ] \ , \\
S^{-}_{\bj 2} = \frac {1}{\sqrt{2}} [K^{\dagger}_{\bj} c^{\;}_{\bj
\uparrow} +  c^{\dagger}_{\bj \downarrow} K^{\;}_{\bj}] \ , \\
S^{z}_{\bj 2} = \frac {1}{2} [\hat{n}_{\bj \uparrow} - \hat{n}_{\bj
\downarrow} -  c^{\;}_{\bj \downarrow} c^{\;}_{\bj \uparrow} -
c^{\dagger}_{\bj \uparrow} c^{\dagger}_{\bj \downarrow} ] \ .
\end{cases}
\label{sdd3}
\end{eqnarray}
In this way we have mapped spin operators acting on a given bond onto
fermionic operators acting on the corresponding site (see
Fig.~\ref{ladder}). This dictionary will be exploited in section
\ref{sec5} to map ladder $S$=1/2 magnets onto 1$d$ Hubbard-like
models. 

\section{Bridging the Languages of Quantum Mechanics}
\label{sec4}

The purpose of this section is to illustrate with examples the
algebraic framework developed in the previous three sections. In the
first part we describe different $SU(N)$ and $SO(N)$ spin-particle mappings which
are a direct consequence of our fundamental theorem. In these cases we
are connecting bosonic languages and, therefore, it is not necessary to
transmute the statistics. In contrast, in the second part we describe
the generalization of the JW transformation where the transmutation of
the statistics plays a fundamental role. In this case, we connect the
different representations of $SU(2)$-spins to constrained fermions
\cite{ours1}. We devote the third part of this section to show how the
fractional exclusion statistics algebras emerge from the present
formalism in a natural way. We close the section showing that the
notion of parastatistics introduced by Green \cite{green1} is
associated to partial transmutations of the canonical bosonic and
fermionic languages. In section \ref{sec7} we connect the gauge-field
to spin languages through the quantum link model relation. In this way
one envisions formal relations between all the different languages used
to describe the properties of quantum systems. In the course of this
section we will see that many well-known transformations are particular
examples of applications of both the fundamental theorem and the
transmutation of statistics.

\subsection{$SU(N)$ spin-particle mappings}
\label{sec4a}

\subsubsection{Schwinger-Wigner bosons}

The connection between the Schwinger-Wigner bosons and the generators
of $su(N)$ in different irreducible representations results from a
simple application of the fundamental theorem. We will consider a
system of Schwinger-Wigner bosons with $N_f=N$ different flavors
($\alpha \in [1,N]$). These bosons obey canonical commutation relations
\begin{eqnarray}
\begin{cases}
[ \hat{b}^{\;}_{\bi\alpha},\hat{b}^{\;}_{\bj\beta} ] = 
[ \hat{b}^\dagger_{\bi\alpha},\hat{b}^\dagger_{\bj\beta} ]=0  
 \ , \\
{[}\hat{b}^{\;}_{\bi\alpha},\hat{b}^\dagger_{\bj\beta}{]}= \delta_{\bi\bj} 
\delta_{\alpha\beta}  \ , 
\end{cases}
\label{crsw}
\end{eqnarray}
and are characterized by the constraint
\begin{equation}
\sum_{\alpha=1}^{N}\hat{b}^\dagger_{\bj\alpha}\hat{b}^{\;}_{\bj\alpha}=
M \ ,
\label{defswb}
\end{equation} 
which sets the dimension of the local Hilbert space ${\cal H}_\bj$, 
$D=\binom {N+M-1}{M}$. For $N=2$ we get $D=M+1$, which means that we
can get any value of $D$ by varying $M$. We also know that the Lie
algebra $su(2)$ has one irreducible  representation for each value of
$D$. In this case $D=2S+1$, where $S$ is the spin of the
representation. According to the fundamental theorem, if $2S+1=M+1$
we can write down the generators of $su(2)$ ($S^x,S^y,S^z$) in the
$S=M/2$ representation as a polynomial function of the
Schwinger-Wigner bosons and vice versa. The explicit form of these
relations is the well-known connection between $SU(2)$-spins and
the two-flavor Schwinger-Wigner bosons ($\alpha=1,2$)
\begin{eqnarray}
S^{z}_{\bj}&=& \frac{1}{2} (\hat{b}^\dagger_{\bj1}\hat{b}^{\;}_{\bj1}
-\hat{b}^\dagger_{\bj2}\hat{b}^{\;}_{\bj2}) ,
\nonumber \\
S^{+}_{\bj}&=& \hat{b}^\dagger_{\bj1} \hat{b}^{\;}_{\bj2} , 
\nonumber \\
S^{-}_{\bj}&=& \hat{b}^\dagger_{\bj2} \hat{b}^{\;}_{\bj1}  .
\label{set2sw}
\end{eqnarray}
According to the fundamental theorem the inverse transformation also
exists. The creation and annihilation operators for Schwinger-Wigner
bosons $\hat{b}^\dagger_{\bi\alpha},\hat{b}^{\;}_{\bi\alpha}$ can be 
written as a polynomial function of $S^x_\bj,S^y_\bj,S^z_\bj$. However,
since the Schwinger-Wigner bosons are not directly associated to the
description of any physical system the inverse transformation is,
apparently, not very useful. For this reason, the only application of
the Schwinger-Wigner bosons is to provide a different framework for
solving spin Hamiltonians. For instance, it is possible to
approximately solve the isotropic Heisenberg model by a mean-field
approximation which preserves the $SU(2)$ invariance of the model. This
symmetry is violated in the usual spin-wave approximation where the
spins are described by Holstein-Primakoff bosons. Through this simple
example we  see yet another important advantage of having different
languages to  describe the same problem: {\it Different languages
provide different frameworks to obtain approximate or exact solutions.}
Physics provides innumerable examples of applications of this
statement. It is not our intention to enumerate all of them in this
paper, but to show the operative way to build the dictionaries
connecting all of those different languages. 

To find the connection between the generators of the $su(N)$ algebra
and the Schwinger-Wigner bosons we just need to consider the general
case of arbitrary number of flavors. To simplify the analysis we can
start by considering the minimum value of $M$, i.e., $M=1$ (i.e., the
fundamental representation). In this case the dimension of the local
Hilbert space is $D=\binom {N}{1}=N$. Since $N$ is also the dimension
of the fundamental representation of $su(N)$, the fundamental theorem
states that the generators of $su(N)$ can be written as a function of
the Schwinger-Wigner bosons. Again, this connection is very well-known
and can be explicitly written as
\begin{equation}
{\cal S}^{\alpha\beta}(\bj)=\hat{b}^\dagger_{\bj\alpha}
\hat{b}^{\;}_{\bj\beta} \ .
\label{sw}
\end{equation}
Using Eqs. (\ref{crsw}) and (\ref{defswb}) we can verify that the 
${\cal S}^{\alpha \beta}$'s are the generators of the $su(N)$ algebra in 
the fundamental representation, i.e., they satisfy the following 
commutation relations
\begin{equation}
[{\cal S}^{\alpha \alpha'}(\bi),{\cal S}^{\beta \beta'}(\bj)]=
\delta_{\bi\bj}[
\delta_{\alpha' \beta} {\cal S}^{\alpha \beta'}(\bj)-\delta_{\alpha \beta'} 
{\cal S}^{\beta \alpha'}(\bj)] ,
\label{conm2}
\end{equation}
and operate in an $N$-dimensional vector space. As in the $N=2$ case,
by considering larger values of $M$ we obtain higher-order
representations of $su(N)$. The relation between the $su(N)$ generators
and the Schwinger-Wigner bosons is the one given by Eq.~(\ref{sw})
independently of the representation.

\subsubsection{Fundamental irrep of $su(N)$ and Jordan-Wigner particles}
 
The fundamental (quark) representation of $su(N)$ can be mapped onto an
algebra of constrained fermions ($\bar{c}^\dagger_{\bj\alpha}=
\bar{a}^\dagger_{\bj\alpha}(\theta=\pi)$) or hard-core bosons 
($\bar{b}^\dagger_{\bj\alpha}=\bar{a}^\dagger_{\bj\alpha}(\theta=0)$) 
with $N_f=N-1$ flavors 
\begin{eqnarray}
{\cal S}^{\alpha\beta}(\bj)&=&{\bar{a}^{\dagger}_{\bj\alpha} 
\bar{a}^{\;}_{\bj\beta}-\frac{\delta_{\alpha\beta}}{N}} 
\nonumber \\
{\cal S}^{\alpha 0}(\bj)&=&\bar{a}^{\dagger}_{\bj\alpha} K_\bj^\theta
\;,\;\;  {\cal S}^{0 \beta}(\bj)= (K_\bj^\theta)^{\dagger}
\bar{a}^{\;}_{\bj\beta} \nonumber \\
{\cal S}^{0 0}(\bj)&=&\frac{N_f}{N}-\sum^{N_f}_{\alpha=1} 
\bar{n}_{\bj\alpha}= -\sum^{N_f}_{\alpha=1} {\cal
S}^{\alpha\alpha}({\bf j}) \ ,
\label{fund}
\end{eqnarray}
where $1\leq \alpha,\beta \leq N_f$ runs over the set of particle
flavors, and $\bar{a}^{\dagger}_{\bj\alpha}=
\tilde{a}^{\dagger}_{\bj\alpha} \prod^{N_f}_{\beta=1}
(1-\tilde{n}_{\bj\beta})$.  ${\cal S}^{\mu\nu}(\bj)$ (with $0\leq
\mu,\nu \leq N_f$) are the components of the $SU(N)$-spin (i.e., there
are $N^2-1$ linear independent components). It is easy to verify that
these are generators of an $su(n)$ Lie algebra satisfying the
commutation relations 
\begin{equation}
[{\cal S}^{\mu \mu'}(\bj),{\cal S}^{\nu \nu'}(\bj)]=
\delta_{\mu'\nu} {\cal S}^{\mu \nu'}(\bj)-\delta_{\mu \nu'} 
{\cal S}^{\nu \mu'}(\bj). 
\label{conm3}
\end{equation}
For instance, for $N=3$ we have \cite{cmt24} ($\alpha=1,2$)
\begin{equation}
{\cal S}(\bj)= \begin{pmatrix}
\frac{2}{3}-\bar{n}_{\bj} &(K_\bj^\theta)^{\dagger} \bar{a}^{\;}_{\bj1}& 
(K_\bj^\theta)^{\dagger}\bar{a}^{\;}_{\bj2}\\  
\bar{a}^\dagger_{\bj1} K_\bj^\theta &\bar{n}_{\bj1}-\frac{1}{3}
& \bar{a}^\dagger_{\bj1} \bar{a}^{\;}_{\bj2} \\  
\bar{a}^\dagger_{\bj2} K_\bj^\theta&\bar{a}^\dagger_{\bj2} 
\bar{a}^{\;}_{\bj1}&\bar{n}_{\bj2}-\frac{1}{3}
\end{pmatrix} \ .
\label{spinsu3}
\end{equation}
We can immediately see that the $2 \times 2$ block  matrix ${\cal
S}^{\alpha\beta}(\bj)$ ($1\leq \alpha, \beta \leq 2$) contains the
generators of $su(2)$. In general, from the commutation relations
Eq.~(\ref{conm3}), we can verify that if ${\cal S}^{\mu\nu}(\bj)$ are
the generators of $su(N)$, then ${\cal S}^{\alpha\beta}(\bj)$ are the
generators of the subalgebra $su(N-1)$. This will be useful in section
\ref{sec5}.

\subsubsection{Generalization to other irreps of $su(N)$}

Here we show that the hard-core bosons $\bar{b}_{\bj\alpha}$ can be 
connected to other irreducible representations of $su(N)$. The
dimension of an $su(N)$ representation of spin $S=M/N$ is $\binom
{N+M-1}{M}$. The number of flavors is then $N_f= \binom {N+M-1}{M}-1$.
In this case, it is more convenient to  adopt the following notation
for the flavors: Each flavor $\alpha$ will be denoted by an array of
$N$ integer numbers $\left \{
m_{1}^{\alpha},m_{2}^{\alpha},\cdots,m_N^{\alpha}\right\}$  satisfying
$\sum_{\sigma=1}^N m_{\sigma}^{\alpha}=M$. There are  $\binom
{N+M-1}{M}$ arrays satisfying this condition.  However, there is a
particular one, $\left \{ m_{1}^{0},m_{2}^0,\cdots,m_{N}^0\right\}$, 
associated to the vacuum state. 
\begin{eqnarray}
{\cal S}^{\mu\nu}(\bj)&=&- \delta_{\mu\nu} S 
+\sum_{\alpha=1}^{N_f} (g^{\alpha
0}_{\mu\nu}\bar{b}^{\dagger}_{\bj\alpha}+g^{0 \alpha}_{\mu\nu}
\bar{b}^{\;}_{\bj\alpha}) \nonumber \\
&+&\sum_{\alpha,\beta=1}^{N_f} g^{\alpha\beta}_{\mu\nu}
\bar{b}^{\dagger}_{\bj\alpha}\bar{b}^{\;}_{\bj\beta}  \ ,
\end{eqnarray}
with
\begin{eqnarray}
g^{\alpha\beta}_{\mu\nu}&=&\delta_{\mu\nu} \delta_{\alpha\beta} \
m^{\alpha}_{\mu} + \delta_{(m^{\alpha}_{\mu}-1)m^{\beta}_{\mu}}
\delta_{(m^{\alpha}_{\nu}+1)m^{\beta}_{\nu}} \nonumber\\
&&\times \prod_{\sigma=1,\sigma \ne \alpha,\beta}^N 
\delta_{m^{\sigma}_{\mu}m^{\sigma}_{\nu}}
\end{eqnarray}
Again, it is easy to verify that these are generators of an $su(n)$
Lie algebra satisfying the commutation relations Eq.~(\ref{conm3}).

This is just an example to show that the generators of a Lie algebra in 
a particular and arbitrary representation of dimension $D$ can be used to
describe a system where $D$ is the dimension of the local Hilbert
space. For each particular case there will be representations which
will be more appropriate (easier to handle) than others. For instance,
this set of $su(N)$ representations can be performed in a more natural
way using Schwinger-Wigner bosons. 

\subsection{$SO(N)$ spin-particle mappings}

Another possible language related to the JW particles is the
$SO(N)$-spin language. By this name we mean the language whose
generators are the identity and the generators of the $so(N)$ Lie
algebra (whose number is $N(N-1)/2$). It is easy to verify that the
following set of operators 
\begin{eqnarray}
{\cal M}^{\alpha\beta}(\bj)&=& 
\bar{a}^{\dagger}_{\bj\alpha} 
\bar{a}^{\;}_{\bj\beta} - \bar{a}^{\dagger}_{\bj\beta} 
\bar{a}^{\;}_{\bj\alpha} , \nonumber \\
{\cal M}^{0 \beta}(\bj)&=&   (K_\bj^{\theta})^{\dagger}
\bar{a}^{\;}_{\bj\beta} - \bar{a}^{\dagger}_{\bj\beta}
K^{\theta}_\bj,\nonumber \\ {\cal M}^{00}(\bj)&=&0  \ ,
\label{son}
\end{eqnarray}
obey the $SO(N)$-spin commutation relations 
\begin{eqnarray}
\left [{\cal M}^{\mu\mu'}(\bi), {\cal M}^{\nu\nu'}(\bj) \right ]\!\!\!
&=& \!\!\!
\delta_{\bi\bj} [\delta_{\mu'\nu} {\cal
M}^{\mu\nu'}(\bj)-\delta_{\mu'\nu'} {\cal M}^{\mu\nu}(\bj) \nonumber \\
&-\delta_{\mu\nu}& \!\!\!{\cal M}^{\mu'\nu'}(\bj)+ \delta_{\mu\nu'} {\cal
M}^{\mu'\nu}(\bj)] ,
\label{crson}
\end{eqnarray}
with $0\leq \mu,\nu\leq N$ and $1\leq \alpha,\beta \leq N_f=N-1$ The
antisymmetric relation ${\cal M}^{\mu\nu}$=$-{\cal M}^{\nu\mu}$ is
immediately derived from the definition of ${\cal M}^{\mu\nu}$,
Eq.~(\ref{son}). In this way, we have constructed another possible
language connected to hard-core particles. Like in the previous case,
the subset of operators ${\cal M}^{\alpha\beta}$ is a set of generators
for an $so(N-1)$ subalgebra of $so(N)$. 

In the following, for simplicity, we will only consider the connection
between $SO(N)$-spins and hard-core bosons
$\bar{b}_{\bj\alpha}=\bar{a}_{\bj\alpha}(\theta=0)$ with non-local 
transmutator $K^{\theta}_\bj=\one$. For instance, for $N=3$, the
generators are ($\alpha=1,2$)
\begin{equation}
{\cal M}(\bj)= 
\begin{pmatrix}
0& \bar{b}^{\;}_{\bj1} \!\!-\! \bar{b}^{\dagger}_{\bj1}
& \bar{b}^{\;}_{\bj2} \!\!-\! \bar{b}^{\dagger}_{\bj2} \\
\bar{b}^{\dagger}_{\bj1} \!\!-\! \bar{b}^{\;}_{\bj1}
& 0 & \bar{b}^\dagger_{\bj1} \bar{b}^{\;}_{\bj2}\!\!-\!\bar{b}^\dagger_{\bj2} 
\bar{b}^{\;}_{\bj1} \\  
\bar{b}^{\dagger}_{\bj2} \!\!-\! \bar{b}^{\;}_{\bj2}
&\bar{b}^\dagger_{\bj2} \bar{b}^{\;}_{\bj1}\!\!-\!\bar{b}^\dagger_{\bj1} 
\bar{b}^{\;}_{\bj2} &0
\end{pmatrix} .
\label{spinso3}
\end{equation}

\subsection{Generalized Jordan-Wigner Particles}
\label{jordanw}

In 1928 Jordan and Wigner related the spin quantum mechanical degree
of freedom to spinless particles with fermion statistics. This
transformation involves the $S$=1/2 irreducible representation of the
Lie algebra $su(2)$. We have recently \cite{ours1} generalized this
one-to-one mapping to any irreducible spin representation of $su(2)$,
spatial dimension of the lattice and particle exchange statistics. From
the physical viewpoint the JW particles are essentially hard-core
independently of their flavor index (i.e., satisfy the Pauli exclusion 
principle with $p=2$ (see below)). 

\begin{figure}[t]
\includegraphics[angle=0,width=8.2cm,scale=1.0]{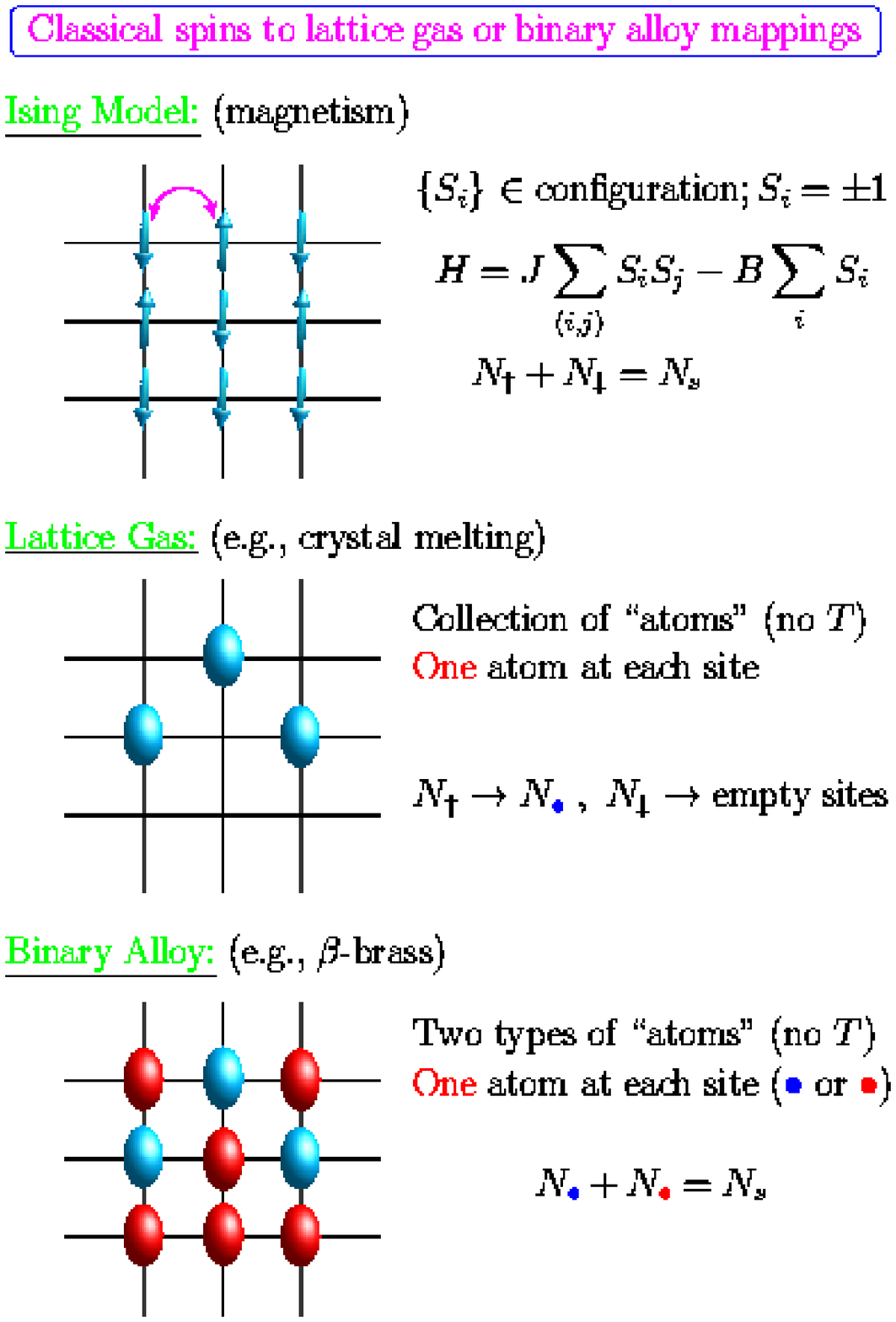}
\caption{Classical version of the Jordan-Wigner particles defined in 
\cite{ours1}. Here the mapping is performed between the simple lattice of 
Ising spins $S_i=\pm 1$ and a lattice gas of one type of atom or a
binary alloy. $N_\sigma$, with $\sigma=\uparrow,\downarrow$, is the number
of Ising spins of type $\sigma$ while $N_{\color{blue} \bullet}$ is the
number of {\it hard-core} atoms of flavor ${\color{blue} \bullet}$ with no
kinetic energy $T$.}
\label{fig1}
\end{figure}

The generalized JW mappings constitute a kind of quantum version of the 
well-known classical spin-lattice-gas transformations \cite{huang} (see
Fig. \ref{fig1}). In a classical lattice gas (or binary alloy) each 
site can be occupied by at most one atom, a hard-core condition equivalent 
to its quantum counterpart. On the other hand, since the spin system is 
classical there is no intrinsic dynamics and the kinetic energy 
of the gas ($T$) must be included a posteriori in an ad-hoc fashion. 
This contrasts the quantum case where the dynamics of the JW particles 
is intrinsic.

We have seen in  section  \ref{sec3} that the canonical fermions can 
be transformed into bosons which are hard-core in each flavor. We will
consider now another type of fermions which naturally emerge from the
strong coupling limit of models for interacting electrons. If the short
range component of the Coulomb repulsion is much larger than the
kinetic energy, the repulsion can be effectively replaced by a
constraint of no double occupancy. This perturbative approach is
usually implemented by a canonical transformation, which leads to an
effective Hamiltonian acting on the subspace of states with no double
occupancy. The fermionic subalgebra used to describe this effective
model is generated by the so-called constrained fermions. Therefore,
the constrained fermions are obtained by imposing to the canonical 
fermions a local constraint of no more than one particle per orbital
(or site). This constraint may be incorporated into the fermionic algebra
by defining the following creation and annihilation operators for the
constrained fields
\begin{equation}
\bar{c}^\dagger_{\bj \sigma} = {c}^\dagger_{\bj \sigma} \!\!\! 
\prod_{\substack{
\tau \in {\cal F}_\eta}} \!\! (1-\hat{n}_{\bj
\tau}) \ , \ 
\bar{c}^{\;}_{\bj \sigma} = \!\!\!\! \prod_{\substack{\tau \in {\cal
F}_\eta
}}  \!\!(1-\hat{n}_{\bj \tau}) \ {c}^{\;}_{\bj \sigma}
\ ,
\end{equation}
where ${\cal F}_\eta$ is the set of flavors, with $\eta=\frac{1}{2},1$ 
depending upon the spin character of the irreducible representation. It
is easy to check that the particles generated by this fermionic algebra
satisfy the constraint of single occupancy, i.e., the eigenvalues of  
${\bar n}_\bj=\sum_{\sigma=1}^{N_f} {\bar n}_{\bj\sigma}$ are either
$0$ or $1$. The most well-known context where these fermions appear in 
condensed matter physics is the strong coupling limit of the Hubbard
model, which leads to the $t$-$J$ Hamiltonian. While the Hubbard model
is described in terms of spin-$1/2$ canonical fermions (electrons), the
fermionic language for the $t$-$J$ model is generated by creation and
annihilation operators of spin-$1/2$ constrained fermions ($\sigma,
\sigma'=\uparrow,\downarrow$)
\begin{eqnarray}
\{ \bar{c}^{\;}_{\bi \sigma}, \bar{c}^{\;}_{\bj \sigma'} \} &=&
\{ \bar{c}^\dagger_{\bi \sigma}, \bar{c}^\dagger_{\bj \sigma'} \} = 0  
\ ,\nonumber \\
\{ \bar{c}^{\;}_{\bi \sigma},\bar{c}^\dagger_{\bj \sigma'} \}
&=& \delta_{\bi\bj} [\bar{c}^\dagger_{\bj \sigma'} \bar{c}^{\;}_{\bj \sigma}
+\delta_{\sigma \sigma'} (1-{\bar n}_\bj)] \ .
\label{cf1}
\end{eqnarray}
In this section we will consider the general case of constrained 
fermions with $N_f=2S$ different flavors. In the context of our 
previous example, this generalization can be interpreted as the
natural language for a $t$-$J$ model with more than one orbital per
site $\bj$ or larger-spin fermions. The set of commutation relations
for only two different flavors, Eq.~(\ref{cf1}), is generalized in the
following way
\begin{eqnarray}
\{ \bar{c}^{\;}_{\bi \sigma}, \bar{c}^{\;}_{\bj \sigma'} \} &=&
\{\bar{c}^\dagger_{\bi \sigma}, \bar{c}^\dagger_{\bj \sigma'} 
\} = 0  \ ,\nonumber \\
\{ \bar{c}^{\;}_{\bi \sigma}, \bar{c}^\dagger_{\bj \sigma'} \}
&=& \delta_{\bi\bj} \begin{cases}
1+\bar{n}_{\bj \sigma}-\bar{n}_{\bj}&
		\text{if $\sigma = \sigma'$}, \\
                \bar{c}^\dagger_{\bj \sigma'} \bar{c}^{\;}_{\bj \sigma} & 
		\text{if $\sigma \neq \sigma'$} . 
		\end{cases} 
\label{crcf}
\end{eqnarray}
Notice that $\prod_{\tau \in {\cal F}_\eta}^{\tau\neq
\sigma}\!(1-\bar{n}_{\bj \tau}) = 1+\bar{n}_{\bj
\sigma}-\bar{n}_{\bj}$ with number operators satisfying $\bar{n}_{\bj
\sigma}\bar{n}_{\bj \sigma'}= \delta_{\sigma \sigma'} \bar{n}_{\bj
\sigma}$.

We may now ask what is the bosonic language obtained from the
constrained fermions when the statistics is transmuted, i.e.,
\begin{equation}
{\bar b}^\dagger_{{\bf j}\alpha}=
{\bar c}^\dagger_{{\bf j}\alpha}  K_\bj^\pi \ .
\label{trans2}
\end{equation}
(We do not need to include the local transmutator since the
constraint does not allow to have more than one particle per site (or
orbital).) The answer to the question becomes clear if we compare the
set of commutation relations for the hard-core bosons,
Eq.~(\ref{conm}), to the set of anticommutation relations for the
constrained fermions, Eq.~(\ref{crcf}).  All {\it anticommutators}
which are not of the form $\{ \bar{c}^{\;}_{\bi \sigma}, 
\bar{c}^\dagger_{\bj \sigma'} \}$ are mapped onto the corresponding
commutators for hard-core bosons, Eq.~(\ref{conm}). This results from the 
transmutator $K_\bj^\pi$. What is the effect of the transmutation on
commutators containing bilinear forms in the creation and annihilation
operators? First, it is easy to check that in this case the commutator
${[}\bar{b}^\dagger_{\bi\alpha}\bar{b}^{\;}_{\bi\beta}, 
\bar{b}^\dagger_{\bj\gamma} {]}=\delta_{\bi\bj} \delta_{\beta \gamma}
\bar{b}^\dagger_{\bi\alpha}$ of bosonic operators is mapped onto the
corresponding commutator of fermionic operators
\begin{equation}
{[}\bar{c}^\dagger_{\bi\alpha}\bar{c}^{\;}_{\bi\beta}, 
\bar{c}^\dagger_{\bj\gamma} {]}=\delta_{\bi\bj}
\delta_{\beta \gamma} \bar{c}^\dagger_{\bi\alpha} \ .
\end{equation}
Finally, to complete this correspondence between commutation and
anticommutation relations we have to consider the effect of the
transmutation on the commutators ${[} \bar{b}^{\;}_{\bj
\sigma},\bar{b}^\dagger_{\bj \sigma}]$ and the anticommutators  $\{
\bar{c}^{\;}_{\bj \sigma},\bar{c}^\dagger_{\bj \sigma}\}$. It is easy
to check that  the transmutation does not change these particular
products, i.e.,
\begin{eqnarray}
\{ \bar{b}^{\;}_{\bi \beta}, \bar{b}^\dagger_{\bi \alpha}\} &=& 
\delta_{\alpha\beta} (1-\bar{n}_{\bi}) + \bar{b}^\dagger_{\bi \alpha}
\bar{b}^{\;}_{\bi \beta}\ .
\end{eqnarray}
We can see that the anticommutator $\{ \bar{b}^{\;}_{\bi \beta}, 
\bar{b}^\dagger_{\bi \alpha}\}$ is a linear combination of generators
of the Lie algebra defined by the set of commutation relations,
Eq.~(\ref{conm}). This is so since the set of generators  $\{I,
\bar{b}^\dagger_{\bi\alpha},\bar{b}^{\;}_{\bi\alpha},
\bar{b}^\dagger_{\bi\alpha} \bar{b}^{\;}_{\bi\beta} \}$ belongs to the
fundamental representation of the Lie algebra.

Therefore, in this case, the generators of the Lie algebra
associated to the bosonic language are transmuted into generators of a
Lie superalgebra associated to the constrained fermions. The Lie
product in the Lie algebra, Eq.~(\ref{conm}), is turned into the graded
Lie product in the superalgebra. The generators of the bosonic Lie
algebra can be separated into two subsets which are mapped onto the odd
and the even generators of the fermionic Lie superalgebra. In this
particular case, the Lie algebra associated to the commutation
relations, Eq.~(\ref{conm}), is $u(1) \bigoplus su(N)$. The creation
and annihilation operators for the hard core bosons,
$\bar{b}^\dagger_{\bi\alpha}$ and $\bar{b}^{\;}_{\bi\alpha}$ are mapped
onto the odd generators of the Lie superalgebra,
$\bar{c}^\dagger_{\bi\alpha}$ and $\bar{c}^{\;}_{\bi\alpha}$, while the
identity and the bilinear forms, $I$ and
$\bar{b}^{\dagger}_{\bi\beta}{\bar b}^{\;}_{\bi\alpha}$  are mapped
onto the even generators $I$ and ${\bar c}^{\dagger}_{\bi\beta}{\bar
c}^{\;}_{\bi\alpha}$. The bosons which are hard-core in each flavor, 
$\tilde{b}^{\;}_{\bi\alpha}$, and the canonical fermions,
${c}^{\;}_{\bi\alpha}$, provide another example of  transmutation of a
Lie algebra into a Lie superalgebra. In this case, the generators of the
Lie algebra $u(1) \bigoplus_{\alpha=1}^{N_f} su(2)$ are transmuted into the
generators of a Heisenberg Lie superalgebra. 
     
In this way, through the transmutation of statistics we have
established a direct connection between the multiflavored constrained
fermions and the multiflavored hard-core bosons. Using this connection
we can see that the MM transformation \cite{matsubara} can be obtained
from the JW transformation \cite{jordan} by a simple transmutation of
the statistics. 

In the course of demonstrating the Corollary I, we have shown that for
each class of languages characterized by the dimension $D$ of the local
Hilbert space, there is one whose generators are also generators of the
Lie algebra $u(1) \bigoplus su(2)$ ($D=2S+1$). In addition, we have
seen that there is another language in the class whose generators are
the identity and the creation and annihilation operators for
multiflavored hard-core bosons ($N_f=D-1$). Then, according to the
fundamental theorem, the three components $\{S^x_\bi,S^y_\bi,S^z_\bi\}$
of a spin $S$ can be written as a polynomial function of $\{I_\bi,
{\bar b}^{\dagger}_{\bi\alpha}, {\bar b}^{\;}_{\bi\alpha}\}$ with
$1\leq \alpha \leq N_f$ and vice versa. This is the generalization of
the MM \cite{matsubara} transformation to any irreducible spin-$S$
representation of $su(2)$. Adding now this result to the mapping
already established between the multiflavored hard-core bosons and the
constrained fermions, we can conclude that it is also possible to write
$\{S^x_\bi,S^y_\bi,S^z_\bi\}$ as a polynomial function of the identity
and the creation and annihilation operators of multiflavored
constrained fermions and vice versa. The later is the generalization of
the JW transformation to any spin $S$ \cite{ours1}. On the other hand,
the generalized JW spin-fermion mapping can be easily extended  to
include a spin-anyon mapping simply by using the anyonic particles 
generated by $\bar{a}^{\dagger}_{\bj\sigma}$ and
$\bar{a}^{\;}_{\bj\sigma}$  (see Eq.~(\ref{amon})) \cite{ours1}. The
explicit form of this generalization is given by \cite{ours1,Noten4p}:

\noindent
\underline{\bf Half-odd integer spin $S$} ($\sigma \in {\cal
F}_{\frac{1}{2}} = \{-S+1, \dots, S \}$): 
\begin{eqnarray}
S^+_\bj &=& \eta_{\bar{S}} \ \bar{a}^\dagger_{\bj \bar{S}+1} \ K_\bj^\theta + 
\sum_{\substack{\sigma \in {\cal F}_{\frac{1}{2}} \\ \sigma \neq S}} \!
\eta_\sigma \ \bar{a}^\dagger_{\bj \sigma+1} \bar{a}^{\;}_{\bj \sigma}
\ , \nonumber \\ 
S^-_\bj &=& \eta_{\bar{S}} \ (K_\bj^\theta)^\dagger \ 
\bar{a}^{\;}_{\bj \bar{S}+1} +
\sum_{\substack{\sigma \in {\cal F}_{\frac{1}{2}} \\ \sigma \neq S}} \!
\eta_\sigma \ \bar{a}^\dagger_{\bj \sigma} \ \bar{a}^{\;}_{\bj \sigma+1}
\ , \nonumber \\ 
S^z_\bj &=& -S + \sum_{\substack{\sigma \in {\cal F}_{\frac{1}{2}}}}
(S+\sigma) \ \bar{n}_{\bj \sigma} \ , \nonumber \\
\bar{a}^\dagger_{\bj \sigma} &=& (K_\bj^\theta)^\dagger 
L_\sigma^{\frac{1}{2}} \left( S^+_\bj
\right)^{\sigma+S} {\cal P}_\bj^{\frac{1}{2}} \ , \nonumber \\
\mbox{where} && {\cal P}_\bj^{\frac{1}{2}} =
\!\! \prod_{\substack{\tau \in {\cal F}_{\frac{1}{2}}}}
\! \frac{\tau -S^z_\bj}{\tau + S} \ , \ L_\sigma^{\frac{1}{2}} =
\!\!\!\!\! \prod_{\tau=-S}^{\sigma-1} \!\!\! \eta_\tau^{-1} \ . \nonumber
\end{eqnarray}

\begin{figure}[t]
\includegraphics[angle=0,width=8.2cm,scale=1.0]{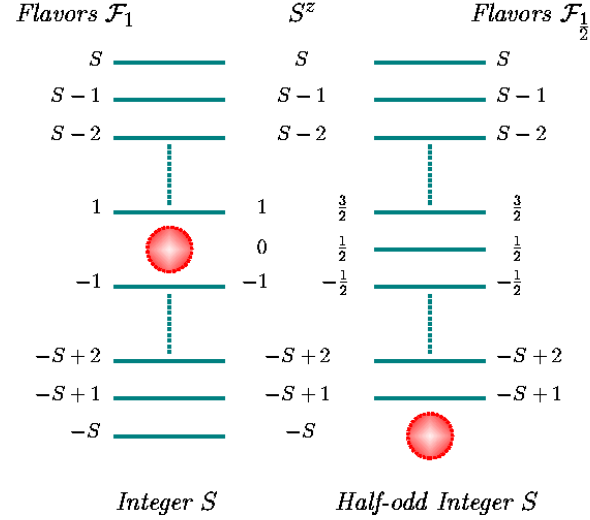}
\caption{Constrained anyon states per site for integer and half-odd
integer spin $S$. In both cases there are $2S$ flavors and the 
corresponding $2S+1$ values of $S^z$ are shown in the middle column.
One degree of freedom is assigned to the anyon vacuum (circle) whose
relative position depends upon the spin being integer or half-odd
integer.}
\label{fig2}
\end{figure}

\noindent
\underline{\bf Integer spin $S$} ($\sigma \in {\cal F}_1 =
\{-S,\dots,-1,1,\dots,S \}$): 
 \begin{eqnarray}
S^+_\bj &=& \eta_0 \ (\bar{a}^\dagger_{\bj 1} \ K_\bj^\theta +
(K_\bj^\theta)^\dagger \
\bar{a}^{\;}_{\bj \bar{1}}) +  \sum_{\substack{\sigma \in {\cal F}_1 \\ 
\sigma \neq -1,S}} \! \eta_\sigma \ \bar{a}^\dagger_{\bj \sigma+1}
\bar{a}^{\;}_{\bj \sigma} \ , \nonumber \\ 
S^-_\bj &=& \eta_0 \ ((K_\bj^\theta)^\dagger \ \bar{a}^{\;}_{\bj 1} +
\bar{a}^\dagger_{\bj \bar{1}} \ K_\bj^\theta) + \sum_{\substack{\sigma 
\in {\cal F}_1 \\ \sigma
\neq -1,S}} \! \eta_\sigma \ \bar{a}^\dagger_{\bj \sigma} \
\bar{a}^{\;}_{\bj \sigma+1} \ , \nonumber \\ 
S^z_\bj &=& \sum_{\substack{\sigma \in {\cal F}_1}} \sigma \
\bar{n}_{\bj \sigma} \ , \nonumber \\
\bar{a}^\dagger_{\bj \sigma} &=& (K_\bj^\theta)^\dagger L_\sigma^1 
\begin{cases}
            \left( S^+_\bj \right)^\sigma {\cal P}_\bj^1 & 
	    \text{if $\sigma > 0$} \ , \\
            \left( S^-_\bj \right)^\sigma {\cal P}_\bj^1 & 
	    \text{if $\sigma < 0$} \ , 
\end{cases} \nonumber \\
\mbox{where} && {\cal P}_\bj^1 =
\!\! \prod_{\substack{\tau \in {\cal F}_1}}
\! \frac{\tau -S^z_\bj}{\tau} \ , \ L_\sigma^1 =
\! \prod_{\tau=0}^{|\sigma|-1}  \eta_\tau^{-1} \ , \nonumber
\label{jwint}
\end{eqnarray}
and $\eta_\sigma = \sqrt{(S-\sigma)(S+\sigma+1)}$ (see Fig. \ref{fig2}). 
[A bar in a subindex means the negative of that number
(e.g., $\bar{\sigma}=-\sigma$).]
The total number of flavors is $N_f=2 S$, and the $S$=1/2 case 
simply reduces to the traditional JW transformation. These
mappings enforce the condition on the Casimir operator ${\bf S}_j^2 =
S(S+1)$. 

\subsection{Fractional Exclusion Statistics Algebras}
\label{sec4d}

What microscopic properties of the fundamental particle constituents 
determine the thermodynamic behavior of matter? It has been known since
the early days of quantum mechanics that particles in nature are either
bosons or fermions in the three-dimensional space. It was immediately
realized that for a given interaction among particles the properties of
matter were completely different depending upon the constituents being
bosons or fermions (for example, normal Coulomb matter is stable thanks
to the fermionic nature of its particles). Moreover, it has been known
for a long time \cite{gentile} that there are other theoretical
possibilities that interpolate between fermions and bosons, i.e., cases
where $p>1$ identical particles occupy one and  the same state. Quantum
statistics was the name given to this phenomenon clearly nonexistent in
the classical physics description of matter. 

In the seventies it was realized that two space dimensions allowed for
other exotic possibilities with particles having fractional spin and
statistics \cite{avinash}. Later on, this concept proved to be not
merely an intellectual exercise since particles obeying such quantum
statistics were experimentally realized in the context of the quantum
Hall effect (QHE). It was Haldane \cite{fract}, to our knowledge, the
first to realize that the Pauli exclusion principle and {\it exchange
statistics} ideas were in principle independent concepts. In this way,
motivated by the properties of quasiparticles in the fractional QHE, he
generalized the Pauli exclusion principle introducing the concept of
fractional {\it exclusion statistics}. 

Basically, two physical features characterize the quantum statistics
of particles: One is related to their indistinguishability and is
characterized by the property that when two identical particles are
exchanged the total wave function acquires a phase factor (exchange or
permutation statistics). Another refers to the ability to accommodate
$p$ particles in the same single-particle quantum state (exclusion
statistics) \cite{Noten3}. The first concept depends upon the space
dimensionality of the system while the second one is independent and,
therefore, unrelated to the notion of anyonic fractional statistics
which is applicable exclusively in two space dimensions. For instance,
from the generalized Pauli principle viewpoint there is no distinction
between particles obeying the algebra of  Eqs.~(\ref{canoa1}) and
(\ref{hca2}) for any statistical angle $\theta$, i.e., all of them
correspond to $p=1$. However, the exchange of two particles $\bar{a}$
does depend upon $\theta$.

In this section we consider the problem of formulating the algebra 
satisfied by these generalized Pauli-exclusion particles and its 
connection to our fundamental theorem. Our definition is a  possible
{\it second-quantized} version of Haldane's definition introduced at
the first quantization level. 

To simplify matters we consider as an example a single flavor (single
site) fractional exclusion statistics algebra
\begin{eqnarray}
\begin{cases}
[ g^{\;},g^{\;} ] = [ g^\dagger,g^\dagger ]=0  
 \ , \\
{[}g^{\;},g^\dagger{]}=1-F  \; , \;
[ g^\dagger g^{\;}, g^\dagger ]=  g^\dagger  \ , 
\end{cases}
\label{canog}
\end{eqnarray}
where the operator $F$ ($F^\dagger=F$) is a polynomial function of
$g^\dagger$ and $g^{\;}$, such that $g^\dagger F=0$. The $F$ term
represents a sort of deformation of the canonical boson algebra
(deformed Heisenberg algebras play an important role in the theory of
representations of quantum groups). For example, $F$ can be
$\frac{p!}{p+1} F = (g^\dagger)^{p}(g^{\;})^{p} = \prod_{j=0}^{p-1}
(g^\dagger g^{\;} -j)$ and $p$ is an arbitrary integer with the
condition $p>0$. For $g$-particles with this $F$ operator it is
very easy to prove that they satisfy the nilpotency condition
\begin{equation}
(g^\dagger)^{p+1} = 0 \ ,
\label{gion}
\end{equation}
which means that one can put up to $p$ particles on each mode. 
A $(p+1)$-dimensional matrix representation of this algebra is
\begin{equation}
g=
\begin{pmatrix}
0&1&0&0&\cdots&0\\
0&0&\sqrt{2}&0&\cdots&0\\
0&0&0&\sqrt{3}&\cdots&0\\
\vdots&\vdots&\vdots&\vdots&\vdots&\vdots\\
0&0&0&0&\cdots&\sqrt{p}\\
0&0&0&0&\cdots&0 
\end{pmatrix} \ .
\end{equation}

Certainly, one can build infinitely many different algebras leading 
to different types of fractional exclusion statistics particles with
the same value of $p$. For example, consider the $su(2)$ algebra 
generated by the operators $\{\bar{g}_\bi^{\dagger}, \bar{g}_\bi^{\;},  
n^g_\bi\}$ (with $(n^g_\bi)^\dagger=n^g_\bi$) in the $S$=1 (triplet)
representation ($(\bar{g}_\bi^{\dagger})^3=0$, i.e., $p=2$)
\begin{eqnarray}
\begin{cases}
[ \bar{g}_\bi^{\;},\bar{g}^{\;}_\bj ] = 
[ \bar{g}_\bi^{\dagger},\bar{g}^{\dagger}_\bj ]=0  
 \ , \\
{[}\bar{g}^{\;}_\bi,\bar{g}^\dagger_\bj{]}=\delta_{\bi\bj}(1-n^g_\bi)  
\; , \; [ n^g_\bi, \bar{g}^\dagger_\bj ]=  
\delta_{\bi\bj} \ \bar{g}^\dagger_\bj  \ . 
\end{cases}
\label{canog1}
\end{eqnarray}
It is straightforward to realize how these particles relate to the usual
spin generators of $su(2)$
\begin{eqnarray}
S^+_\bj &=& \sqrt{2} \ \bar{g}^{\dagger}_\bj  \ , \nonumber \\
S^-_\bj &=& \sqrt{2} \ \bar{g}^{\;}_\bj \ , \nonumber \\
S^z_\bj &=&  n^g_\bj - 1 \ , 
\label{g-on2}
\end{eqnarray}
which can be seen as a possible generalization of the MM 
\cite{matsubara} transformation for $S$=1.

For example, if we consider the $1d$ $X$$Y$ model in the $S$=1 spin
representation
\begin{equation}
H_{\sf xy}=J \sum_\bj (S^x_\bj S^x_{\bj+1}+S^y_\bj S^y_{\bj+1}) \ ,
\end{equation}
it is simply related to the $g$-particle Hamiltonian
\begin{eqnarray}
H_{\sf xy}&=&J \sum_{\bj} (
\bar{g}^\dagger_{\bj} \bar{g}^{\;}_{\bj+1}+ \bar{g}^\dagger_{\bj+1} 
\bar{g}^{\;}_{\bj}) \nonumber \\
&=&2J\sum_{\bk} \cos\bk \ \bar{g}^\dagger_{\bk}\bar{g}^{\;}_{\bk} \ ,
\end{eqnarray}
where $\bar{g}^\dagger_{\bk}=(1/\sqrt{N_s}) \sum_\bj \exp(i \bk\cdot\bj)
\ \bar{g}^\dagger_{\bj}$ is the Fourier-transformed operator.


\subsection{Green's Parastatistics}

Since the very beginnings of quantum mechanics people tried to
understand the depth and consequences of the symmetrization postulate 
which asserts that physical states of identical particles must be
either symmetric (Bose statistics) or antisymmetric (Fermi statistics)
under permutations. In 1953 H.S. Green
\cite{green1} considered the possibility of having non-identical but
dynamically similar particles satisfying what is now known as
parastatistics. The introduction of parastatistics allowed some hope to
represent all particle fields in terms of a fundamental spinor field
and, of course, canonical fermions and bosons represented particular
cases (parastatistics of order $p=1$).

Soon, Greenberg \cite{greenberg} suggested that quarks satisfy
parastatistics of order 3 and Green \cite{green2} reformulated the
neutrino  theory of light identifying the neutrinos with parafermions
of order 2. It is not our intention to discuss the physical aspects of
these assertions but to present parastatistics as another operator
language which allows representation of Lie algebras and to relate
these particles to the ones already introduced in previous Sections. 

Following Green \cite{green1,green2}, let us introduce the auxiliary
modes (upper (lower) sign will be used to define the parafermions
(parabosons))
\begin{eqnarray}
\begin{cases}
[d_\bi^\alpha,d_\bj^\alpha]_\pm=[\check{d}_\bi^\alpha,
\check{d}_\bj^\alpha]_\pm=0  
 \ ,\\
[d_\bi^\alpha,\check{d}_\bj^\alpha]_\pm=\delta_{\bi\bj}  \ ,
\end{cases}
\label{para1}
\end{eqnarray}
labeled by the Green indices $\alpha=1,\cdots,p$ and $\bj=1,\cdots,N$
and where $\check{d}_\bj^\alpha=(d_\bj^\alpha)^\dagger$, together with
the condition $d_\bj^\alpha |{\rm vacuum}\rangle = 0 \ , \ \forall
(\bj,\alpha)$. For $\alpha \neq \beta$ the auxiliary modes satisfy
non-standard relations, i.e., 
\begin{eqnarray}
\begin{cases}
[d_\bi^\alpha,d_\bj^\beta]_\mp=[\check{d}_\bi^\alpha,
\check{d}_\bj^\beta]_\mp=0   \ ,\\
[d_\bi^\alpha,\check{d}_\bj^\beta]_\mp=0 \ .
\end{cases}
\label{para2}
\end{eqnarray}
Parafermion and paraboson creation and annihilation operators are
defined in terms of the auxiliary modes by
\begin{equation}
d^\dagger_\bj=\sum_{\alpha=1}^p \check{d}_\bj^\alpha \ , \ 
d^{\;}_\bj=\sum_{\alpha=1}^p d_\bj^\alpha \ ,
\end{equation}
and consequently satisfy the commutation relations
\begin{eqnarray}
\begin{cases}
[[d_\bi^\dagger,d_\bj^{\;}]_\mp,d_\bl^{\;}]_-=-2\delta_{\bi\bl} \ d_\bj^{\;}
 \ ,\\
[[d_\bi^{\;},d_\bj^{\;}]_\mp,d_\bl^{\;}]_-=0 \ .
\end{cases}
\label{para3}
\end{eqnarray}
It easy to verify that one can define a set of commutative number
operators $n^d_\bj$ ($n^\alpha_\bj=\check{d}_\bj^\alpha d_\bj^\alpha$) by
\begin{equation}
n^d_\bj=\frac{1}{2} ([d_\bj^\dagger,d_\bj^{\;}]_\mp \pm
p)=\sum_{\alpha=1}^p
n^\alpha_\bj \ , 
\end{equation}
which satisfy the commutation rules
\begin{equation}
[n^d_\bi,d_\bj^\dagger]_-=\delta_{\bi\bj} \ d_\bj^\dagger \ .
\end{equation}

Note that in the parafermionic case $(n^\alpha_\bj)^2=n^\alpha_\bj$, which
implies that $n^d_\bj$  has eigenvalues ranging from 0 to $p$. Moreover, 
\begin{equation}
(d_\bj^\dagger)^p= p! \prod_{\alpha=1}^p \check{d}_\bj^\alpha \ ,
\end{equation}
such that $(d_\bj^\dagger)^{p+1}=0$, which in a sense also generalizes
the Pauli exclusion principle. It is clear from these definitions that
canonical fermions (bosons) are parafermions (parabosons) of order
$p=1$. 

To connect these para-particles to the ones already described in
previous sections, one has to realize that (total or partial) local and
non-local transmutations are necessary. It is easy to see that the
parafermionic auxiliary modes are
\begin{equation}
\check{d}_\bi^\alpha =  c_{\bi \alpha}^\dagger \, \exp[i\pi
(\sum_{\bj<\bi}
 \sum_{\beta \neq \alpha} \hat{n}_{\bj \beta} + 
 \sum_{\beta < \alpha} \hat{n}_{\bi \beta})]\ ,
\end{equation}
where a partial non-local transmutation connects $\check{d}_\bj^\alpha$
to the canonical multiflavored fermions $c_{\bj \alpha}^\dagger$,
Eq.~(\ref{canoc}). Similarly, the parabosonic auxiliary fields are
obtained as
\begin{equation}
\check{d}_\bi^\alpha = b_{\bi \alpha}^\dagger \, \exp[i\pi
(\sum_{\bj<\bi}
 \sum_{\beta \neq \alpha} {n}_{\bj \beta} + 
 \sum_{\beta < \alpha} {n}_{\bi \beta})]\ ,
\end{equation}
in terms of canonical multiflavored bosons $b_{\bj \alpha}^\dagger$, 
Eq.~(\ref{canob}).

One can represent all Lie algebras in terms of parafermion or paraboson
creation and annihilation operators. Several authors contributed to
this observation in the early sixties and seventies \cite{liepara}. In
the following we just present as an example the isomorphism between the
algebra of $N$ parafermions and the proper orthogonal complex Lie
algebra $B_N$ (i.e., the complexification of $so(2N+1)$) of dimension
$N (2N+1)$. The set of operators antisymmetric in their indices
($\bmm,\bn \in [1,2N]$)
\begin{eqnarray}
\begin{cases}
L_{\bmm\bn} = -L_{\bn\bmm}=i(\ell_\bn \ell_\bmm -  \ell_\bmm \ell_\bn) \ ,
(\bmm<\bn)\\
L_{\bmm0} = -L_{0\bmm}= \ell_\bmm \ , \\
L_{00} = 0 \ ,
\end{cases}
\end{eqnarray}
where
\begin{equation}
\ell_{2\bj-1}= \frac{d^\dagger_\bj+d^{\;}_\bj}{2} \ , \ 
\ell_{2\bj}= \frac{d^\dagger_\bj-d^{\;}_\bj}{2i} \ ,
\end{equation}
represents a basis which span the Lie algebra isomorphic to $B_N$ 
($\mu,\mu',\nu,\nu'\in[0,2N]$)
\begin{eqnarray}
[L_{\mu\mu'},L_{\nu\nu'}]=& i(-\delta_{\mu'\nu} L_{\mu\nu'} + 
\delta_{\mu'\nu'} L_{\mu\nu} \nonumber
\\ 
+&\delta_{\mu\nu} L_{\mu'\nu'}-\delta_{\mu\nu'} L_{\mu'\nu}) \ .
\end{eqnarray}

\section{Equivalent Classes of Models}
\label{sec5}

In the previous section we have shown how the different languages of
quantum mechanics are connected. How can we exploit these dictionaries
to get a better understanding of different physical phenomena? There
are many different answers to this question which we will develop in
the  next sections. In particular, the present section is devoted to
show the equivalence between models which, in principle, describe
completely different systems \cite{boundary}. In other words, we show
that models describing different physical problems are associated to
the same Hamiltonian written in different languages. In this way, these
dictionaries appear as an essential tool to connect distinct fields of
physics. Preliminary applications of these concepts are given in
Refs.~\cite{ours1,ours2}. In Ref.~\cite{ours1}, we showed explicit
connections between spin $S$=1 and $t$-$J$-like models. In
Ref.~\cite{ours2} we demonstrated the equivalence between $SU(3)$
ferromagnetism, coexistence of magnetism and Bose-Einstein 
condensation in an interacting bosonic gas, and coexistence of a 
spin-nematic phase and ferromagnetism for a spin one system. The choice
of examples in the present section \ref{sec5} is by no means
exhaustive, it is simply illustrative. Table \ref{table1} summarizes
some of the most celebrated mappings known today.

\begin{table}[htb]
\begin{center}
\footnotesize
\hspace*{-0.5cm}
\begin{tabular}{|c|c|c|l|}
\hline
\raisebox{0pt}[13pt][7pt]{\large $d$} &
\raisebox{0pt}[13pt][7pt]{\large Model A} &
\raisebox{0pt}[13pt][7pt]{\large Model B} &
\raisebox{0pt}[13pt][7pt]{\large M - S}\\ 
\hline \hline
1 & Isotropic $XY$ ($S$=1/2) & $t$ & C - E
\\
1 & Anisotropic $XY$ ($S$=1/2) & $t$-$\Delta$ & C - E
\\
1 & $XXZ$ ($S$=1/2)& $t$-$V$   & C - E (BA)
\\ 
1 & $XXZ$ ($S$=1/2)& $t$-$J_z$ & P - E (BA)
\\ 
1 & $XYZ$ ($S$=1/2)& $t$-$\Delta$-$V$ & C - E (BA)
\\
1 & Majumdar-Ghosh ($S$=1/2) & $t$-$t'$-$V$-$V'$ & C - GS
\\
1 & BB $S=1$ ($\phi=\pi/4$, LS)  & $t$-$\bar{J}$-$V$-$\mu$ &
C - E (BA)
\\
1 & BB $S=1$ ($\tan \phi$=1/3, AKLT) &
$t$-$\Delta$-$\bar{J}$-$V$-$\mu$ & C - GS
\\
1 & BB $S=1$ ($\phi=7\pi/4$, TB)  &$t$-$\Delta$-$\bar{J}$-$V$-$\mu$ & C
- E (BA)
\\
1 & BB $S=1$ ($\phi=3\pi/2$, K)  &$\Delta$-$\bar{J}$-$V$-$\mu$ & C
- QE
\\
1 & {$S$=3/2} & {F ($S$=1/2) Hubbard}  & C - E (BA)
\\
2 & $U(1)$ gauge magnet  & F strings  & C - E
\\
2 & Shastry-Sutherland ($S$=1/2) & $t$-$t'$-$V$-$V'$  & C - GS
\\
Any & BB $S=1$ ($\phi=5\pi/4$) & $t$-$\bar{J}$-$V$-$\mu$& C - QE 
\\
Any & Anisotropic $S=1$ Heisenberg &B ($S$=1/2) Hubbard  & P
- U
\\

\hline
\end{tabular}
\end{center}
\caption{Equivalence between spin models (model A)
and particle models (model B), equivalence that can be simply
established by using our fundamental theorem. The meaning of the
different terms is clarified in the text (BB: bilinear-biquadratic, F:
fermionic, B: bosonic). Last column describes the type
of equivalence (C: complete Hilbert space mapping, P: partial Hilbert
space mapping) and type of solutions (E: exact, E (BA): exact
Bethe-ansatz, GS: only ground state, QE: quasi-exact, U: unsolvable).}
\label{table1}
\end{table}

\begin{figure}[htb] \hspace*{0.0cm}
\includegraphics[angle=0,width=8.0cm,scale=1.0]{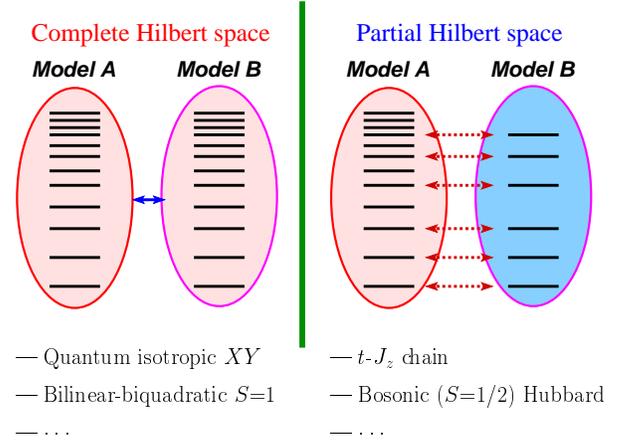}
\caption{Schematics of the two types of mappings that one can perform:
Complete and partial Hilbert space mappings. The figure also
shows examples illustrating each class.}
\label{fig4b}
\end{figure}

Another interesting aspect of these mappings is the possibility of
unveiling hidden symmetries which lead to exact or quasi-exact
solutions. By exact solvability we mean that the full spectral problem
is reduced to an algebraic procedure, while problems which are
quasi-exactly solvable are those which admit only a partial
determination of the spectrum by an algebraic procedure (see Fig.
\ref{fig4b}). Sometimes
there are particular languages which allow us to recognize invariant
subspaces of our Hamiltonian. The restricted action the Hamiltonian in
an invariant subspace can always be described by a class of languages
which are more basic than the original one. By more basic we mean a
lower dimension of the local Hilbert space, or in other words,
languages describing systems with less degrees of freedom. There are
particular cases where the simplification is even deeper, since by
using a more basic language we also recognize hidden symmetries for the
restricted action of the Hamiltonian in the invariant subspace.  The
quasi-exact solution of the $1d$ $t$-$J_z$ \cite{ours4} model, is an
example of successful application of this concept. This quasi-exact
solution led to the exact quantum phase diagram and charge excitations
of the $t$-$J_z$ model \cite{ours4}. 

\subsection{$SU(2)$ Heisenberg Magnets}

The $XYZ$ $S$=1/2 spin model ($D=2$) is the most popular example of a
family of Hamiltonians which is Bethe-ansatz solvable in $d$=1
\cite{Baxter} 
\begin{equation}
H_{\sf xyz}= \sum_{\langle \bi,\bj \rangle} J_x \
S^{x}_{\bi}S^{x}_{\bj} + J_y \ S^{y}_{\bi}S^{y}_{\bj} + J_z \
S^{z}_{\bi}S^{z}_{\bj}\ .
\label{xyz1}
\end{equation} 
Using the MM \cite{matsubara} transformation of Eq.~(\ref{set2m}), the
$XYZ$ model is mapped onto a gas of interacting hard-core bosons with
density-density interactions and particle non-conserving terms 
\begin{eqnarray}
H_{\sf xyz}&=& t \sum_{\langle \bi,\bj \rangle} (\bar{b}^\dagger_\bi
\bar{b}^{\;}_\bj + \bar{b}^\dagger_\bj \bar{b}^{\;}_\bi)  + \Delta
\sum_{\langle \bi,\bj \rangle} (\bar{b}^\dagger_\bi
\bar{b}^{\dagger}_\bj + \bar{b}^{\;}_\bj \bar{b}^{\;}_\bi) 
\nonumber \\
&+&  V
\sum_{\langle \bi,\bj \rangle} (\bar{n}_\bi -\frac{1}{2})
(\bar{n}_\bj-\frac{1}{2})\ ,
\label{xyz2}
\end{eqnarray} 
where $t=(J_x+J_y)/4$, $\Delta=(J_x-J_y)/4$, and $V=J_z$. Note that in
$d$=1 the non-local transmutator is a symmetry and, therefore, we could
have used spinless fermions instead of hard-core bosons in
Eq.~(\ref{xyz2}). 

Another family of spin Hamiltonians ($D=3$) which has been extensively
studied  is the bilinear-biquadratic (BB) $S$=1 Heisenberg model
($J>0$)
\begin{equation} 
H_\phi =  J \sqrt{2} \sum_{\langle {\bf i},{\bf j} \rangle} [ \cos \phi
\ {\bf S}_{\bf i} \cdot {\bf S}_{{\bf j}} + \sin \phi \left ( {\bf
S}_{\bf i} \cdot {\bf S}_{{\bf j}} \right )^2 ] \ .
\label{spin1xx}
\end{equation}
Using the $S$=1 generalization of the JW transformation for
hard-core bosons \cite{ours1},
\begin{eqnarray}
S^+_\bj &=& \sqrt{2} \ (\bar{b}^\dagger_{\bj \uparrow} +  
\bar{b}^{\;}_{\bj \downarrow}) \  \nonumber \\ 
S^-_\bj &=& \sqrt{2} \ (\bar{b}^{\;}_{\bj \uparrow} +
\bar{b}^\dagger_{\bj \downarrow} ) \  \nonumber \\ 
S^z_\bj &=& \bar{n}_{\bj \uparrow} - \bar{n}_{\bj \downarrow} 
\end{eqnarray}
(see section~\ref{jordanw}), $H_\phi$ is mapped onto a $t$-$J$-like
model for $S$=1/2 hard-core bosons including particle non-conserving
terms
\begin{eqnarray}
\lefteqn{\!\!\!\!\!\!\! \!\!H_\phi= \! \sum_{\langle {\bf i},{\bf j}
\rangle,\sigma} (t \ {\bar{b}}^\dagger_{{\bf i} \sigma}
{\bar{b}}^{\;}_{{\bf j} \sigma} +\Delta \ {\bar{b}}^\dagger_{{\bf i}
\sigma} {\bar{b}}^{\dagger}_{{\bf j} \sigma} + {\rm H.c.})+ 4\Delta
\sum_{\langle {\bf i},{\bf j} \rangle} s^z_{\bf i} \cdot  s^z_{\bf j}}
\nonumber \\
\!\!+&&\!\! \!\!\!\!\!\bar{J}\sum_{\langle {\bf i},{\bf j} \rangle} ({\bf
s}_{\bf i} \cdot  {\bf s}_{\bf j} - \! \frac{\bar{n}_{\bf i}
{\bar{n}}_{\bf j}}{4}) \!+ \!V \! \sum_{\langle{\bf i},{\bf j} \rangle}
\! {\bar{n}}_{\bf i}{\bar{n}}_{\bf j}-\!\mu \!\sum_{{\bf j}}(
{\bar{n}}_{\bf j} - 1) ,
\label{spin1xx2}
\end{eqnarray}
where ${\bf s}_{\bf j}=\frac{1}{2} {\bar {b}}^\dagger_{{\bf j} \alpha}
\boldsymbol{\sigma}_{\alpha\beta} {\bar {b}}^{\;}_{{\bf j} \beta}$
($\boldsymbol{\sigma}$ denoting Pauli matrices), $t=J\sqrt{2}\cos
\phi$, $\Delta=J\sqrt{2}(\cos \phi - \sin \phi)$,
$\bar{J}=J2\sqrt{2}\sin\phi$, $V=\bar{J}$, $\mu={\sf z} J\sqrt{2} \sin
\phi$, and ${\sf z}$ is the coordination of the lattice.
Again, for $d$=1 the exchange statistics of the particles is irrelevant
and one could have used $S$=1/2 {\it constrained} fermions instead of
hard-core bosons. From the known solutions in the spin model one can
immediately recognize the solvable cases in the particle model
Eq.~(\ref{spin1xx2}). These are: $\phi=\pi/4$ ($d$=1) Lai-Sutherland
(LS) \cite{lai}, $\phi=7\pi/4$ ($d$=1) Takhtajan-Babujian (TB)
\cite{tatba}, $\tan\phi=1/3$ ($d$=1) Affleck-Kennedy-Lieb-Tasaki (AKLT)
\cite{aklt}, $\phi=3\pi/2$ ($d$=1) Kl\"umper (K) \cite{klum}, and 
$\phi=5\pi/4$ (any $d$) our work in Ref. \cite{ours2}.

\begin{figure}[htb]
\includegraphics[angle=0,width=8.6cm,scale=1.0]{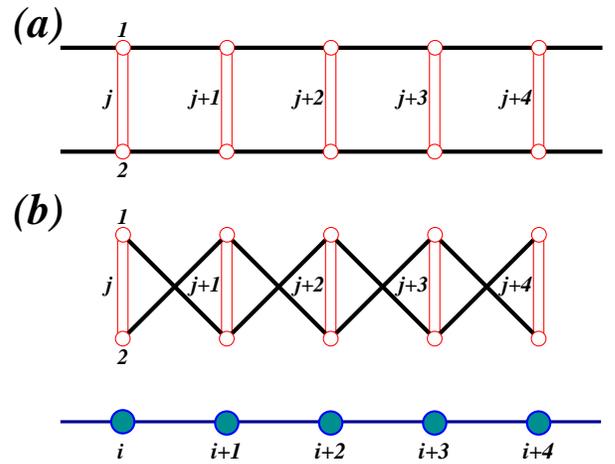}
\caption{A spin $S$=1/2 Heisenberg ladder mapped onto a Hubbard chain
model with correlated hopping. Models $a$ and $b$ are related through
a gauge transformation.}
\label{fig4d}
\end{figure}

Of particular current interest are the $S$=1/2 Heisenberg models on 
ladders. The simplest case corresponds to having only nearest-neighbor
magnetic interactions (see Fig.~\ref{fig4d}$a$)
\begin{eqnarray}
\lefteqn{ 
H_{\sf Heis}^{\sf ladd} = J_1 \sum_{\bf j} (\Delta_1 S^z_{\bj
1}S^z_{\bj 2} +  S^x_{\bj 1}S^x_{\bj 2} + S^y_{\bj 1}S^y_{\bj 2})}
&&\nonumber \\
&+& \!\!\! J_2 \!\!\! \sum_{{\bf j},\nu=1,2}  \!\!\! (\Delta_2 S^z_{\bj
\nu}S^z_{\bj+1 \nu} +  S^x_{\bj \nu}S^x_{\bj+1 \nu} + S^y_{\bj \nu}
S^y_{\bj+1 \nu}).
\label{laddH}
\end{eqnarray}
Using the mapping given in Eqs.~(\ref{sdd3}) which connects $S$=1/2
spins with canonical fermions, $H_{\sf Heis}^{\sf ladd}$ can be
rewritten as a Hubbard-like model on a linear chain (up to an
irrelevant constant). For instance, if $\Delta_1=1$ and $\Delta_2=0$ we
get
\begin{eqnarray} 
\lefteqn{ 
H_{\sf Heis}^{\sf ladd}=   t \sum_{\bf j, \sigma}  c^{\dagger}_{\bj
\sigma} c^{\dagger}_{\bj+1 {\bar \sigma}} (1-\hat{n}_{\bj \bar{\sigma}}
- \hat{n}_{\bj+1 \sigma}) + {\rm H.c.}}
&&\nonumber \\
&+& \!\! t \sum_{\bf j, \sigma}  c^{\dagger}_{\bj \sigma} c^{\;}_{\bj+1
\sigma} [(1-\hat{n}_{\bj {\bar \sigma}})(1-\hat{n}_{\bj+1 {\bar
\sigma}})-  \hat{n}_{\bj {\bar \sigma}}\hat{n}_{\bj+1 {\bar \sigma}}] +
{\rm  H.c.} \nonumber \\
&-& \!\! U \sum_{\bf j} \hat{n}_{\bj \uparrow} \hat{n}_{\bj \downarrow},
\label{laddH1}
\end{eqnarray}
where $U=J_1$ and $t=J_2$. This is a correlated Hubbard model with zero
two-body hopping terms plus a superconducting term. In the absence of
the superconducting term, this model has been exactly solved by
Arrachea and Aligia \cite{Arrachea}. To eliminate the superconducting
terms the original spin model has to be modified in the following way
\begin{eqnarray} 
\lefteqn{ 
\tilde{H}_{\sf Heis}^{\sf ladd} = J_1 \sum_{\bf j} (\Delta_1 S^z_{\bj
1}S^z_{\bj 2} +  S^x_{\bj 1}S^x_{\bj 2} + S^y_{\bj 1}S^y_{\bj 2})}
&&\nonumber \\
&+& \!\!\! J_2 \!\!\! \sum_{{\bf j},\nu=1,2}  \!\!\!P_{\bf j} (\Delta_2
S^z_{\bj \nu}S^z_{\bj+1 \nu} +  S^x_{\bj \nu}S^x_{\bj+1 \nu} + S^y_{\bj
\nu} S^y_{\bj+1 \nu})P_{\bf j},
\nonumber
\label{laddH2}
\end{eqnarray}
where $P_\bj= \sum_{\nu=1,2}(S^{z}_{\bj \nu} + S^{z}_{\bj+1
\nu})=\hat{n}_{\bj \uparrow}-\hat{n}_{\bj \downarrow}+\hat{n}_{\bj+1
\uparrow}-\hat{n}_{\bj+1 \downarrow}=2(s^z_{\bj}+s^z_{\bj+1})$.
$\tilde{H}_{\sf Heis}^{\sf ladd}$ for $\Delta_1=1$ and $\Delta_2=0$ is
equivalent to the Hubbard model with correlated hopping solved in
Ref.~\cite{Arrachea}. If the sites 1 and 2 are interchanged in one
sublattice (see Fig.~\ref{fig4d}$b$), the sign of the three-body
hoppings changes from positive to negative.

\begin{figure}[htb]
\includegraphics[angle=0,width=8.6cm,scale=1.0]{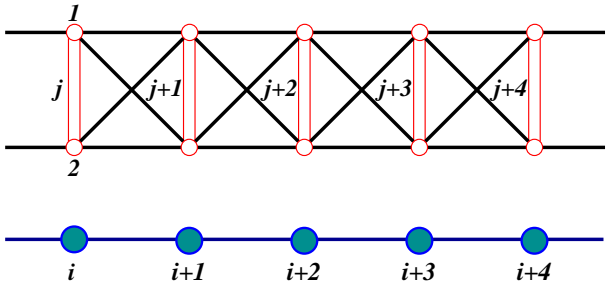}
\caption{Sutherland $S$=1/2 model and its mapping onto a Hubbard-like
chain model.}
\label{fig4c}
\end{figure}

Let us consider now the Heisenberg spin-ladder model illustrated in
Fig.~\ref{fig4c}
\begin{eqnarray} 
\lefteqn{ 
{H}_{\sf SS} = J_1 \sum_{\bf j} (\Delta_1 S^z_{\bj
1}S^z_{\bj 2} +  S^x_{\bj 1}S^x_{\bj 2} + S^y_{\bj 1}S^y_{\bj 2})}
&&\nonumber \\
&+& \!\!\!\! J_2 \!\!\! \sum_{{\bf j},\nu,\nu'}  \!\!(\Delta_2
S^z_{\bj \nu}S^z_{\bj+1 \nu'} +  S^x_{\bj \nu}S^x_{\bj+1 \nu'} + S^y_{\bj
\nu} S^y_{\bj+1 \nu'}).
\label{laddH3}
\end{eqnarray}
The isotropic limit ($\Delta_1=\Delta_2=1$) of this model has been
considered by Sutherland \cite{bill} who established that for
$J_1>2J_2$ the ground state is a product state of bond singlets. Again,
using Eqs.~(\ref{sdd3}), $H_{\sf SS}$ can be rewritten as a 1$d$ 
Hubbard-like model (up to an irrelevant constant)
\begin{eqnarray} 
\lefteqn{ 
{H}_{\sf SS} = t \sum_{\bf j, \sigma}  c^{\dagger}_{\bj \sigma}
c^{\dagger}_{\bj+1 {\bar \sigma}}  (1-\hat{n}_{\bj \bar{\sigma}})(1-
\hat{n}_{\bj+1  \sigma}) + {\rm H.c.}}
&&\nonumber \\
&+&  t \sum_{\bf j, \sigma}  c^{\dagger}_{\bj \sigma} c^{\;}_{\bj+1
\sigma} (1-\hat{n}_{\bj {\bar \sigma}})(1-\hat{n}_{\bj+1 {\bar
\sigma}}) +  {\rm H.c.}
\nonumber \\
&+& J_z \sum_{\bf j} s^z_{\bj} s^z_{\bj+1} - U \sum_{\bf j}
\hat{n}_{\bj \uparrow} \hat{n}_{\bj \downarrow}  - \mu \sum_{\bf j}
\hat{n}_{\bj},
\label{laddH4}
\end{eqnarray}
where $U=J_1\Delta_1$, $t=J_2$, $J_z= 4 J_2 \Delta_2$, and
$\mu=(1-\Delta_1) J_1/2$.  This model is quasi-exactly solvable in the
isotropic case $\Delta_1=\Delta_2=1$. In addition, if we cancel the
superconducting terms, the resulting model is also quasi-exactly
solvable because there are invariant subspaces for which the action of
the Hamiltonian can be mapped onto a $t$-$J_z$ chain model
\cite{ours4} (remember that a constrained $S$=1/2 fermion is related to
the canonical ones through the relation $\bar{c}^{\dagger}_{\bj \sigma}=
c^{\dagger}_{\bj \sigma}(1-\hat{n}_{\bj \bar{\sigma}})$).

For correlated hopping model Hamiltonians, the previous
mapping, Eq. (\ref{sdd3}), works well. However, for standard
(non-correlated) Hamiltonians it does not.  Let us introduce a new
transformation that is expected to work in the latter case
\begin{eqnarray}
\begin{cases}
S^{+}_{\bj 1}= c^{\dagger}_{\bj \uparrow} {\bar K}_{\bj \uparrow},
\\
S^{z}_{\bj 1}= \hat{n}_{\bj \uparrow} - \frac {1}{2},
\\
S^{+}_{\bj 2}= c^{\dagger}_{\bj \downarrow} {\bar K}_{\bj \downarrow},
\\
S^{z}_{\bj 2}= \hat{n}_{\bj \downarrow} - \frac {1}{2},
\end{cases}
\label{mn}
\end{eqnarray}
where the nonlocal operator ${\bar K}_{\bj \sigma}$ is now defined as
\begin{eqnarray}
{\bar K}_{\bj \uparrow}&=&\exp[i\pi (\sum_{\bl} \hat{n}_{\bl
\downarrow}+\sum_{{\bl} <{\bj}} \hat{n}_{\bl \uparrow})] \ ,
\nonumber \\
{\bar K}_{\bj \downarrow}&=&\exp[
i\pi \sum_{{\bl}<{\bj}} \hat{n}_{\bl \downarrow}] \ .
\end{eqnarray}
This transformation was originally introduced by Mattis and Nam \cite{Nam}
to solve the following Hubbard-like model
\begin{eqnarray}
H_{\sf MN}= \frac{\epsilon}{4} \sum^{N-1}_{\bj, \sigma} 
(c^{\dagger}_{\bj \sigma}-c^{\;}_{\bj \sigma})
(c^{\dagger}_{\bj+1 \sigma}+c^{\;}_{\bj+1 \sigma})
\nonumber \\
+ U \sum^{N}_{\bj=1} (\hat{n}_{\bj \uparrow}-\frac{1}{2})(\hat{n}_{\bj
\downarrow}-\frac{1}{2}),
\end{eqnarray}
which after the spin-particle transformation, Eq. (\ref{mn}), becomes
($\nu=1,2$)
\begin{equation}
H_{\sf MN}= \epsilon \sum^{N-1}_{\bj, \nu} S^{x}_{\bj \nu} S^{x}_{\bj+1
\nu}  + U \sum^{N}_{\bj=1} S^{z}_{\bj 1} S^{z}_{\bj 2}.
\end{equation}
Using the same transformation, the one-dimensional Hubbard Hamiltonian
\begin{equation}
H^{1d}_{\sf Hubb}=t \!\!\sum^{N-1}_{{\bj}, \sigma}\!
(c^{\dagger}_{{\bj} \sigma} c^{\;}_{{\bf j}+1\sigma}+ 
c^{\dagger}_{{\bj+1} \sigma} c^{\;}_{{\bj}\sigma})+U\sum^{N}_{\bj=1}
(\hat{n}_{\bj\uparrow}-\frac{1}{2})(\hat{n}_{\bj\downarrow}-\frac{1}{2}),
\end{equation}
can be transformed into
\begin{equation}
H^{1d}_{\sf Hubb}=2t \sum^{N-1}_{\bj, \nu} (S^{x}_{\bj \nu} S^{x}_{\bj+1 \nu} 
+ S^{y}_{\bj \nu} S^{y}_{\bj+1 \nu})
+ U \sum^{N}_{\bj=1} S^{z}_{\bj 1} S^{z}_{\bj 2},
\end{equation}
which represents a two-leg ladder made out of two $XY$-chains coupled
by an Ising interaction.

Other interesting examples of quasi-exactly solvable models in 2$d$ of
relevance for strongly correlated matter will be presented in a
separate publication \cite{bos}.

\subsection{Single-band fermionic Hubbard Model}

The Hubbard model is the most popular model of a strongly interacting
system in condensed matter physics. It contains a kinetic energy term
represented by a hopping integral $t$ plus a local on-site Coulomb
repulsion $U$. The single-band Hubbard Hamiltonian is
($\sigma=\uparrow,\downarrow$)
\begin{equation}
H_{\sf Hubb}=t \!\!\sum_{\langle {\bf i,j} \rangle, \sigma}\!
(c^{\dagger}_{{\bf i} \sigma} c^{\;}_{{\bf j}\sigma}+ c^{\dagger}_{{\bf
j} \sigma} c^{\;}_{{\bf i}\sigma})+U\sum_\bi
(\hat{n}_{\bi\uparrow}-\frac{1}{2})(\hat{n}_{\bi\downarrow}-\frac{1}{2}).
\end{equation}
The dimension of the local Hilbert space is $D=4$ since per lattice
site $\bj$ we can only have the states: $\{\ket{0}, c^{\dagger}_{{\bf j} 
\uparrow}\ket{0}, c^{\dagger}_{{\bf j}  \downarrow}\ket{0},
c^{\dagger}_{{\bf j} \uparrow}c^{\dagger}_{{\bf j}
\downarrow}\ket{0}\}$, and $\ket{0}$ is the vacuum state. Replacing the
$c$-operators by the expression for transmutation of statistics given
in Eq.~(\ref{trans}) we get
\begin{eqnarray}
H_{\sf Hubb}&=&t\sum_{\langle {\bf i,j} \rangle, \sigma} (  {\tilde
b}^{\dagger}_{{\bf i}\sigma} \hat{\cal T}^{\;}_{{\bf i}\sigma} 
\hat{\cal T}^{\;}_{{\bf j}\sigma} K^{\dagger}_{\bf i} K^{\;}_{\bf j}
{\tilde b}^{\;}_{{\bf j}\sigma} + {\rm H.c.}  )
\nonumber \\
&+&U\sum_{\bf i} ({\tilde n}_{\bi\uparrow}-\frac{1}{2}) ({\tilde
n}_{\bi\downarrow}-\frac{1}{2}) \ ,
\end{eqnarray}
where $\hat{\cal T}^{\;}_{\bi\uparrow}=\exp[i\pi {\tilde
n}_{\bi\downarrow}]$,  $\hat{\cal T}^{\;}_{\bi\downarrow}=\one$,
$K_{\bf j} = \exp[i \sum_{\bf l}  a({\bf l},{\bf j})  \ \tilde{n}_{\bf
l}]$ , with ${\tilde n}_{\bf j}= {\tilde n}_{\bj\uparrow}+{\tilde
n}_{\bj\downarrow}$, and we have used the result
\begin{equation}
[ \hat{\cal T}^{\;}_{\bj\sigma},K^{\dagger}_{\bf i}]=0 \ .
\end{equation}
The (statistical gauge field) vector potential associated to this 
transmutation of statistics is
\begin{eqnarray} \hspace*{-0.2cm}
A_{\nu}({\bf j})\!=\!\pi({\tilde n}_{{\bf
j}\downarrow} \! - \!\! {\tilde n}_{{\bf j+e_\nu}\downarrow}) \!
-\!\!\sum_{\bf l}[a({\bf l},{\bf j})\!-\!a({\bf l},{\bf j}+{\bf
e}_\nu)] \ {\tilde n}_{\bf l} .
\end{eqnarray}
Using this expression for the vector potential we can rewrite  
$H_{\sf Hubb}$ in the following way
\begin{eqnarray}
H_{\sf Hubb}&=&t\sum_{\langle {\bf i,j} \rangle, \sigma}
( \tilde{b}^{\dagger}_{{\bf i} \sigma} \exp[iA_{\nu}({\bf i})]
\ \tilde{b}^{\;}_{{\bf j}\sigma}+ {\rm H.c}  ) \nonumber \\
&+& U\sum_\bi(\tilde{n}_{\bi\uparrow}-\frac{1}{2})
(\tilde{n}_{\bi\downarrow}-\frac{1}{2}) \ .
\end{eqnarray}
In this way we see that the Hubbard Hamiltonian can be written in a
bosonic representation but with interactions which are non-local when
$d>1$. When $d=1$ the vector potential acts as a correlated hopping
term and the interactions become local 
\begin{eqnarray}
H^{1d}_{\sf Hubb}\!&=&\!\!t\!\sum_{\bi}
[ \ \tilde{b}^{\dagger}_{{\bf i} \uparrow} 
(1-2\tilde{n}_{{\bf i}+1 \downarrow})
\tilde{b}^{\;}_{\bi+1\uparrow}+\tilde{b}^{\dagger}_{{\bf i} \downarrow}
(1-2\tilde{n}_{{\bf i} \uparrow})\tilde{b}^{\;}_{\bi+1\downarrow} 
\nonumber \\
&+& {\rm H.c} \ ]+U\sum_\bi(\tilde{n}_{\bi\uparrow}-\frac{1}{2})
(\tilde{n}_{\bi\downarrow}-\frac{1}{2}) \ .
\end{eqnarray}

The $SU(2)$-spin $S$=3/2 is another physical object for which the 
dimension of the local Hilbert space is $D=4$. From our fundamental 
theorem it is possible to write down these constrained bosons ${\tilde
b}^{\dagger}_{\bj\sigma}$ in terms of spins $S$=3/2. In this way, we can
find a $S$=3/2 representation for the Hubbard model. A possible
mapping between these bosons and a spin $S$=3/2 is
\begin{eqnarray}
S^{+}_\bj&=&\sqrt{3} \ {\tilde b}^{\dagger}_{\bj\downarrow} + 2 \ {\tilde
b}^{\dagger}_{\bj\uparrow}  {\tilde b}^{\;}_{\bj\downarrow} \ ,
\nonumber \\
S^{-}_\bj&=&\sqrt{3} \ {\tilde b}^{\;}_{\bj\downarrow} + 2 \ {\tilde
b}^{\dagger}_{\bj\downarrow} {\tilde b}^{\;}_{\bj\uparrow}\ ,
\nonumber \\
S^{z}_\bj&=& \frac{1}{2}({\tilde n}_{\bj\uparrow}-{\tilde
n}_{\bj\downarrow})+
\frac{3}{2} ({\tilde n}_{\bj\uparrow}+{\tilde n}_{\bj\downarrow}-1) \ .
\label{s32}
\end{eqnarray}
This mapping can be inverted to get the bosonic operators as a function
of the spin operators
\begin{eqnarray}
{\tilde b}^{\dagger}_{{\bf j} \uparrow}&=&\frac{1}{2\sqrt{3}}(S^{+}_\bj)^2
\ , \nonumber \\
{\tilde b}^{\dagger}_{{\bf j}
\downarrow}&=&\frac{1}{\sqrt{3}}S^{+}_\bj(S^z_\bj+\frac{1}{2})^2 \ .
\end{eqnarray}

Similarly, the expression for the spin $S$=3/2 operators in terms of
canonical fermions is
\begin{eqnarray}
S^{+}_\bj&=&\sqrt{3} \ {c}^{\dagger}_{\bj\downarrow} 
{K}^{\;}_{\bj}+ 2 \ 
{c}^{\dagger}_{\bj\uparrow}  {c}^{\;}_{\bj\downarrow} \ ,
\nonumber \\
S^{-}_\bj&=&\sqrt{3} \ {K}^\dagger_{\bj}
{c}^{\;}_{\bj\downarrow} + 2 \ 
{c}^{\dagger}_{\bj\downarrow} {c}^{\;}_{\bj\uparrow}\ ,
\nonumber \\
S^{z}_\bj&=& \frac{1}{2}({\hat n}_{\bj\uparrow}-{\hat
n}_{\bj\downarrow})+
\frac{3}{2} ({\hat n}_{\bj\uparrow}+{\hat n}_{\bj\downarrow}-1) \ .
\label{s32f}
\end{eqnarray}
Again, we can write down the fermionic operators in terms of the spin
operators
\begin{eqnarray}
c^{\dagger}_{{\bf j} \uparrow}&=&\frac{1}{2\sqrt{3}}(S^{+}_\bj)^2
\ {\cal K}^{\dagger}_{\bj\uparrow}
\nonumber \\
c^{\dagger}_{{\bf j}
\downarrow}&=&\frac{1}{\sqrt{3}}S^{+}_\bj(S^z_\bj+\frac{1}{2})^2 \ {\cal
K}^{\dagger}_{\bj\downarrow} \ .
\end{eqnarray}
Using these expressions we can write down the Hubbard model in terms of
$S$=3/2 spins as
\begin{eqnarray}
H_{\sf Hubb}&=& \frac{t}{3} \sum_{\langle {\bf i,j} \rangle} [ S_{\bf
i}^{+}(S^z_{\bf i}+\frac{1}{2})^2 \exp[iA_{\nu}({\bf i})] (S_{\bf
j}^z+\frac{1}{2})^2S_{\bf j}^{-} \nonumber \\
&+& \frac{(S_{\bf i}^{+})^2}{2} \exp[iA_{\nu}({\bf i})]\frac{(S_{\bf
j}^{-})^2}{2} + {\rm H.c.} ] \nonumber \\
&+& \frac{U}{4} \sum_{\bi} [(S^z_{\bf i})^2-\frac{5}{4}]\ .
\end{eqnarray}

The $S$=3/2 representation of the 1$d$ Hubbard model is
\begin{eqnarray}
H^{1d}_{\sf Hubb}&=&  -\frac{t}{3} \sum_{{\bf j}} [S_{\bf
j}^{+}(S^z_{\bf j}+\frac{1}{2}) (S_{{\bf j}+1}^z+\frac{1}{2})^2S_{{\bf
j}+1}^{-} \nonumber \\
&+& \frac{1}{2}(S_{\bf j}^{+})^2  (S^z_{{\bf j}+1}+1)(S_{{\bf
j}+1}^{-})^2 + {\rm H.c.}] \nonumber \\
&+& \frac{U}{4} \sum_{\bj} [(S^z_{\bf j})^2-\frac{5}{4}]\ .
\end{eqnarray}
Since the 1$d$ Hubbard model is Bethe ansatz solvable \cite{lieb}, we
are providing a new spin $S$=3/2 Hamiltonian which is also Bethe ansatz
solvable through the isomorphic mapping between the two languages
(fermions and spins).

We can also write down the 1$d$ Hubbard model using the hierarchical 
$SU(4)$ language. To do so, we fist need to find the dictionary 
connecting the components of an $SU(4)$-spin in the fundamental 
representation with the creation and annihilation operators of the
particle language 
\begin{equation}
{\cal S}({\bf j})\!= \!\begin{pmatrix} {\cal S}^{00}(\bj)&
\tilde{b}^{\;}_{{\bf j} \uparrow}(1-\tilde{n}_{{\bf j} \downarrow})& 
\tilde{b}^{\;}_{{\bf j} \downarrow}(1-\tilde{n}_{{\bf j} \uparrow})& 
\tilde{b}^{\;}_{{\bf j} \downarrow}\tilde{b}^{\;}_{{\bf j} \uparrow} \\
(1-\tilde{n}_{{\bf j} \downarrow})\tilde{b}^{\dagger}_{{\bf j}
\uparrow}&{\cal S}^{11}(\bj)&
\tilde{b}^{\dagger}_{{\bf j} \uparrow}\tilde{b}^{\;}_{{\bf j}
\downarrow} & 
\tilde{n}_{{\bf j} \uparrow}\tilde{b}^{\;}_{{\bf j} \downarrow}\\
(1-\tilde{n}_{{\bf j} \uparrow})\tilde{b}^{\dagger}_{{\bf j}
\downarrow}&
\tilde{b}^{\dagger}_{{\bf j} \downarrow}\tilde{b}^{\;}_{{\bf j}
\uparrow} &{\cal S}^{22}(\bj)&
\tilde{n}_{{\bf j} \downarrow}\tilde{b}^{\;}_{{\bf j} \uparrow}\\
\tilde{b}^{\dagger}_{{\bf j} \uparrow}\tilde{b}^{\dagger}_{{\bf j}
\downarrow} &
\tilde{b}^{\dagger}_{{\bf j} \downarrow}\tilde{n}_{{\bf j} \uparrow}&
\tilde{b}^{\dagger}_{{\bf j} \uparrow}\tilde{n}_{{\bf j} \downarrow}&
{\cal S}^{33}(\bj)
\end{pmatrix} \ ,
\label{spinsu4}
\end{equation}
where ${\cal S}^{00}(\bj)=(1-\tilde{n}_{{\bf j}\uparrow}) 
(1-\tilde{n}_{{\bf j} \downarrow}) -  \frac {1}{4}$, ${\cal
S}^{11}(\bj)=\tilde{n}_{{\bf j} \uparrow}(1-\tilde{n}_{{\bf j}
\downarrow})-\frac{1}{4}$, ${\cal S}^{22}(\bj)=\tilde{n}_{{\bf j}
\downarrow}(1-\tilde{n}_{{\bf j} \uparrow})-\frac{1}{4}$, and ${\cal
S}^{33}(\bj)=\tilde{n}_{{\bf j} \uparrow}\tilde{n}_{{\bf j} \downarrow}
-\frac{1}{4}$.
For a general $SU(N)$ group there are actually two kinds of spinors:
{\it upper} and {\it lower}. The upper spinors transform according to
the conjugate representation. For the particular case of $SU(2)$, the
conjugate representation is equivalent to the original one, i.e., the
conjugation is equivalent to a similarity transformation. However, in
general for $N>2$ the conjugate representation is not equivalent to
the original one. Consequently, the $SU(N>2)$ ferromagnetic and
antiferromagnetic Heisenberg Hamiltonians are essentially different
operators (in the case of $SU(2)$, they just differ by an overall sign)
\cite{Auerbach}. In the same way we wrote in Eq.~(\ref{spinsu4}) the
generators of $su(4)$ in the fundamental representation, we can write
down the corresponding expressions for the generators in the
conjugate representation
\begin{equation}
\tilde{\cal S}({\bf j})\!= \!\begin{pmatrix} -\tilde{\cal S}^{00}(\bj)&
-\tilde{b}^{\;}_{{\bf j} \uparrow} \tilde{n}_{{\bf j} \downarrow}& 
-\tilde{b}^{\;}_{{\bf j} \downarrow} \tilde{n}_{{\bf j} \uparrow}& 
-\tilde{b}^{\;}_{{\bf j} \downarrow}\tilde{b}^{\;}_{{\bf j} \uparrow} \\
-\tilde{n}_{{\bf j} \downarrow} \tilde{b}^{\dagger}_{{\bf j}
\uparrow}& -\tilde{\cal S}^{11}(\bj)&
-\tilde{b}^{\dagger}_{{\bf j} \uparrow}\tilde{b}^{\;}_{{\bf j}
\downarrow} & 
(\tilde{n}_{{\bf j} \uparrow}-1)\tilde{b}^{\;}_{{\bf j} \downarrow}\\
-\tilde{n}_{{\bf j} \uparrow} \tilde{b}^{\dagger}_{{\bf j}
\downarrow}&
-\tilde{b}^{\dagger}_{{\bf j} \downarrow}\tilde{b}^{\;}_{{\bf j}
\uparrow} & -\tilde{\cal S}^{22}(\bj)&
(\tilde{n}_{{\bf j} \downarrow}-1)\tilde{b}^{\;}_{{\bf j} \uparrow}\\
-\tilde{b}^{\dagger}_{{\bf j} \uparrow}\tilde{b}^{\dagger}_{{\bf j}
\downarrow} &
\tilde{b}^{\dagger}_{{\bf j} \downarrow}(\tilde{n}_{{\bf j} \uparrow}-1)&
\tilde{b}^{\dagger}_{{\bf j} \uparrow}(\tilde{n}_{{\bf j} \downarrow}-1)&
-\tilde{\cal S}^{33}(\bj)
\end{pmatrix} \ ,
\label{spinsu4p}
\end{equation}
where $\tilde{\cal S}^{00}(\bj)= \tilde{n}_{{\bf j}
\uparrow}\tilde{n}_{{\bf j} \downarrow} -\frac{1}{4}$, $\tilde{\cal
S}^{11}(\bj)=\tilde{n}_{{\bf j} \downarrow}(1-\tilde{n}_{{\bf j}
\uparrow})-\frac{1}{4}$, $\tilde{\cal S}^{22}(\bj)=\tilde{n}_{{\bf j}
\uparrow}(1-\tilde{n}_{{\bf j} \downarrow})-\frac{1}{4}$, and
$\tilde{\cal S}^{33}(\bj)=(1-\tilde{n}_{{\bf j}\uparrow})
(1-\tilde{n}_{{\bf j} \downarrow}) -  \frac {1}{4}$.

The expression of the 1$d$ Hubbard model in terms of the 
$SU(4)$ hierarchical language is ($J_{\mu\nu}=J_{\nu\mu},
J'_{\mu\nu}=J'_{\nu\mu}$)
\begin{eqnarray}
H^{1d}_{\sf Hubb}\!\!\!&=&\!\!\!\sum_{{\bf j}} [ J_{\mu \nu} {\cal
S}^{\mu \nu}({\bf j}){\cal S}^{\nu \mu}({\bf j}+1) + J'_{\mu \nu} {\cal
S}^{\mu \nu}({\bf j}) \tilde{\cal S}^{\nu \mu}({\bf j}+1) \nonumber \\
&+& \frac {U}{2} ({\cal S}^{00}(\bj)+{\cal S}^{33}(\bj))]
\label{su4hubb}
\end{eqnarray}
where the non-zero {\it magnetic} interactions are: 
$J_{01}=J_{02}=-J_{13}=-J_{23}=t$  and
$J'_{01}=-J'_{02}=-J'_{13}=-J'_{23}=-t$. We can see from Eq.~(\ref{su4hubb})
that in this representation the $U$ term plays the role of a magnetic
field. The case $J'_{\mu \nu}=0$ corresponds to the Hubbard-like model
which was exactly solved by Arrachea and Aligia \cite{Arrachea}.

\subsection{Single-band bosonic Hubbard Model}

We consider now a model of hard-core bosons including two degenerate
orbitals $\alpha=\{1,2\}$ per site
\begin{eqnarray}
H&=&-t \!\! \!\!\sum_{\langle {\bf i,j} \rangle,\alpha,\alpha'}  ({\tilde
b}^{\dagger}_{{\bf i}\alpha} {\tilde b}^{\;}_{{\bf j}\alpha'} +{\tilde
b}^{\dagger}_{{\bf j}\alpha'} {\tilde b}^{\;}_{{\bf i}\alpha}) + U
\sum_{\bf j} {\tilde n}_{{\bf j}1} {\tilde n}_{{\bf j}2}
\nonumber \\
&+&  V \sum_{\langle {\bf i,j} \rangle} ({\tilde n}_{\bf i}-1) ({\tilde
n}_{\bf j}-1)- \mu \sum_{\bf j} {\tilde n}_{\bf j},
\label{2hc}
\end{eqnarray}
where $\tilde{n}_{{\bf j}\alpha}= {\tilde b}^{\dagger}_{{\bf j}\alpha}
{\tilde b}^{\;}_{{\bf j}\alpha}$ and $\tilde{n}_{\bf j}=\tilde{n}_{{\bf
j}1}+\tilde{n}_{{\bf j}2}$. The first term represents a hopping
connecting pair of orbitals belonging to nearest-neighbor sites
$\langle {\bf i,j} \rangle$. The second is a local Coulomb repulsion
term between particles occupying different orbitals of the same site.
The third term corresponds to a nearest-neighbor density-density
repulsion $V$. The dimension of the local Hilbert space is $D=4$ since
per lattice site $\bj$ we can only have the states: $\{\ket{0}, {\tilde
b}^{\dagger}_{{\bf j}1}\ket{0}, {\tilde b}^{\dagger}_{{\bf j}2}\ket{0},
{\tilde b}^{\dagger}_{{\bf j}1}{\tilde b}^{\dagger}_{{\bf
j}2}\ket{0}\}$, and $\ket{0}$ is the vacuum state. 

As mentioned in section \ref{sec3b}, the hard-core boson operators in
each site generate an $su(2)$ algebra:  $\{{\tilde b}^{\dagger}_{{\bf
i}\alpha},{\tilde b}^{\;}_{{\bf i}\alpha}, \tilde{n}_{{\bf i}\alpha}-1/2
\}$. The associated representation is the fundamental one, i.e., 
$S$=1/2. Since there are two orbitals per site, the local algebra
associated in each site is the direct sum of two $su(2)$ algebras, and
the representation is the direct sum of both $S$=1/2 representations.
It is well known that by reducing the direct sum we get two irreducible
representations: the singlet and the triplet representations. The
singlet is associated to the antisymmetric (under the permutation
of both orbitals) state 
\begin{equation}
\ket{\psi_{{\bf i}{\sf A}}}=\frac{1}{\sqrt{2}}  ({\tilde
b}^{\dagger}_{{\bf i}1}-{\tilde b}^{\dagger}_{{\bf i}2}) \ket{0} = 
{\tilde b}^{\dagger}_{{\bf i}{\sf A}} \ket{0}\ .
\end{equation} 
The three remaining states, which belong to the triplet representation,
are symmetric. These states can be generated by the creation operator
\begin{equation}
{\tilde b}^{\dagger}_{{\bf i}{\sf S}}=\frac{1}{\sqrt{2}}
({\tilde b}^{\dagger}_{{\bf i}1}+{\tilde b}^{\dagger}_{{\bf i}2}) \ .
\end{equation}
By applying this operator to the vacuum state $\ket{0}$ one can get a
particular  basis for the triplet representation
\begin{equation}
\{\ket{0}, {\tilde b}^{\dagger}_{{\bf i}{\sf S}}\ket{0}, {\tilde
b}^{\dagger}_{{\bf i}{\sf S}}{\tilde b}^{\dagger}_{{\bf i}{\sf
S}}\ket{0}\} \ .
\label{triplet}
\end{equation}
The singlet plus the triplet states generate another basis for the 
local Hilbert space of $H$ with $D=4$ (${\tilde n}_\bj=  {\tilde
b}^{\dagger}_{{\bf j}{\sf A}} {\tilde b}^{\;}_{{\bf j}{\sf A}}+{\tilde
b}^{\dagger}_{{\bf j}{\sf S}} {\tilde b}^{\;}_{{\bf j}{\sf S}}$). If we
apply the MM \cite{matsubara} transformation to these hard-core bosons,
we get a spin $S$=1/2 for each orbital. After that transformation it
becomes clear that $\ket{\psi_{{\bf i}{\sf A}}}$ is the singlet state,
while the other three (Eq.~(\ref{triplet})) are the triplet states with
$S_z=-1,0,1$. Note that the local algebra satisfied by ${\tilde
b}^{\dagger}_{{\bf i}{\sf A}}$ and ${\tilde b}^{\dagger}_{{\bf i}{\sf
S}}$ is not the same as the one  satisfied by ${\tilde
b}^{\dagger}_{{\bf i}1}$ and  ${\tilde b}^{\dagger}_{{\bf i}2}$. In
particular, ${\tilde b}^{\dagger}_{{\bf i}{\sf A}} {\tilde
b}^{\dagger}_{{\bf i}{\sf S}}={\tilde b}^{\dagger}_{{\bf i}{\sf S}}
{\tilde b}^{\dagger}_{{\bf i}{\sf A}}=0$.

Going back to our model, Eq.~(\ref{2hc}), we immediately notice that
the local singlet state $\ket{\psi_{{\bf i}{\sf A}}}$
is invariant under the application of $H$. In other words, if there
is one particle in an antisymmetric state at site $\bi$, that local state
will be conserved. There is a local $U(1)$ symmetry associated to this
conservation
\begin{equation}
\ket{\psi_{{\bf i}{\sf A}}} \rightarrow  e^{\Phi_\bi}\ket{\psi_{{\bf
i}{\sf A}}} \ ,
\end{equation}
which leaves $H$ invariant. Consequently, one can identify invariant
subspaces of $H$. These subspaces are classified according to the set
of sites ${\cal N}_{\sf A}$ which are in a singlet state. We can now take
advantage of the invariance of these subpaces, and project $H$ onto 
each of them to reduce the effective number of degrees of freedom. The
expression for $H$ (up to a constant) restricted to the subspace
associated to the set ${\cal N}_{\sf A}$ is
\begin{eqnarray}
H_{{\cal N}_{\sf A}}&=& -2t \sum_{\overline{\langle {\bf i,j}\rangle}}  
({\tilde b}^{\dagger}_{{\bf i}{\sf S}} 
{\tilde b}^{\;}_{{\bf j}{\sf S}} +{\tilde b}^{\dagger}_{{\bf j}{\sf
S}}  {\tilde b}^{\;}_{{\bf i}{\sf S}}) - \mu 
\sum_{\bar{\bf j}} {\tilde n}_{{\bf j}{\sf S}} \nonumber \\
&+& V \sum_{\overline{\langle {\bf i,j}\rangle}}
 ({\tilde n}_{{\bf i}{\sf S}}-1) ({\tilde n}_{{\bf j}{\sf S}}-1)
\nonumber \\
&+& \frac{U}{2} \sum_{\bar{\bf j}}  {\tilde
n}_{{\bf j}{\sf S}} ({\tilde n}_{{\bf j}{\sf S}}-1) \ ,
\label{hf}
\end{eqnarray}
where $\overline{\langle {\bf i,j} \rangle}$ and $\bar{\bf j}$ means
sites $\bi,\bj \in \!\!\!\!\!/ \ {\cal N}_{\sf A}$ and
$\tilde{n}_{\bj{\sf S}} = 1-[\tilde{b}^{\;}_{{\bf j}{\sf S}},\tilde{
b}^{\dagger}_{{\bf j}{\sf S}}]$. Having a singlet at a given site $\bj$
is equivalent to take the site out of the lattice since that local
singlet is {\it frozen} at $\bj$ (Pauli blocking). The action of $H$ on each invariant
subspace can be described by the operators ${\tilde b}^{\dagger}_{{\bf
i}{\sf S}}$ and ${\tilde b}^{\;}_{{\bf i}{\sf S}}$, which are
generators of an $su(2)$ algebra in the $S$=1 (triplet) representation.
This is easily seen using the mapping of Eq.~(\ref{g-on2}) after
replacing the ${\bar g}^{\;}_{{\bf j}}$'s by the ${\tilde b}^{\;}_{{\bf
j}{\sf S}}$'s operators acting only on sites  $\bj \in \!\!\!\!\!/ \
{\cal N}_{\sf A}$
\begin{eqnarray}
S^+_\bj &=& \sqrt{2} \ \tilde{b}^{\dagger}_{\bj{\sf S}}  \ , \nonumber \\
S^-_\bj &=& \sqrt{2} \ \tilde{b}^{\;}_{\bj{\sf S}} \ , \nonumber \\
S^z_\bj &=&  \tilde{n}_{\bj{\sf S}} - 1 \ . 
\label{g-on3}
\end{eqnarray}

A natural consequence of this transformation is the possibility of
mapping $H_{{\cal N}_{\sf A}}$ onto an $S$=1 spin Hamiltonian. This is
done by inverting Eq. (\ref{g-on3})
\begin{eqnarray} 
{\tilde b}^{\dagger}_{{\bf j}{\sf S}}&=& \frac{1}{\sqrt{2}}S_{\bf j}^{+}
\ , \nonumber \\
{\tilde b}^{\;}_{{\bf j}{\sf S}}&=& \frac{1}{\sqrt{2}}S_{\bf j}^{-}
\ , \nonumber \\
{\tilde n}_{{\bf j}{\sf S}}&=&S_{\bf j}^{z}+1 \ ,
\label{bspin}
\end{eqnarray}
and replacing these expressions into Eq.~(\ref{hf}) to get
\begin{eqnarray}
H_{{\cal N}_{\sf A}}&=& -t \sum_{\overline{\langle {\bf i,j}\rangle}} 
(S^{+}_{\bf i}  S^{-}_{\bf j} +S^{-}_{\bf i}  S^{+}_{\bf j}) + V
\sum_{\overline{\langle {\bf i,j}\rangle}} S^{z}_{\bf i} S^{z}_{\bf j}
\nonumber \\
&+& \frac{U}{2} \sum_{\bar{\bj}}  (S^{z}_{\bj})^2+ (\frac{U}{2}-\mu)
\sum_{\bar{\bj}} S^{z}_{\bf j} \ .
\label{shf}
\end{eqnarray}
This is an anisotropic Heisenberg model ($J_{z}=V$ and
$J_{x}=J_{y}=-2t$) with a magnetic field $B_z=U/2-\mu$ applied along
the $z$ axis. In addition, there is an easy-plane single-ion
anisotropy. Therefore, each invariant subspace of $H$ gives rise to an
anisotropic Heisenberg model $H_{{\cal N}_{\sf A}}$  acting  on a
partially depleted lattice (i.e., the sites in ${\cal N}_{\sf A}$ are
removed from the lattice). From the variational principle it becomes
evident that the lowest energy subspace is the one containing no local
singlet  states, i.e., ${\cal N}_{\sf A}$ is the empty set $\emptyset$
(the original lattice is not depleted).

\begin{figure}[htb]
\includegraphics[angle=0,width=8.6cm,scale=1.0]{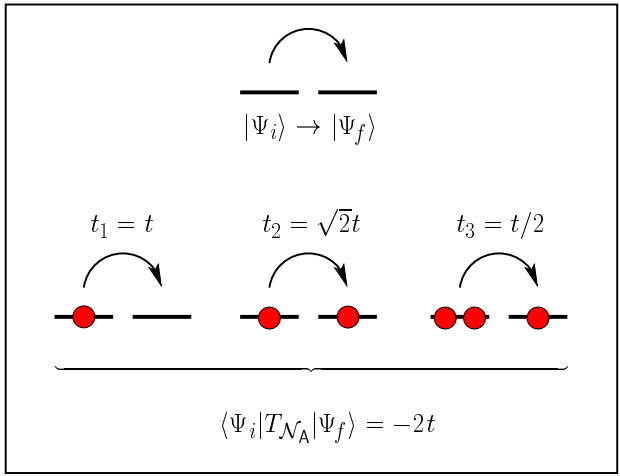}
\caption{Hopping processes for the bosonic Hubbard model written in the
$g$-particles language, Eq.~(\ref{bos-hub-g}). There are three different
hopping processes $t_1$, $t_2$, and $t_3$. Note, however, that the
non-vanishing matrix elements of the kinetic energy $T_{{\cal N}_{\sf
A}}$ are the same (-2$t$) regardless of the hopping process.}
\label{fig4}
\end{figure}

We know from our fundamental theorem that there are many other possible
languages that may be used to describe the present system. In
particular, since $({\tilde b}^{\dagger}_{{\bf i}{\sf S}})^3=0$  it is
clear that these modes satisfy a generalized Pauli exclusion principle
with $p=2$ and, therefore, one can establish a mapping to 
$g$-particles which satisfy the algebra written in Eq. (\ref{canog}). 
The explicit form for this mapping is
\begin{eqnarray} 
{\tilde b}^{\dagger}_{{\bf j}{\sf S}}&=& g^{\dagger}_{\bf
j} \ [1+(\frac{1}{\sqrt{2}}-1) n^{g}_{\bf j}] \ ,
\nonumber \\
{\tilde b}^{\;}_{{\bf j}{\sf S}}&=& [1+(\frac{1}{\sqrt{2}}-1) 
n^{g}_{\bf j}] \ g^{\;}_{\bf j} \ ,
\nonumber \\
{\tilde n}_{{\bf j}{\sf S}}&=& n^{g}_{\bf j},
\label{bspin1}
\end{eqnarray}
where $n^{g}_{\bf j}=g^{\dagger}_{\bf j}g^{\;}_{\bf j}$. Again, by
replacing these expressions into Eq.~(\ref{hf}), we can write down 
$H_{{\cal N}_{\sf A}}$ in terms of the $g$-particles as
\begin{eqnarray}
H_{{\cal N}_{\sf A}}&=& -\sum_{\overline{\langle {\bf i,j}\rangle}}  
(g^{\dagger}_{\bf i} g^{\;}_{\bf j} 
+g^{\dagger}_{\bf j} g^{\;}_{\bf i})(h_1+h_2+h_3)\nonumber \\
&+& \frac{U}{2} \sum_{\bar{\bf j}}  {n}_{{\bf j}}^g 
({n}_{{\bf j}}^g-1) + V \sum_{\overline{\langle {\bf i,j}\rangle}}
({n}_{{\bf i}}^g-1) ({n}_{{\bf j}}^g-1) \nonumber \\
&-&\mu \sum_{\bar{\bf j}} {n}_{{\bf j}}^g \ ,
\label{bos-hub-g}
\end{eqnarray}
where $h_1=t_1(n^g_{\bf i}+n^g_{\bf j}-2)(n^g_{\bf i}+n^g_{\bf j}-3)$, 
$h_2=-t_2(n^g_{\bf i}+n^g_{\bf j}-1)(n^g_{\bf i}+n^g_{\bf j}-3)$, and
$h_3=t_3(n^g_{\bf i}+n^g_{\bf j}-1)(n^g_{\bf i}+n^g_{\bf j}-2)$ with
$t_1=t$, $t_2=\sqrt{2}t$, and $t_3=t/2$. The correlated hopping terms
values are such that the matrix elements of the three possible hopping
processes, illustrated in Fig.~\ref{fig4}, are the same.

\subsection{BCS reduced Hamiltonian}

These general mappings between languages are not restricted to real 
space lattices, indeed one can find mappings between modes in Fourier
space, for instance. A very well-know example is the one introduced by
P.W. Anderson \cite{anderson} for the case of the
Bardeen-Cooper-Schrieffer (BCS) reduced Hamiltonian
\begin{eqnarray}
H_{\sf BCS}&=&\sum_{\bf k} \epsilon_{\bf k}(\hat{n}_{{\bf k}\uparrow}+
\hat{n}_{{\bf -k}\downarrow}-1) \nonumber \\
&-& V \sum_{\bf k,k'} \! ' \ c^{\dagger}_{\bf
k'\uparrow}c^{\dagger}_{-\bf k'\downarrow} c^{\;}_{-\bf
k\downarrow}c^{\;}_{\bf k\uparrow} \ ,
\end{eqnarray}
where the (canonical) fermionic modes (electrons) are defined on a
momentum space lattice with momentum index $\bk$ and $\hat{n}_{{\bf
k}\sigma}= c^{\dagger}_{\bf k\sigma} c^{\;}_{\bf k\sigma}$ (with
$\sigma=\uparrow,\downarrow$). Again, this is a case  where the
dimension of the local Hilbert space is $D=4$ with a possible basis
$\{\ket{0}, c^{\dagger}_{{\bf k} \uparrow}\ket{0}, c^{\dagger}_{{\bf
k}  \downarrow}\ket{0}, c^{\dagger}_{{\bf k} \uparrow}c^{\dagger}_{{\bf
k} \downarrow}\ket{0}\}$, and it turns out it is exactly solvable.

It is clear from this Hamiltonian that if the total charge of the pair
(${\bf k}\uparrow,{\bf -k}\downarrow$) is one (i.e., the pair state is
occupied by a single electron, $\hat{n}_{{\bf k}\uparrow}=1$ and
$\hat{n}_{{\bf -k}\downarrow}=0$, or $\hat{n}_{{\bf k}\uparrow}=0$ and
$\hat{n}_{{\bf -k}\downarrow}=1$) it will remain equal to one after
application of $H_{\sf BCS}$ (Pauli blocking). We will show now that this conservation
is derived from an $SU(2)$ gauge symmetry of the BCS reduced
Hamiltonian which becomes explicit when we write $H_{\sf BCS}$ in terms
of the generators of $su(2)\bigoplus su(2)$. This constitutes another
example of the use of alternative languages to unveil hidden symmetries
and, to our knowledge, this has not been revealed in the way we will
present below.

We just need to note that the following sets of operators 
\begin{eqnarray}
{\tau}^{z}_{\bf k}&=&\frac{1}{2} (\hat{n}_{{\bf k}\uparrow}+\hat{n}_{{\bf
-k}\downarrow}-1) \ ,\nonumber \\
{\tau}^{+}_{\bf k}&=& c^{\dagger}_{{\bf k}\uparrow} c^{\dagger}_{{\bf
-k}\downarrow} \ ,\nonumber \\
{\tau}^{-}_{\bf k}&=& c^{\;}_{{\bf -k}\downarrow} c^{\;}_{{\bf k}\uparrow}
\ ,
\label{set1}
\end{eqnarray}
and
\begin{eqnarray}
S^{z}_{\bf k}&=&\frac{1}{2} (\hat{n}_{{\bf k}\uparrow}-\hat{n}_{{\bf
-k}\downarrow})\ , \nonumber \\
S^{+}_{\bf k}&=& c^{\dagger}_{{\bf k}\uparrow} c^{\;}_{{\bf
-k}\downarrow}\ , \nonumber \\
S^{-}_{\bf k}&=& c^{\dagger}_{{\bf -k}\downarrow} c^{\;}_{{\bf
k}\uparrow} \ ,
\label{set2}
\end{eqnarray}
satisfy the spin $su(2)$ commutation relations ($\mu,\nu,\lambda=x,y,z$)
\begin{eqnarray}
\left [ S_\bk^\mu, S_{\bk'}^\nu \right ] &=& i \delta_{\bk\bk'} 
\epsilon_{\mu \nu\lambda} S_\bk^\lambda \ , \nonumber \\
\left [ \tau_\bk^\mu, \tau_{\bk'}^\nu \right ] &=& i \delta_{\bk\bk'} 
\epsilon_{\mu \nu\lambda} \tau_\bk^\lambda \ , \nonumber \\
\left [ S_\bk^\mu, \tau_{\bk'}^\nu \right ] &=& 0\ . 
\label{core}
\end{eqnarray}
The last set of local $su(2)$ generators is associated to the
conservation of the parity of the total charge in each pair $({\bf
k}\uparrow,{\bf -k}\downarrow$). Since these are local transformations,
the symmetry generated by them is an $SU(2)$ gauge symmetry. To see
this we just need to re-write $H_{\sf BCS}$ in the pseudospin language
\cite{anderson}
\begin{equation}
H_{\sf BCS}=2 \sum_{\bf k} \epsilon_{\bf k} \tau^{z}_{\bf k} -V
\sum_{\bf k,k'}\! ' \ (\tau^{x}_{\bf k'}{\tau}^{x}_{\bf
k}+\tau^{y}_{\bf k'}{\tau}^{y}_{\bf k}) \ .
\end{equation}
Since $H_{\sf BCS}$ is only a function of the set of local $su(2)$
generators given by Eq. (\ref{set1}) it commutes with the set of
generators given by Eq. (\ref{set2}). 

In this way, this $SU(2)$ gauge symmetry splits the total Hilbert space
into invariant subspaces labelled by the set of momenta ${\cal N}_{\sf
bp}$ which have single occupancy of electrons. Within each subspace
(which corresponds to a depleted momentum space lattice) the operators
$\bar{b}^\dagger_\bk=c^{\dagger}_{{\bf k}\uparrow} c^{\dagger}_{{\bf
-k}\downarrow}$ are the single-flavor hard-core bosons whose algebra
was defined in Eq.~(\ref{conm}), and the Hamiltonian becomes 
\begin{equation}
H_{\sf BCS}^{{\cal N}_{\sf bp}}=2 \sum_{\overline{\bk}} \epsilon_{\bf
k} \bar{b}^\dagger_\bk\bar{b}^{\;}_\bk -V \sum_{
\overline{\bk},\overline{\bk'}}\! ' \
\bar{b}^\dagger_{\bk'}\bar{b}^{\;}_\bk \ ,
\end{equation}
where $\overline{\bk}$ and $\overline{\bk'}$ mean sites in a momentum
lattice $\bk,\bk' \np \ {\cal N}_{\sf bp}$. (This model is exactly
solvable.) Thus, a superconducting BCS state is a Bose-Einstein
condensate, in momentum space, of hard-core bosons (i.e., spinless JW
particles). 

\subsection{Bosonic $t$-$J$ model}

In the previous section \ref{sec4a} we introduced an $su(N)$ language for
the  hard-core bosons (see Eq. (\ref{fund})). In other words, we showed
that the  generators of $su(N)$ in the fundamental representation can
be used to describe a system of hard-core bosons with $N_f=N-1$
different flavors. To illustrate a possible application of this result,
we will consider here the case of spin $S$=1/2 hard-core bosons (i.e.,
$N_f=2$ and $D=N=3$). In particular, we will show that the bosonic
$t$-$J$ model \cite{Noten4}
\begin{eqnarray}
H^b_{t \mbox{\small -}J}&=&t \! \sum_{\langle {\bf i},{\bf j}
\rangle,\sigma} ( {\bar{b}}^\dagger_{{\bf i} \sigma}
{\bar{b}}^{\;}_{{\bf j} \sigma} + {\bar{b}}^\dagger_{{\bf j} \sigma}
{\bar{b}}^{\;}_{{\bf i} \sigma} ) + J \sum_{\langle
{\bf i},{\bf j} \rangle} ({\bf s}_{\bf i} \cdot  {\bf s}_{\bf j} - 
\frac{\bar{n}_{\bf i} {\bar{n}}_{\bf j}}{4})\nonumber \\ 
&-&  \mu \sum_{{\bf j}} {\bar{n}}_{\bf j} \ ,
\label{hamiltJb}
\end{eqnarray}
with ${\bf s}_{\bf j}=\frac{1}{2} {\bar {b}}^\dagger_{{\bf j} \alpha}
\boldsymbol{\sigma}_{\alpha\beta} {\bar {b}}^{\;}_{{\bf j} \beta}$
($\boldsymbol{\sigma}$ denoting Pauli matrices), 
can be rewritten as an anisotropic $SU(3)$ Heisenberg model. To this end 
we just need to use the identities given in Eq.~(\ref{fund}) to replace 
the bosonic operators by the corresponding $su(3)$ generators 
${\cal S}^{\mu \nu}({\bf j})$ to get
\begin{eqnarray}
H^b_{t \mbox{\small -}J} \!\!&=&\! \sum_{\langle {\bf i},{\bf j}
\rangle} J_{\mu \nu} {\cal S}^{\mu \nu}({\bf i}) {\cal S}^{\nu
\mu}({\bf j}) + B \sum_{{\bf j}}{\cal S}^{00}({\bf j}) \ ,
\label{newform}
\end{eqnarray}
with ($J_{\mu\nu}=J_{\nu\mu}$)
\begin{eqnarray}
J_{00}&=&-\frac{J}{2}, \;\;\;\;J_{10}=J_{20}=t,
\nonumber \\
J_{12}&=&J_{11}=J_{22}=\frac{J}{2},\;\; B=\mu+\frac{{\sf z}J}{6} \ ,
\end{eqnarray}
and ${\sf z}$ is the coordination of the lattice. In this way we have
established an exact mapping between the $t$-$J$ model and an
anisotropic $SU(3)$ Heisenberg model with an applied `magnetic field'
$B$. 

We have also seen in section \ref{sec4a} that the $su(N)$ generators 
can be represented in terms of Schwinger-Wigner bosons. Therefore, we
can use Schwinger-Wigner bosons to get a mean-field solution of the 
Hamiltonian (Eq. (\ref{newform})). In this way we see how the change
of language in the bosonic $t$-$J$ model opens the possibility to 
get a simple and original solution \cite{ours5}. 

These examples illustrate the process of unveiling hidden connections
between different problems just by changing the language. We also saw
in the bosonic Hubbard model example that the same process which
unveils hidden symmetries reduces the whole space to a direct sum of
invariant subspaces. The effective number of degrees of freedom in each
subspace is lower than the original one. Consequently, we can use a
new  class of languages ${\cal A} \wedge \Gamma$ with a lower ${\rm
dim} \ \Gamma$ to describe the original problem. This reduction not only
simplifies the resolution of the problem, but also establishes new
connections with other physical systems.

\subsection{$SO(3)$ Heisenberg Model}

As it was shown in Eq.~(\ref{spinso3}), the generators of $so(3)$ are
the components of an antisymmetric tensor of order three. By
contracting this tensor with the completely antisymmetric one
$\epsilon_{\alpha \mu \nu}$, we can see that the three generators of
$so(3)$ transform like a vector $\bf L$
\begin{equation}
L_{\alpha}({\bf j})= \epsilon_{\alpha \mu \nu} {\cal M}^{\mu \nu}({\bf
j}) \ .
\end{equation}
It is well known that the $so(3)$ and $su(2)$ algebras are isomorphic.
As a consequence of this, we can establish a simple mapping between
${\bf L}$ and the three generators of $su(2)$ ${\bf S}$ in the spin one
representation
\begin{equation}
{\bf S}({\bf j})= i {\bf L}({\bf j}) \ .
\end{equation}
Using this basic result of group theory, we can map the
$SO(3)$ Heisenberg model
\begin{eqnarray}
\lefteqn{H^{\sf SO(3)}_{\sf Heis}=J \sum_{\langle \bi,\bj \rangle} 
{\bf L}({\bf i}) \cdot {\bf L}({\bf j})}  \\
&=& -4J \sum_{\langle \bi ,\bj \rangle} s^{y}_\bi s^{y}_\bj+J
\sum_{\langle \bi,\bj \rangle \sigma} ({\bar
b}^{\dagger}_{\bi\sigma}{\bar b}^{\dagger}_{\bj\sigma} -{\bar
b}^{\dagger}_{\bi\sigma}{\bar b}^{\;}_{\bj\sigma} +{\rm H.c.})\nonumber
\end{eqnarray}
into the $S=1$ $SU(2)$ Heisenberg model:
\begin{equation}
H^{\sf SO(3)}_{\sf Heis}= - H^{\sf SU(2)}_{\sf Heis}= - J \sum_{\langle
\bi,\bj \rangle} {\bf S}({\bf i}) \cdot {\bf S}({\bf j}) \ .
\end{equation}
In other words, the $S=1$ $SU(2)$ Heisenberg model is equivalent  to a
model consisting of a gas of $S=1/2$ hard-core bosons with a magnetic
interaction along the $y$-axis and a superconducting term.

\section{Broken versus Emergent Symmetry}
\label{sec5-6}

Matter organizes in different manners depending upon the nature of its
constituents and interactions. Symmetry and topology are fundamental
guiding principles behind this organization. A symmetry transformation
is a modification of the observer's point of view that does not change
the outcome of an experiment performed on the same system. In
mathematical terms it is a transformation that takes the Hilbert space
of states $\cal H$ into an equivalent one. Wigner's theorem asserts
that any transformation $\hat{T}$ which conserves the transition
probability between rays in $\cal H$
\begin{equation}
|\langle \hat{T}^\dagger \Psi | \hat{T} \Phi \rangle|^2= |\langle \Psi
| \Phi \rangle|^2
\end{equation}
can be represented by a linear unitary or antilinear antiunitary map
$\cal O$ (${\cal O}^\dagger={\cal O}^{-1}$) on $\cal H$. Since the
operation of time-reversal is one of the few relevant examples in
physics which involves an antiunitary operator, we will only consider
unitary mappings in the following.  

Symmetries may be classified as external or {\it space-time} (e.g.,
the  Poincar\'e group) and internal. The latter refers to the set of
transformations that leaves the  Hamiltonian of the system $H$
invariant; i.e., these are the symmetries of the physical laws. This
set forms a group which is named the internal symmetry group $\cal G$
and is defined as 
\begin{equation}
{\cal G} = \{\fg_\alpha\} \ , \ \mbox{with group elements} \ \fg_\alpha  , \
\mbox{that satisfy } [H,\fg_\alpha]=0 \  ,
\end{equation}
with $\fe$ representing the identity element, and where the number of
elements defines its order, which may be finite, denumerable infinite
(discrete), or non-denumerable infinite (continuous). In general,
groups of symmetries in physics are either finite or Lie groups
(non-denumerable infinite). Besides, the group $\cal G$ may be Abelian
(i.e., $[\fg_\alpha,\fg_{\alpha'}]=0, \forall \alpha,\alpha'$) or
non-Abelian, and local (also called gauge, meaning that the symmetry
applies to subsystems of the original physical system) or global.
Invariant physical observables, $O$, are those Hermitian operators
which remain invariant under the symmetry group $\cal G$, i.e.,
$[O,\fg_\alpha]=0$. Every observable which is a function of the groups
elements $\fg_\alpha$ is a constant of motion since it commutes with
$H$. Table \ref{tablemod} shows representative examples of physical
models displaying different kinds of symmetries.
\begin{table}[htdp]
\begin{center}
\begin{tabular}{|c||c|c|}
\hline
 &        &        \\
{\sf Symmetry}&          Discrete & Continuous  \\
 &        &        \\\hline \hline
 &           Ising ($Z_2$)&  classical $XY$ ($O(2)$) \\ 
Global  &        &        \\ 
 & $XYZ$ ($D_{2h}$)       & Heisenberg ($SU(2)$)       \\\hline
 &  Gauge magnet ($Z_2$)      &  Gauge magnet ($U(1)$)      \\
Local   &      &         \\
&        &   BCS (hidden $SU(2)$)    \\\hline
\end{tabular}
\end{center}
\caption{Examples of models displaying different kinds of symmetries. The group
(or subgroup) of symmetries involved is written in parenthesis.}
\label{tablemod}
\end{table}
For each group element $\fg_\alpha$ there is a unitary operator (see
above) that will be denoted as ${\cal O}_\alpha={\cal O}(\fg_\alpha)$
which maps $\cal H$ into an equivalent Hilbert space. The set $\{{\cal
O}_\alpha\}$ forms a representation of the group $\cal G$. A
representation is an homomorphic mapping of the group $\cal G$ onto a
set of linear operators ${\cal O}$ such that: ${\cal O}(\fe)=\one$, and
${\cal  O}_\alpha {\cal  O}_\beta= {\cal O}(\fg_\alpha\fg_\beta)$. The
dimension of the representation, dim$(\cal O)$, is the dimension of the
(vector) space on which it acts. By a representation we will mean a
non-singular (in particular, unitary) dim$(\cal O)$$\times$dim$(\cal
O)$ matrix representation. A representation is irreducible if its
invariant subspaces under the action of all the elements of the group
are only ${\bf 0}$ and the full space. A completely reducible
representation can be written as a direct sum of irreducible
representations (irreps). The eigenstates of $H$ that have the same
eigenvalue $E_n$ form an invariant subspace
\begin{eqnarray}
H {\cal O}^{\;}_\alpha \ket{\Psi_n}&=& {\cal O}^{\;}_\alpha H \ket{\Psi_n}=E_n
{\cal O}_\alpha \ket{\Psi_n}
\end{eqnarray}
meaning $\ket{\Psi'_n}= {\cal O}_\alpha \ket{\Psi_n}$ is also an
eigenstate with the same eigenvalue. When the dimension of this
invariant subspace is larger than one, the energy eigenvalue $E_n$ is
{\it degenerate}. The dimension of the {\it degenerate} subspace is
equal to the dimension of the representation of $\cal G$ associated
with the eigenstate $\ket{\Psi_n}$. If the group $\cal G$ is Abelian
all the irreps are one-dimensional and there is no degeneracy induced
by $\cal G$.


Lie groups play a fundamental role in physics. There is a notion of
continuity or {\it closeness} imposed on the elements of the group
manifold $\fg_\alpha$ such that a finite transformation of the group
can be generated by a series of infinitesimal ones. For a one-parameter
continuous group the representations (the homomorphic mapping must be
continuous) of its elements can be written
\begin{eqnarray}
{\cal O}_\alpha(\theta)=\exp[i \theta X_\alpha] \ ,
\end{eqnarray}
where $\theta$ is a continuous parameter and $X_\alpha$'s are the
generators of the Lie algebra. The representations of the group elements
are defined such that $\theta=0$ represents the identity operator
$\one$ and an infinitesimal transformation $\delta\theta$ is expressed
as
\begin{eqnarray}
{\cal O}_\alpha(\delta\theta)=\one+i \ \delta\theta \  X_\alpha \ ,
\end{eqnarray}
where the generators form a Lie algebra,
\begin{eqnarray}
[X_\alpha, X_\beta]= i C_{\alpha\beta}^\gamma \ X_\gamma \ ,
\end{eqnarray}
with $C_{\alpha\beta}^\gamma$ representing the structure constants of
the algebra. Notice that the generators themselves are conserved
quantities; i.e., $[H,X_\alpha]=0$.

Let us provide an example to show how these ideas are applied. Suppose
we have the following model Hamiltonian representing interacting
spinless fermions
\begin{equation}
H=-t\sum_{\langle \bi,\bj \rangle} (c^\dagger_\bi c^{\;}_\bj + c^\dagger_\bj
c^{\;}_\bi)+ V  \sum_{\langle \bi,\bj \rangle} (\hat{n}_\bi-\frac{1}{2})
(\hat{n}_\bj-\frac{1}{2}) \ ,
\label{tV2d}
\end{equation}
where $\langle \bi,\bj \rangle$ represents nearest neighbors in an
otherwise bipartite lattice (i.e., the union of two interpenetrating
sublattices $A$ and $B$). Among the elements of $\cal G$, there are two
Abelian symmetries: one continuous and global $U(1)$ related to charge
conservation, and another discrete and local (staggered) $Z_2$ related
to a particle-hole transformation. The continuous symmetry is realized
by the unitary mappings
\begin{equation}
{\cal O}^{\;}_\theta = \exp[i \theta \sum_\bj \hat{n}_\bj] \ , \ \ {\cal
O}^{\;}_\theta \  c^\dagger_\bj \ {\cal O}^\dagger_\theta = \exp[i \theta]
c^\dagger_\bj \ ,
\end{equation}
while the discrete one is realized by the identity and the
particle-hole transformation  
\begin{eqnarray}
{\cal O}^{\;}_{\sf p-h}&=& \prod_\bj \exp[i \pi \delta_{\bj B}\hat{n}_\bj]\exp[i
\frac{\pi}{2}(c^\dagger_\bj + c^{\;}_\bj)] \ , \nonumber \\
\ {\cal O}^{\;}_{\sf p-h} \  c^\dagger_\bj \ {\cal O}^\dagger_{\sf p-h}&=& 
\begin{cases} 
\ \ c^{\;}_\bj   \ \ \  \mbox{sublattice $A$} \\ 
-c^{\;}_\bj  \ \ \      \mbox{sublattice $B$}   
\end{cases} \ ,
\end{eqnarray}
where $\delta_{\bj B}$ is one if $\bj$ belongs to sublattice $B$ and
zero otherwise.

In some instances the states of matter display the symmetries
compatible with the quantum equations of motion (i.e., symmetries of
$H$)
\begin{equation}
H \ket{\Psi(t)} =  i \partial_t \ket{\Psi(t)} \ ,
\end{equation}
while other more interesting situations are characterized by states
with less (broken-symmetry scenario) or with more symmetries ({\it
emergent symmetry} scenario). In subsection \ref{sec5-6B}  we will
introduce and expand on the latter concept.

\subsection{Broken Symmetry}
\label{sec5-6A}

To put the concept of emergent symmetry in context let us start
summarizing the, in principle independent but very  powerful,
broken-symmetry scenario. The broken-symmetry phenomenon, which is
manifested by a lowest-energy state not having the full symmetry of the
Hamiltonian $\cal G$ but less, has been beautifully described in P. W.
Anderson's {\it Basic Notions of Condensed Matter Physics} book
\cite{andersonrev}. Here we will simply restate the main known
results. 

To say that the ground state of $H$, $\ket{\Psi_0}$, is invariant under $\cal
G$ means that 
\begin{equation}
{\cal O}_\alpha \ket{\Psi_0} =  \ket{\Psi_0} \ , \ \forall \alpha \ .
\end{equation}
If $\ket{\Psi_0}$ is {\it not} invariant under a given symmetry
operation ${\cal O}_\alpha$, we say that the symmetry is {\it
spontaneously broken}. (Notice that if such a symmetry does not exist
and $\ket{\Psi_0}$ is not invariant that means that the symmetry is
{\it explicitly broken}.) The broken-symmetry state is also called the
{\it ordered state}. In general, a subset of $\cal G$, ${\cal
G}_{BS}$, contains the transformations that do not leave $\ket{\Psi_0}$
invariant, while a residual symmetry subgroup ${\cal G}_{R}$ remains. 

To make a quantitative distinction between the symmetric and the
broken-symmetry phases we need to introduce the concept of {\it order
parameter}, representing the supplementary variable needed to describe
the lower symmetry state. By definition, the order parameter is a
physical quantity which is zero for the symmetric phase and non-zero
for the broken-symmetry state. However, this simple requirement is
still too general since there are many different quantities that can
satisfy this condition. The most natural choice is dictated by the
symmetry which is spontaneously broken. The order parameter is then a
physical quantity  that transforms like a non-trivial representation of
the symmetry group. The adequate choice of representation depends on
the  particular problem under consideration. For instance, out of a
broken $SU(2)$  symmetry, a system may have dipolar (usual magnetic
ordering), nematic, or more sophisticated multipolar orderings. Each
transforms like a different representation of the $SU(2)$ group. One of
the most important experimental challenges is the design of physical
measurements that distinguish between the different representations of
the broken-symmetry groups.

Spontaneous symmetry breaking can be associated with collective
phenomena, which is the relevant case for matter organization, or with
a trivial phenomenon that occurs when part of the one-particle spectrum
of $H$ is degenerate. The latter scenario is not of interest here
because it does not lead to the important concept of {\it ergodicity
breaking}. As an example, consider the ground state of an odd number of
noninteracting spin-1/2 fermions.  The ground state is not $SU(2)$
invariant (i.e, it breaks that symmetry) but as soon as we add one
particle to the system the $SU(2)$ symmetry is restored. On the other
hand, when the phenomenon is collective, spontaneous symmetry breaking
is in general related to ergodicity breaking meaning that for {\bf
given initial conditions} the equations of motion, although symmetric,
cannot connect (because of kinematic or dynamical reasons) states that
would otherwise restore those symmetries. The phenomenon of  ergodicity
breaking is not exclusive to systems with a macroscopically  large
number of degrees of freedom (thermodynamic limit), it may happen in
systems with a finite number of degrees of freedom as well (e.g., a
single particle in a double well with an infinite barrier in between).
The first case is a result of the existence of a multitude of
inequivalent representations of the observables of an infinite system.
In other words, it is a result of the non-commutativity of the limits
\begin{equation}
\lim_{\Omega \rightarrow \infty} \lim_{F \rightarrow 0} \psi(\Omega,F) \neq 
\lim_{F \rightarrow 0} \lim_{\Omega \rightarrow \infty} \psi(\Omega,F) \ ,
\end{equation}
where $\Omega$ is the volume of the system, and $F$ is the generalized
external field that linearly couples to the order parameter $\psi$.
Notice that when the symmetry breaking is induced by ergodicity
breaking, the subspaces which get {\it disconnected} in the dynamics of
the system are still {\it connected} by symmetry operations of ${\cal
G}_{BS}$. This means that ergodicity breaking in a system with a
Hamiltonian invariant under ${\cal G}_{BS}$ requires the ground state
to be degenerate. 

Our last observation raises the following question: Is it possible to
have spontaneous symmetry breaking when the ground state is
non-degenerate? The answer to this question is {\bf yes}. Symmetry
breaking implies that the ground state is not invariant under the
action of ${\cal G}_{BS}$. Non-degeneracy means that the representation
associated to this state is one-dimensional. Let's consider an example
of this situation in which ${\cal G}$ is the Abelian group of
translations. The system consists of an $N_s$-site ring of
noninteracting spinless fermions in the presence of a uniform magnetic
field threading the ring whose total flux is $N_s \phi$
\begin{equation}
H=-t\sum_{\bj=1}^{N_s} (c^\dagger_\bj \exp[i \phi] \, c^{\;}_{\bj+1} +
c^\dagger_{\bj+1} \exp[-i \phi] \, c^{\;}_\bj) \ ,
\label{tV1d}
\end{equation}
The ground state of this system is non-degenerate and has a non-zero
total momentum (non-zero current); i.e., it does not belong to the
trivial representation of the group of translations.

As usual, mathematical concepts have physical relevance whenever there are
observable consequences. So, what are the consequences of symmetry breaking?
They are:

\noindent
1. {\it Generalized rigidity and long-range order}

With the appearance of a broken-symmetry state an order parameter
$\psi$ emerges which represents a measure of the degree of asymmetry in
the broken-symmetry phase. The general problem of how to explicitly
define an order parameter is deferred for a later section, here we will
simply assume that we know $\psi$ and it is defined as the expectation
value of some space-dependent {\it local} observable ({\it order
field}) $\hat{\psi}({\bf x})$. Since the state is translationally
invariant (in the thermodynamic limit) it happens that
\begin{equation}
\lim_{|{\bf x}-{\bf x'}|\rightarrow \infty} \langle \hat{\psi}({\bf
x})\hat{\psi}({\bf x'})\rangle \rightarrow \psi^2 \ .
\end{equation}
In other words, the broken symmetry state carries long-range
correlations in the order field. Conversely, long-range correlations
implies a broken-symmetry state, therefore, long-range correlations is
a necessary and sufficient condition for the existence of an ordered
state. P.W. Anderson has called this phenomenon {\it generalized
rigidity}. In his words \cite{andersonrev} it is an ``{\it emergent
property} not contained in the simple laws of physics, although it is a
consequence of them.'' 

Given a physical Hamiltonian it is not straightforward, in general, to
determine whether its ground state is invariant or not. However, under
certain conditions (e.g., low space dimensionality) one can certainly
establish that there is no broken-symmetry phase and, therefore, no
long-range order. Those conditions constitute the hypothesis of the
Mermin-Wagner-Hohenberg theorem. The idea behind this theorem is that
(thermal or quantum) fluctuations may destroy long-range order, so the
conditions of the theorem are directly tied to the reasons that may
cause strong fluctuations to the ordered state. Continuous
symmetries, low space dimensionality, short-range (constituent)
interactions cause stronger fluctuations. Thus at finite temperature,
$T>0$, short-range spin models with a continuous symmetry in $d\leq 2$
do not show spontaneous ordering. At $T=0$, the existence of a gap in
the excitation spectrum (of a system with a continuous symmetry)
precludes long-range order, but the presence of gapless excitations
does not necessarily imply this order. (Note that the
Mermin-Wagner-Hohenberg and Goldstone (see below) theorems are two
complementary aspects of the same fact, and both follow in a unified
way from a clever use of the Bogoliubov's inequality \cite{phmartin}.)


%

\noindent
2. {\it New massless particles: Nambu-Goldstone modes}

As mentioned above, continuous symmetries play a special role since
degeneracies are non-countably infinite. If a {\it continuous} symmetry
is spontaneously broken, the spectrum of $H$ generally has gapless
collective excitations (soft modes). These emergent excitations
(quasiparticles), which can be interpreted as new particles with zero
mass, are called Nambu-Goldstone modes or bosons. They are the
quantized  excitations associated with a spatial twist of the order
parameter. The energy of the twist is typically proportional to the
inverse of the system's length, i.e., it increases linearly with the
wave vector $\bf k$ of the twist. (The energy of the resulting state
vanishes as ${\bf k} \rightarrow \bf 0$.) The number of Nambu-Goldstone
bosons is at most ${\rm dim} \; {\cal G} - {\rm dim} \; {\cal G}_R$.
The Anderson-Higgs mechanism in gauge theories provides an exception to
this theorem, i.e., some (or all, depending upon the symmetry group of
the gauge theory) Nambu-Goldstone bosons {\it do not emerge} even
though the continuous symmetry is broken. The idea behind this
mechanism is that after the symmetry is broken, the coupling of our
ungauged system to a massless gauge field generates a mass for the
gauge field giving rise to massive bosonic excitations.

\noindent
3. {\it Topological defects}

The breakdown of long-range order (with temperature, for instance)
carries the formation of defect structures such as vortices and domain
walls, each characterized by the type of singularity in the order
parameter. The topological stability of these defects is defined by the
homotopy class of the manifold where the order parameter lives, and
work has been developed to mathematically classify these defect
structures. The book by P.W. Anderson \cite{andersonrev} provides an
excellent introduction to the subject.

\subsection{Emergent Symmetry}
\label{sec5-6B}

In the previous subsection, we revised the fundamental aspects and the 
deep physical consequences of the concept of broken symmetry. We will
introduce now another notion which complements the previous one and
also plays a central role in the description of physical systems. This
is the notion of {\it emergent symmetry}. In broken symmetry 
phenomenon, the symmetry of the considered system is lowered below  a
critical temperature. But is it possible to have the opposite 
situation where the symmetry of our system increases when the
temperature is lowered? In a sense that will become clear below, the
answer to this question is  yes. However, we will see that the
emergence of a new symmetry  is not signaled by a phase transition but
by a crossover between two  different energy scales. This new concept
is useful  to characterize the {\it relevant} degrees of freedom that
dominate the low-energy physics and  the nature of the quasiparticles
that result from the interactions between these degrees of freedom. 


The isolation of the relevant degrees of freedom for the description of
complex systems plays a central role in physics. In general, this
process results from a careful separation of the different energy
scales involved in the Hamiltonian $H$ under consideration. The
elimination of  the irrelevant degrees of freedom can be achieved by
different methods like perturbation theory, algebraic approaches, or
renormalization group. The result is a new low-energy effective
Hamiltonian, $H_{\sf eff}$, that is a valid description of the physical
system below some characteristic energy  $E_c$. The Hilbert space
${\cal M}_0$ of $H_{\sf eff}$ is then a subspace of the Hilbert space
${\cal H}$ of the original Hamiltonian $H$. Sometimes the spectrum  of
$H_{\sf eff}$ and the low-energy spectrum of $H$ are {\it exactly} the
same. However, in the more general case, the difference between both
spectra is finite and can be made arbitrarily small by increasing the
distance between the corresponding energy scales in $H$. 

We will say that a new symmetry {\it emerges} whenever there is a
symmetry group $G$ of transformations (which is not a group of
symmetries of $H$) with elements of the form $g_\alpha=\sum_{\bk} a_\bk
\prod_{\bj=1}^{n_\bk} U_\bj$ ($a_\bk$ is a $c$-number and $n_\bk$ an
integer), where the unitary operator $U_\bj$ acts on the local Hilbert
space ${\cal H}_{\bj}$, and 
\begin{equation}
[g_\alpha,H_{\sf eff}]=0 \; \;\;\; \mbox{for any}
\;\;\; g_\alpha \in G \ .
\label{comm}
\end{equation}
The transformations in $G$ are defined in the original Hilbert space
${\cal H}$ and leave the subspace ${\cal M}_0$ invariant. [${\cal
H}=\bigotimes_\bj {\cal H}_{\bj}$ is an arbitrary decomposition of
${\cal H}$.] The condition $g_\alpha=\sum_{\bk} a_\bk
\prod_{\bj=1}^{n_\bk} U_\bj$ means that the $g_\alpha$'s are linear
combinations of $n_\bk$-local operators (product of $n_\bk$  operators
each of them acting on the local space ${\cal H}_{\bj}$). In this way,
we are excluding {\it non-local} transformations. This is because
for each $g_\alpha$ that commutes with $H_{\sf eff}$, it is always
possible  to find a non-local operator that commutes with $H$ and is
identical to $g_\alpha$ when restricted to the  subspace ${\cal M}_0$. 
Note that if $H_{\sf eff}$ provides an  exact description of the
low-energy spectrum of $H$, we say that the emergent symmetry is exact.
In contrast, if $H_{\sf eff}$ is only a very good approximation to the
low-energy spectrum of $H$, the emergent symmetry is only an
approximate concept. In other words, suppose that $H_{\sf eff}+H'_{\sf
eff}$ is the Hamiltonian that reproduces the  exact low-energy spectrum
of $H$. If, for instance,  we have derived $H_{\sf eff}$ within
perturbation theory  to order $n$ in the small parameter $\eta$ (ratio
between the small and the large energy scales), $H'_{\sf eff}$ is of
order $\eta^{n+m}$, where $m$ is a positive integer. Therefore,
\begin{equation}
[g_\alpha,H_{\sf eff}+H'_{\sf eff}]=[g_\alpha,H'_{\sf eff}]={\cal
O}(\eta^{n+m}) \;\;\;\; \mbox{for any} \;\;\; g_\alpha \in G \ .
\label{comm2}
\end{equation}
Although in these cases the emergent symmetry is an approximate
notion, the concept is still useful to identify the nature and
properties of the low-energy quasiparticles. This is illustrated in
the last examples of this section. 

Let us start considering cases in which the notion of emergent symmetry
is exact. During last few decades, exact dimer ground states were found
for different quantum spin models \cite{Shastry1,kumar}. The most
famous example is the Majumdar-Ghosh solution  \cite{Majumdar} of the
Heisenberg spin-1/2 chain with nearest and second-nearest neighbor
interactions $J_1$ and $J_2$, respectively. The common characteristic
of these quantum spin models is the emergence of a local $U(1)$ gauge
symmetry that gives rise to the conservation of each local dimer. In
general, the dimerized ground states can  be written as
\begin{equation}
|\Psi_0 \rangle = \bigotimes_{({\bf i}_1,{\bf i}_2)} |\phi_s({\bf
i}_1,{\bf i}_2)\rangle \ ,
\label{dimer}
\end{equation}
where $({\bf i}_1,{\bf i}_2)$ denotes a pair of sites and each site
belongs to one and only one pair. The wave function $|\phi_s({\bf
i}_1,{\bf i}_2)\rangle$ is  the singlet state for the pair $({\bf
i}_1,{\bf i}_2)$. Let us now introduce the following local $U(1)$
transformation
\begin{equation}
\hat{T}({\bf i}_1,{\bf i}_2)= \exp[i a_{12}   P({\bf i}_1,{\bf i}_2)] ,
\end{equation}
where $P({\bf i}_1,{\bf i}_2)= |\phi_s({\bf i}_1,{\bf i}_2)\rangle
\langle  \phi_s({\bf i}_1,{\bf i}_2)|$ is a local projector on the
singlet state of the bond $({\bf i}_1,{\bf i}_2)$. It is clear that
$\hat{T}({\bf i}_1,{\bf i}_2)|\Psi_0 \rangle=\exp[i a_{12}]|\Psi_0 \rangle$.
The generators of this $U(1)$ gauge group  are the projectors $P({\bf
i}_1,{\bf i}_2)$ which count the number of singlet  states on each bond
$({\bf i}_1,{\bf i}_2)$. Then, the set of bonds $({\bf i}_1,{\bf i}_2)$
provides the natural subsytem decomposition of ${\cal H}$. It is
important to note that in most of these cases the emergent symmetry is
only present in the ground state which is separated from the lowest
energy excitations by an energy gap; i.e., ${\cal M}_0$ is a
one-dimensional subspace. Nevertheless, this emergent property of the
ground state determines the nature of the low-energy excitations, which
are local spin triplets on each bond, that propagate on a sea of
singlet dimers. For models like the one of Eq.~(\ref{laddH3}) or
Ref.~\cite{Lin}, there is an invariant low-energy subspace given by a
dimerized ground state and an arbitrary  number of triplet excitations,
with the condition that triplets cannot be created on dimers which are
nearest-neighbors. The action of the Hamiltonian restricted to this
subspace is invariant under a $U(1)$ gauge transformation and, again,
the exact low-energy theory has an infinite number of symmetries not
present in the original models.  

What is the origin of the emergent $U(1)$ gauge symmetry of quantum
dimer ground states of the form of Eq. (\ref{dimer})? For most of the
cases, if not all, the origin is {\it geometrical frustration}. The
particular connectivity of the lattice (or lattice topology) produces
negative interference between the different links that connect two
dimers. When the interference is complete, the dimers become {\it
disconnected} and the gauge symmetry emerges. This principle is by no
means restricted  to quantum dimer systems. We can imagine different
local structures (instead  of dimers) that get {\it disconnected} due
to the same reason. For instance, currents localized in
plaquettes can also result from an emergent $SU(2)$ gauge symmetry in
frustrated lattices. As shown in Ref.~\cite{batshas}, this occurs for
particular fillings of the Hubbard model on a family of frustrated
lattices. The guiding principle is always the same. In a certain region
of parameters, local currents with two possible orientations  become
conserved quantities. The local chirality can be described with a
pseudospin-1/2 variable that is the relevant degree of freedom to
build a low-energy effective theory. Any weak physical interaction that
breaks the emergent gauge symmetry can introduce a finite coupling
between the local currents and produce different orderings of the
chiral degrees of freedom. In the case of Ref.~\cite{batshas}, the
interaction that breaks the gauge symmetry is an inter-site Coulomb
repulsion. This interaction induces an $XY$-like ordering of the local
currents. The low-energy  excitations are chiral-waves that are
described by magnons in the pseudospin language. Following the same
strategy, one can find physical Hamiltonians that give rise to
different and, sometimes, unusual low-energy degrees of freedom (see the
example of Eq. (\ref{kekulon}) below). 

However {\it frustration} is not a requirement to have an exact
emergent gauge symmetry. For instance, the {\it Fermi liquid} or the
{\it band insulator} are two examples of emergent symmetries that do
not involve any geometrical frustration. These phases of matter are not
characterized by {\it any} broken symmetry since their corresponding
ground states are non-degenerate. According to Landau's theory of Fermi
liquids \cite{Nozieres}, the quasiparticles of the  Fermi liquid have
an infinite lifetime if they are right at the Fermi surface. This means
that the low-energy spectrum of the Fermi liquid is invariant under
local $U(1)$ transformations of the form $\exp[i \phi_{{\bf k}_F}
c^{\dagger}_{{\bf k}_F}c^{\;}_{{\bf k}_F}]$ (where $c^\dagger_{{\bf
k}_F}$ is the creation operator of a quasiparticle with Fermi wave
vector ${\bf k}_F$). The band insulator can also be characterized by a
local $U(1)$ emergent symmetry. In this case, the $U(1)$ symmetry can
be factorized in real space and is generated by the local
transformation:  $\exp[i a c^{\dagger}_{{\bf j}}c^{\;}_{{\bf j}}]$.
This is the mathematical  expression for charge localization that
characterizes the insulating state: the charge is locally conserved.
Like in the case of the spin dimers, only the ground state of the band
insulator exhibits the property of emergent symmetry. In contrast, the
dimension of ${\cal M}_0$ for the Fermi liquid phase is equal to the
number of wave vectors ${\bf k}_F$ that are on the Fermi surface. Note
that the Fermi surface is the manifold associated with the group of
emergent symmetries whose topology characterizes the universality
class of the Fermi liquid. 

Exactly and quasi-exactly solvable models provide also examples for
exact emergent symmetries. A model is quasi-exactly solvable when only
part of the spectrum can, in a purely algebraic form, be exactly
diagonalized. Let us call ${\cal S}_0$ the subspace generated by the
exactly solvable part of the spectrum. Since $H$ is exactly solvable
when restricted to ${\cal S}_0$, there is a set of operators 
$g_\alpha$ that commute with $H: {\cal S}_0 \rightarrow {\cal S}_0$. If
${\cal S}_0$ is also the lowest energy subspace and the operators
$g_\alpha$ can be factorized as $g_\alpha=\sum_\bk a_\bk
\prod_{\bj=1}^{n_\bk} U_\bj$,  we have another case of emergent
symmetry. As an example of a quasi-exactly solvable model that  has
also an emergent symmetry, we will consider the $t$-$J_z$ chain
\cite{ours4}. The lowest  energy subspace of this model can be mapped
into the exactly solvable $S$=1/2 $XXZ$ chain. This means that 
$H_{\mbox{$t$-$J_z$}}:{\cal S}_0 \rightarrow {\cal S}_0$ has an
infinite number of symmetries $g_n$  that are linear combinations of
$n$-local operators (products of $n$-body spin variables) \cite{Ha}.
These are the quantum symmetries that make the $XXZ$ chain an
integrable problem. The relevant  low-energy degrees of freedom of
$H_{\mbox{$t$-$J_z$}}$ are holes which are attached to an anti-phase 
domain for the staggered magnetization. For $J_z < 8t$, the system is a
Luttinger  liquid of particles that carry both electrical and
topological charges.

We will consider now the cases in which the emergent symmetry is an
approximate, albeit important, concept. In the simple example that we
describe below, a global $SU(2)$ invariance emerges at low energies.
The model is a $d$-dimensional hyper-cubic Kondo lattice with an
attractive ($U>0$) Hubbard interaction for the conduction band, an
anisotropic Kondo interaction between the magnetic impurities and the
conduction electrons, and a Heisenberg antiferromagnetic interaction
($J>0$) between
the localized spins ($\sigma=\uparrow,\downarrow$)
\begin{eqnarray}
H^{\sf KA}\!\!\!&=&\!\!\! -t \!\!\sum_{\langle {\bf i,j} \rangle, \sigma} 
(c^{\dagger}_{{\bf i} \sigma} c^{\;}_{{\bf j} \sigma} +
c^{\dagger}_{{\bf j} \sigma} c^{\;}_{{\bf i} \sigma}) - \mu \sum_{\bf
i} \hat{n}_{{\bf i}} - U \sum_{\bf i} \hat{n}_{{\bf i} \uparrow} \hat{n}_{{\bf i}
\downarrow} \nonumber \\
&+&\!\!\! J_{K} \! \sum_{\bf i} (\gamma S^z_{\bf i}s^z_{\bf
j}+S^x_{\bf i}s^x_{\bf j}+S^y_{\bf i}s^y_{\bf j}) + \! J \! \sum_{\langle
{\bf i,j} \rangle}  {\bf S}_{\bf i} \cdot {\bf S}_{\bf j},
\label{KA}
\end{eqnarray} 
where ${\bf S}_{\bf i}$ is the spin operator for the localized moment
at  the site ${\bf i}$, and  $s^{\nu}_{\bf i}= 1/2 \sum_{\tau,\tau'}
{c}^{\dagger}_{{\bf i} \tau}  \sigma^{\nu}_{\tau\tau'}{c}^{\;}_{{\bf i}
\tau'}$ with $\nu=\{x,y,z\}$. The symmetry group of $H^{\sf KA}$ is
$U(1) \times U(1)$. The corresponding generators or conserved
quantities are the total number of particles $\sum_{\bf i} \hat{n}_{\bf i}$
and the $z$-component of the total spin $\sum_{\bf i} (S^z_{\bf
i}+s^z_{\bf i})$. This model describes the competition between a  
Kondo-like system ($J_K \gg U$) and an $s$-wave superconductor
coexisting with antiferromagnetic ordering of the localized magnetic
moments ($J_K \ll U$). We will consider here the large $U$ limit $U \gg
J_K, t$. In this limit, the low-energy subspace ${\cal M}_0$ is
generated by states in which the sites of the conduction band are
either empty or double occupied. In other words, the conduction states
are non-magnetic and the Kondo interaction is therefore quenched. The
low-energy effective model in ${\cal M}_0$ is
\begin{eqnarray}
H^{\sf KA}_{\sf eff}&=&  J \sum_{\langle {\bf i,j} \rangle}  {\bf S}_{\bf
i} \cdot {\bf S}_{\bf j} + {\tilde t} \sum_{\langle {\bf i,j} \rangle,
\sigma}  ({\bar b}^{\dagger}_{\bf i} {\bar b}^{\;}_{\bf j} + {\bar
b}^{\dagger}_{\bf j} {\bar b}^{\;}_{\bf i}) + 2 {\tilde t}
\sum_{\langle {\bf i,j} \rangle, \sigma} {\bar n}_{\bf i} {\bar n}_{\bf
j} \nonumber \\
&-& {\tilde \mu} \sum_{\bf i} {\bar n}_{\bf i} \ ,
\label{hkaeff}
\end{eqnarray} 
where ${\tilde t}=2 t^2/U$, and ${\tilde \mu}=2\mu+{\sf z}{\tilde t}$. The
hard-core bosons represent the local Cooper pairs 
\begin{equation}
{\bar b}^{\dagger}_{\bf i}= c^{\dagger}_{{\bf i} \uparrow}
c^{\dagger}_{{\bf i} \downarrow} \;\;\;\;\;\;\;\;\; {\bar
b}^{\;}_{\bf i}= c^{\;}_{{\bf i} \downarrow} c^{\;}_{{\bf i} 
\uparrow} \ .
\end{equation}
The first observation is that the localized spins and the conduction 
electrons are decoupled in $H^{\sf KA}_{\sf eff}$. The original Kondo
interaction is suppressed by the competing $U$ term. As a consequence,
the symmetry group of $H^{\sf KA}_{\sf eff}$ is $U(1) \times SU(2)$. The $U(1)$
symmetry is again associated with the conservation of the number of
particles that now are Cooper pairs. The {\it emergent} spin rotational
$SU(2)$ invariance is explicit from the expression  $H^{\sf KA}_{\sf
eff}$ of Eq. (\ref{hkaeff}). This is a simple example of a  global
$SU(2)$ symmetry that only appears at low energies. The higher-order
terms in $J_K/U$ will remove this $SU(2)$ invariance leaving the
original $U(1)$ symmetry of rotations around the $z$-axis. This means
that one of the two Goldstone modes associated to the spontaneously
broken $SU(2)$ symmetry will acquire a small mass of order $(J_K/U)^n$
with $n \geq 2$.

In the same way an emergent global symmetry is helpful for identifying
the nature of the quasiparticles, an emergent gauge symmetry provides a
guiding principle for identifying the {\it relevant} degrees of freedom
at low energies. In general, a noninteracting theory is characterized
by local symmetries that express the independence of each particle. For
instance, the translation of only one particle is a symmetry for a 
noninteracting gas. The inclusion of the interactions removes this
local symmetry and the many-body problem becomes, in general,
non-trivial. However, in many cases it is possible to find another
locally gauge-invariant limit for the interacting problem. In this
case, the local gauge invariance emerges only at low energies and
signals the appearance of effective degrees of freedom that become
decoupled. For instance, the origin of antiferromagnetism is more
transparent in the strong coupling limit of the half-filled Hubbard
model than in the weak or intermediate coupling regimes. Before giving
a formal expression for this statement, it is convenient to illustrate
its meaning with this simple example. Let us consider a repulsive
Hubbard Hamiltonian at half-filling ($U>0$)
\begin{eqnarray}
H_{\sf Hubb}&=& -t \sum_{\langle {\bf i,j} \rangle, \sigma} 
(c^{\dagger}_{{\bf i} \sigma} c^{\;}_{{\bf j} \sigma} +
c^{\dagger}_{{\bf j} \sigma} c^{\;}_{{\bf i} \sigma}) + U \sum_{\bf i}
\hat{n}_{{\bf i} \uparrow} \hat{n}_{{\bf i} \downarrow} \nonumber \\
&-& \mu \sum_{\bf i} \hat{n}_{{\bf i}} \ .
\label{hubb}
\end{eqnarray}
In the infinite-$U$ limit, the low-energy subspace ${\cal M}_0$ is
generated  by states having one particle per site, i.e., there is one
spin $S=1/2$ localized on each site. All states in the manifold ${\cal
M}_0$ have the same energy,  and this massive degeneracy is associated
with an emergent $SU(2)$ gauge symmetry. In other words, in this limit
the local spins get {\it decoupled}  and are free to rotate without
changing the energy of the system. This tells us that the natural
degrees of freedom to describe  the system in the strong coupling limit
are localized spins, instead of itinerant fermions. The presence of an
emergent gauge symmetry indicates that there is a limit in our original
interacting theory for which some degrees of freedom become
noninteracting. The internal structure of these degrees of freedom is
determined by the gauge symmetry group. If the system in our example is
close to the gauge invariant limit, $U \gg t$,  but finite, the
relevant degrees of freedom are still the same but they become weakly
interacting. This is the origin of antiferromagnetism in the strongly
interacting Mott insulators. Since the large-$U$ limit of the Hubbard
model is very well known, it may seem to the reader that the use of the
new concept of emergent symmetry just provides a complicated way to
describe a simple phenomenon. However, this is not the case for the
non-trivial examples that are described below. In addition, we will see
that the concept of emergent gauge symmetry is important as a guiding
principle to find new states of matter that result from strongly
interacting systems. One has to keep in mind that even
antiferromagnetism remained as a hidden phase for a long period of
time.

Let us consider the case of a gauge symmetry group that is a direct
product of local symmetry groups, $G=\bigotimes_{\bk} G_\bk$, where each
local group $G_\bk$ acts on the local space ${\cal H}_\bk$. If the
reduced Hilbert space ${\cal M}_0$ admits the decomposition ${\cal
M}_0= \bigotimes_\bk {\cal H}_\bk$, this will provide the natural basis to
write down $H_{\sf eff}$. In particular, if $G_\bk$ is a continuous
group, the generators of $G_\bk$ are conserved quantities at low
energies because they commute with $H_{\sf eff}$. These generators are
physical degrees of freedom and their conservation imply that they are
noninteracting. Note that these effective degrees of freedom are the
{\it bricks} to build new phases out of interacting systems. As soon as
we move away  (but not too far) from the gauge invariant limit, these
degrees of freedom will interact, producing, in some cases, novel types
of orderings. The main goal in the rest of this section will be 
illustrating this phenomenon with different examples.

\begin{figure}[htb]
\includegraphics[angle=0,width=8.0cm,scale=1.0]{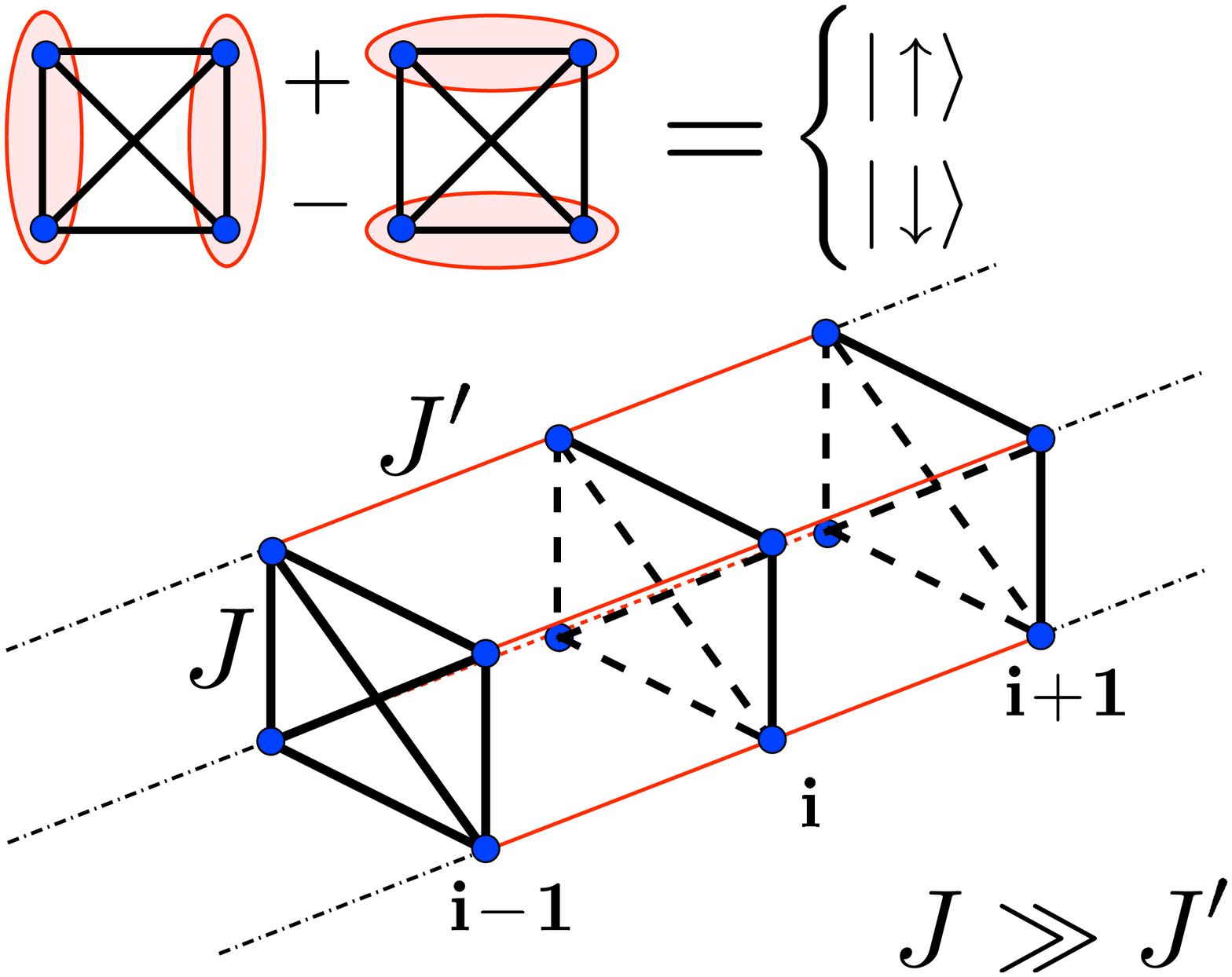}
\caption{Four-leg spin ladder. In each plaquette $\bi$, there are 4
spin-1/2 which interact through a Heisenberg exchange term of strength
$J$. The interaction between plaquettes is $J'$, and the links are
indicated in the figure. }
\label{ladd}
\end{figure}

We will consider now the spin-1/2 ladder of Fig.~\ref{ladd},  described by
the following Hamiltonian
\begin{equation}
H^{\sf SL}= J \sum_{\bi, \nu,\nu' } 
{\bf S}_{\bi \nu} \cdot {\bf S}_{\bi \nu' }
+ J' \sum_{\bi, \nu=1,4 } 
{\bf S}_{\bi \nu} \cdot {\bf S}_{\bi+1 \nu},
\label{kekulon}
\end{equation}
with $J , J' > 0$ and $1 \leq \nu, \nu' \leq 4$. In the limit $J' \ll
J$, the low-energy subspace of $H^{\sf SL}$ only contains  states in
which each square plaquette ${\bi}$ is in a singlet state. This is
clear  when we analyze the spectrum of an isolated plaquette. The
energy of the eigenstates only depends on the total spin $S_T$:
$E(S_T)= S_T(S_T+1) J$. Thus, the two possible singlets ($S_T=0$) are
the lowest energy states, with the $S_T=1$ and $S_T=2$ states having
eigenvalues $2J$ and $6J$, respectively. 

We will classify the two local singlet states according to the
eigenvalues of the reflection symmetry plane along the diagonal of  the
plaquette (see Fig.~\ref{ladd}). One of the singlet states
$|S_{\bi,s}\rangle$ is symmetric under this reflection while the other
one, $|S_{\bi,a}\rangle$, is antisymmetric. These two singlet states
can be described with an effective pseudospin ${\tau}=1/2$ variable. We
will represent the symmetric state $|S_{\bi,s}\rangle=\ket{\uparrow}$
by the eigenvector of $\tau^z$ with eigenvalue 1/2 and the
antisymmetric one, $|S_{\bi,a}\rangle=\ket{\downarrow}$, by the other
eigenvector of $\tau^z$. In the infinite-$J$ limit, the low-energy
degrees of freedom ${\boldsymbol \tau}_{\bi}$ are completely decoupled.
This situation is similar to the infinite-$U$ limit of the  Hubbard
model. However, in the present case, the pseudospin variable
${\boldsymbol \tau}_{\bi}$ does not represent a magnetic degree of
freedom. By means of a canonical transformation  to second order in
$J'/J$, we can derive the following low-energy effective Hamiltonian
\begin{equation}
H^{\sf SL}_{\sf eff}= \frac {J_{\sf eff}}{4} \sum_{\bi} (2 \tau^z_{\bi}
\tau^z_{\bi+1} -  \tau^x_{\bi} \tau^x_{\bi+1} - \tau^y_{\bi} \tau^y_{\bi+1}),
\end{equation}
where $J_{\sf eff}= -(J')^2/J$. $H^{\sf SL}_{\sf eff}$ is an Ising-like
anisotropic Heisenberg  model. Since $J_{\sf eff}<0$, the ground state
is a fully polarized ferromagnet that has two possible orientations
($Z_2$ broken symmetry). In terms of the original variables, the fully
polarized states correspond to the direct product of the symmetric and
the antisymmetric singlets: $\bigotimes_\bi |S_{\bi,s}\rangle$ and
$\bigotimes_\bi |S_{\bi,a}\rangle$. The $Z_2 \times U(1)$ symmetry of
$H^{\sf SL}_{\sf eff}$ is an emergent symmetry because $H$ is not invariant
under the corresponding transformations. These symmetries are
explicitely removed when higher order terms in $J'/J$ are added to
$H^{\sf SL}_{\sf eff}$. These terms  will stabilize one of the fully
polarized ferromagnetic solutions. However, the presence of these terms
does not modify the nature of the quasiparticles of $H^{\sf SL}_{\sf
eff}$. These novel quasiparticles, which we will name {\it kekulons}
\cite{kekule}, are magnons in the pseudospin language whose spectrum is
gapped due to the Ising-like character of $H^{\sf SL}_{\sf eff}$. In
the original language, these magnons are massive singlet waves. The
main effect of the higher-order corrections to $H^{\sf SL}_{\sf eff}$
is a renormalization of the  mass (gap) of these quasiparticles.
Therefore, even though the emergent symmetry is approximate in the
present problem, its most important physical consequences remain
valid. 

We have seen that there is a systematic procedure for obtaining new
types of orderings. The procedure starts from some limit in parameter
space, in general the strong coupling limit, in which effective
low-energy degrees of freedom are simultaneously stabilized and
decoupled. The interactions between these degrees of freedom appear
when we move away from this limit and the symmetries or approximate
symmetries of these effective interactions have important consequences
for the properties of the ground state and the low-energy excitations.
These degrees of freedom are the {\it bricks} for the novel orderings
that result from the effective interactions and simplify the
description of the new phase. These novel orderings, that are in
general associated  with the presence of competing interactions, pose
a challenge for experimental physicists to develop probes that are
sensitive to the new phases.

\section{Broken Symmetry Phase Transitions}
\label{sec6}

It is widely recognized that the resolution of interacting systems at a
microscopic level requires sophisticated techniques which go beyond the
traditional approaches. Much of the current understanding on quantum
phases of matter and their corresponding phase transitions is due to
renormalization group analysis or the existence of a few exact
solutions. The latter, the most desirable scenario, is also the hardest
since one seldom encounters solutions in spatial dimensions larger than
one. For strongly interacting systems the situation worsens because it
is difficult, if not impossible, to identify a small parameter: Their
ground states are susceptible to different quantum orderings and low
energy excitations because of competing interactions that abound in
nonlinearities. Often these systems are near quantum criticality which
makes their study even more challenging. 

A fundamental notion of universality (or equivalence) naturally emerges
from our dictionaries. This notion refers to the fact that many
apparently different problems in nature have the same underlying
algebraic structure and, therefore, the same physical behavior. In this
way, as a result of unveiling the symmetry structure of the problem,
there is a concept of physical equivalence hidden in those
dictionaries. If the complete Hilbert space of the problem maps onto
another in a different language, the concept of universality applies to
all length and time scales. On the other hand, if only certain
invariant subspaces of the original Hamiltonian map onto another
physical system, then, universality will only manifest at certain
energy scales.

It was Landau \cite{Landau} who first noticed the crucial role of
symmetry for phase transitions. Since the symmetry of a given state
cannot be changed continuously, different symmetries must be associated
to states of matter which are qualitatively different. To characterize
the symmetry of a given state, Landau introduced the notion of an order
parameter. This parameter is zero in the symmetric state and has a
nonzero average when the symmetry is broken. Another property of this
parameter is that it is non-invariant under at least one of the 
transformations of the system Hamiltonian. As Landau recognized long
time ago, the order parameter is the relevant physical quantity to
build a macroscopic theory of thermodynamical phase transitions.
However, the search for the order parameter characterizing a broken
symmetry state can be a highly non-trivial task. 

There are examples in nature of systems entailing hidden microscopic
order parameters which do not correspond to macroscopic variables,
i.e., they cannot be coupled to any external physical field. To find
out these hidden order parameters we cannot avoid a microscopic
description of our physical system. It is important to notice the
practical consequences of this search: A new microscopic order
parameter amounts to predicting a new possible state of matter. The
glassy materials are good examples of systems belonging to this class.
Since the order parameters characterizing a broken continuous symmetry
are associated in general to the generators of the symmetry group, one
can imagine that a language based on symmetry generators can shed some
light into the search of the possible order parameters for a given
system. It is the purpose of this section to illustrate this idea with
different examples and to introduce the notion of {\it hierarchical
language}. One of our goals is to show the fundamental role played by
the hierarchical languages in the classification of orders in matter.

\subsection{Classification of Order Parameters}

To illustrate the general procedure of getting and classifying the
possible order parameters of a given model, we start discussing the
family of Heisenberg models with $SU(N)$ spins in the fundamental
representation
\begin{equation}
H^{\sf SU(N)}_{\sf Heis}= J \sum_{\langle {\bf i},{\bf j} \rangle}
{\cal S}^{\mu \nu}({\bf i}) {\cal S}^{\nu \mu}({\bf j}) \ .
\label{suNheis}
\end{equation}
(A summary of the methodology to obtain and classify order parameters
is presented at the end of section \ref{sec6b}.) 
For $J<0$, the ground states of this family of Hamiltonians can be
exactly computed and it is a fully polarized $SU(N)$ ferromagnet. The
order parameter is the $SU(N)$ magnetization ${\cal S}^{\mu
\nu}=\sum_\bj {\cal S}^{\mu \nu}(\bj)$ which has $N^2-1$ (number of
generators of $su(N)$) components. If $J>0$, and the ground state has
antiferromagnetic long range order, the order parameter is the
staggered  $SU(N)$  magnetization ${\cal S}^{\mu \nu}_{\sf
ST}=\sum_{{\bf j}} e^{i{\bf Q \cdot \bj}}{\cal S}^{\mu \nu}({\bf j})$
(${\bf Q}$ is the antiferromagnetic wave vector). By adding terms which
break the $SU(N)$ symmetry explicitly, we can get lower symmetry order
parameters associated to the subgroups of $SU(N)$. 

We can ask now what is the expression of these order parameters in a
spin $SU(2)$ or in a bosonic representation of $H^{\sf SU(N)}_{\sf
Heis}$. The answer to this question will allow us to explore new phases
which can appear in the spin or the bosonic equivalent models. In
addition, we will find that the high symmetry points of these models
correspond to situations where two or more different phases are
coexisting. 

We start with the spin $SU(2)$ representations of $H^{\sf SU(N)}_{\sf
Heis}$. The local Hilbert space of this $SU(N)$ Heisenberg Hamiltonian
has dimension $N$ so we can use an $SU(2)$ spin $S=(N-1)/2$ to
represent this model. In this representation the Hamiltonian
Eq.~(\ref{suNheis}) has the following form
\begin{equation}
 H^{\sf SU(N)}_{\sf Heis}= J \sum_{\langle {\bf i},{\bf j} \rangle} 
\sum^{N-1}_{l=0} \alpha_{l} ({\bf S}_{\bf i} \cdot {\bf S}_{{\bf
j}})^l \ ,
\label{suNheiss}
\end{equation}
with the values of $\alpha_{l}$'s determined from the following system
of equations
\begin{eqnarray} \hspace*{-0.0cm}
\sum^{N-1}_{l=0} \!\alpha_{l} \{[1-S(S+1)]^l-(-1)^l[S(S+1)]^l
\}\!\!&=&\!\! 2(-1)^N  \ , \nonumber \\ \hspace*{-.0cm}
\sum^{N-1}_{l=0} \!\alpha_{l} \{ [{\cal Y}_1 - S(S+1) ]^l-
(-1)^l[S(S+1)]^l \}\!\!&=&\!\!0 \ , \nonumber \\ \hspace*{-.0cm}
\sum^{N-1}_{l=0} \!\alpha_{l} \{ [{\cal Y}_2- S(S+1)]^l-
[1-S(S+1)]^l \}\!\!&=&\!\!0 \ ,
\label{sist}
\end{eqnarray}
where ${\cal Y}_1=m(2m+1)$, ${\cal Y}_2=n(2n-1)$ and $m$, $n$ are
integers. When $N$ is an odd integer (i.e., $S$ is integer) $m$ and $n$
satisfy $0 \leq m \leq S$, and $1\leq n \leq S$, while for even $N$
(i.e., $S$ is half-odd integer), $0 \leq m \leq S-1/2$, and $1\leq n
\leq S+1/2$. It is easy to check that there are $N-1$ linearly
independent equations and $N$ variables $\alpha_{l}$. The undetermined
variable corresponds to an additive constant. 

The simplest case corresponds to $SU(2)$. Here we get the spin 1/2
($S$=(2-1)/2) version 
\begin{equation} \hspace*{-0.9cm}
H^{\sf SU(2)}_{\sf Heis} = 2J \sum_{\langle {\bf i},{\bf j} \rangle} 
{\bf S}_{\bf i} \cdot {\bf S}_{{\bf j}} \ .
\label{spin1o2}
\end{equation}
To understand the general procedure we continue with the first
non-trivial example, i.e., $SU(3)$. This particular example has been
analyzed in Ref. \cite{ours2}. By solving the above system of equations
we get that the spin one ($S$=(3-1)/2) version of the $SU(3)$
Heisenberg Hamiltonian is (up to a constant)
\begin{equation} \hspace*{-0.9cm}
H^{\sf SU(3)}_{\sf Heis} = J \sum_{\langle {\bf i},{\bf j} \rangle} [ \
{\bf S}_{\bf i} \cdot {\bf S}_{{\bf j}} +  \left ( {\bf S}_{\bf i}
\cdot {\bf S}_{{\bf j}} \right )^2 ] \ .
\label{spin1}
\end{equation}
To fix ideas let us start with the simple ferromagnetic case $J<0$.
As mentioned above, the ground state of the ferromagnetic $SU(3)$
Heisenberg model is the state with maximum total $SU(3)$ spin ${\cal
S}$. The order parameter associated to this broken symmetry is the
total $SU(3)$ magnetization ${\cal S}^{\mu \nu }= \sum_\bj {\cal
S}^{\mu \nu }(\bj)$ which has eight independent components. We now
raise the following question: what is the order parameter of the 
equivalent model Eq.~(\ref{spin1}) written in the ($SU(2)$) $S$=1 
language? To answer this question we need to write down the relation
between the components of ${\cal S}^{\mu \nu}$ and the $S$=1
generators. The fundamental theorem guarantees the existence of these
mappings. From Eq.~(\ref{spinsu3}) and the generalized JW
mappings (section \ref{jordanw}), we get (the site index $\bj$ is
omitted)
\begin{eqnarray}
S^x&=&\frac{1}{\sqrt{2}} ({\cal S}^{0 1}+{\cal S}^{2 0}+ {\cal S}^{0
2}+{\cal S}^{1 0}) \ , \nonumber \\
S^y&=&\frac{-1}{\sqrt{2}i} ({\cal S}^{0 1}+{\cal S}^{2 0}-{\cal S}^{0
2}-{\cal S}^{1 0}) \ , \nonumber \\
S^z&=&{\cal S}^{1 1}-{\cal S}^{2 2} \ ,
\nonumber \\
(S^x)^2-\frac{2}{3}&=&\frac{1}{2}({\cal S}^{1 2}+{\cal S}^{2 1}+{\cal
S}^{0 0}) \ , \ (S^z)^2-\frac{2}{3}= -{\cal S}^{0 0}  \ , \nonumber
\\
\left \{ S^x,S^y \right \}&=& i ({\cal S}^{2 1}-{\cal S}^{1 2}) \ ,
\nonumber \\
\left \{ S^x,S^z \right \}&=&\frac{1}{\sqrt{2}} ({\cal S}^{0 1}-{\cal
S}^{2 0}-{\cal S}^{0 2}+{\cal S}^{1 0}) \ , \nonumber \\
\left \{ S^y,S^z \right \}&=&\frac{-1}{\sqrt{2}i} ({\cal S}^{0 1}-{\cal
S}^{2 0}+{\cal S}^{0 2}-{\cal S}^{1 0}) \ .
\label{gene}
\end{eqnarray}
The first three operators are the components of the $S$=1
ferromagnetic order parameter, while the second five are the components
of the spin-nematic order parameter (components of the bilinear
symmetric traceless tensor). The traceless condition implies that 
$(S^y)^2-\frac{2}{3}=-((S^z)^2-\frac{2}{3})-((S^x)^2-\frac{2}{3})$. In
this way, we see that by rotating the $SU(3)$ ferromagnetic order
parameter it is possible to evolve from a ferromagnetic to a
spin-nematic ground state and vice versa. This means that the $SU(3)$
invariance of the spin Hamiltonian gives rise to coexistence of both
phases. Since the algebra $su(3)$ provides a fundamental
representation  when the local Hilbert space has $D=3$, any local and
linear operator can be  written as a linear combination of the identity
and the $su(3)$ generators. The $SU(3)$ {\it magnetization} is the
highest-dimensional order parameter for the $S$=1 problem. By reducing
the symmetry of the Hamiltonian to any subgroup $G$ of $SU(3)$ we can
obtain lower symmetry order parameters. The general procedure consists
in building a new basis for the generators where each operator
transforms according to an irreducible representation of $G$. For
instance, if we take $G=SU(2)$ we can do this classification by
inverting the Eqs. (\ref{gene})
\begin{eqnarray}
{\cal S}^{0 0}&=&\frac{2}{3}-(S^z)^2,\;\;{\cal S}^{1
1}=\frac{S^z(S^z+1)}{2}-\frac{1}{3} \ , \nonumber \\
{\cal S}^{1 0}&=&\frac{1}{2\sqrt{2}} [ S^+ + \left \{ S^+,S^z \right \}
] \, \nonumber \\
{\cal S}^{0 1}&=&\frac{1}{2\sqrt{2}} [ S^- + \left \{ S^-,S^z \right \}
] \, \nonumber \\
{\cal S}^{2 0}&=&\frac{1}{2\sqrt{2}} [ S^- - \left \{ S^-,S^z \right \}
] \, \nonumber \\
{\cal S}^{0 2}&=&\frac{1}{2\sqrt{2}} [ S^+ - \left \{ S^+,S^z \right \}
] \, \nonumber \\
{\cal S}^{1 2}&=& \frac{i}{2} \left \{ S^x,S^y \right \}+(S^x)^2 +
\frac{1}{2}(S^z)^2-1 \ ,  \nonumber \\
{\cal S}^{2 1}&=& \frac{1}{2i} \left \{ S^x,S^y \right \}+(S^x)^2+
\frac{1}{2}(S^z)^2-1 \ .
\label{igene}
\end{eqnarray}
In this way we see that the ferromagnetic and the spin-nematic order
parameters form a particular basis of $su(3)$ generators in the
fundamental representation. In other words, we can write down any
linear operator as a linear combination of the components of both order
parameters. Therefore, if a $S$=1 Hamiltonian is $SU(2)$ invariant the
local order parameter can be either ferromagnetic or spin-nematic since
it can be written as a linear combination of both. This exhausts all
possible order parameters for an $SU(2)$ invariant $S$=1 Hamiltonian. 
  
Following the previous procedure we can now classify the possible local
order parameters of an $SU(2)$ spin $S$ Hamiltonian. The ordinary 
magnetization is the only phase that can be derived from $S$=1/2 
spins. It is well-known that a spin-nematic phase like the one
described above cannot exist for $S$=1/2. For $S$=1, we have seen that
there is an additional spin-nematic phase which appears in a natural
way from the connection with the $SU(3)$ group. In general, we can
write down a spin $S$ Hamiltonian in terms of the generators of
$SU(2S+1)$ in the fundamental representation. To determine the
possible local order parameters of an $SU(N)$ invariant spin $S$
Hamiltonian we have to reduce this space of generators according to the
irreducible representations of $SU(2)$. It is easy to check that those
representations correspond to the totally symmetric tensors of rank
$l \leq 2S$, i.e., the possible order parameters are generated by
application of the $SU(2)$ transformations to the set: $\{
S^z,(S^z)^2,\cdots,(S^z)^{2S} \}$. $(S^z)^{l}$ is the highest weight
operator of the $SU(2)$ representation associated to the totally
symmetric tensor of rank $l$. The dimension of this representation is
$2l+1$. Therefore, the dimension of the space spanned by these
operators is $\sum^{2S}_{l=1}(2l+1)=(2S+1)^2-1$ which coincides with
the dimension of the space of generators of $SU(2S+1)$. In this way we
see that there are $2S$ independent local order parameters for a spin
$S$ problem. The first two correspond to the local magnetization and
the local spin-nematic order parameters. To our knowledge, there
is no special name for the other multipolar orderings. 

This last example illustrates the general procedure to follow in order
to determine and classify the possible local order parameters of a
given system. It is important to remark that this classification can be
made in any representation \cite{ours2}. For instance, we can re-write
the $SU(3)$ Heisenberg Hamiltonian in terms of $S$=1/2 hard-core
bosons. In that language the local order parameters may be the local
magnetization and the Bose-Einstein condensation order parameter
$\bar{b}^\dagger_\sigma$. In this way our algebraic procedure allows
one to identify the possible local order parameters of a given system.
In addition, the coexistence of more than one order parameter (more
than one phase) can be described in a unified way by using an adequate
language. 

Another connection, which is useful to illustrate the purpose of this
section, is the one relating the $SU(N)$ Heisenberg Hamiltonians to
$t$-$J$-like Hamiltonians for hard-core bosons with spin $S=(N-2)/2$
(i.e., $N_f=N-1$ different flavors). To this end we have to use the
bosonic expressions for the generators of $SU(N)$ (Eq.~(\ref{fund}))
introduced in section \ref{sec4}. Using these expressions we can
re-write the $SU(N)$ Heisenberg Hamiltonian in the following way
\begin{eqnarray}
H^{\sf SU(N)}_{\sf Heis}\!\!&=& \!\!
\tilde{H}^{\sf SU(N-1)}_{\sf Heis} + J \!\!\sum_{{\langle {\bf i},{\bf j}
\rangle},\alpha} [ {\bar b}^{\dagger}_{\bi\alpha} {\bar
b}^{\;}_{\bj\alpha}+  \bar{b}^{\dagger}_{\bj\alpha} {\bar
b}^{\;}_{\bi\alpha} ] \nonumber \\
&+&J \sum_{\langle {\bf i},{\bf j} \rangle} \bar{n}_{\bf i}
\bar{n}_{\bf j}-{\sf z}J \frac{N-1}{N} \sum_{\bj} \bar{n}_{\bj} \ ,
\label{suNtj}
\end{eqnarray}
where $\bar{n}_{\bj}=\sum_{\alpha=1}^{N_f}\bar{n}_{\bj \alpha}$, and
$\tilde{H}^{\sf SU(N-1)}_{\sf Heis}$ is given by
\begin{equation}
\tilde{H}^{\sf SU(N-1)}_{\sf Heis}= J \!\!\!\!\sum_{{\langle {\bf i},{\bf
j}\rangle},(\mu,\nu)=1,N_f} \!\!\!\!{\cal S}^{\mu \nu}({\bf i}) {\cal
S}^{\nu\mu}({\bf j}) \ .
\end{equation}
This mapping is valid for any spatial dimension $d$. If we write down the
same Hamiltonian using fermions instead of hard-core bosons, a gauge
field appears in $d=2$ due to the presence of the $K_\bj$ operators
which transmute the statistics. When $d=1$ the Hamiltonian is exactly
the same for fermions and hard-core bosons (and anyons, in general). In
this particular language we can check that the ferromagnetic $SU(N)$
order parameter describes the coexistence of all the magnetic phases
associated to the $SU(N-1)$ Heisenberg Hamiltonian (for instance, 
ferromagnetic and spin-nematic phases for $N=4$), a Bose-Einstein
condensate for each of the $N-1$ different flavors, and a homogeneous
(${\bf k}=0$) charge density wave. 

Let us start analyzing the simplest $N=2$ case (Eq.~(\ref{spin1o2})).
This case corresponds to spinless bosons with a kinetic energy term and
a nearest-neighbor density-density interaction. There are no spin
degrees of freedom for this particular case and the $SU(2)$ invariance
is then associated to the charge degrees of freedom. The two phases 
which coexist in this case are the Bose-Einstein condensate and the
uniform charge density wave order. For $N=3$ we have the usual $t$-$J$
Hamiltonian for spin $S$=1/2 hard-core bosons (or fermions in $1d$). In
this case the $SU(3)$ symmetry gives rise to a coexistence between
$S$=1/2 ferromagnetism, Bose-Einstein condensation in both flavors, and
the uniform charge density wave. These orders are different components
of the same $SU(3)$ order parameter (see Eq.~(\ref{spinsu3})). In this
particular language, $\bar{b}^{\dagger}_{\uparrow}={\cal S}^{1 0}$,
$\bar{b}^{\dagger}_{\downarrow}={\cal S}^{2 0}$,
$\bar{b}^{\;}_{\uparrow}={\cal S}^{0 1}$, and
$\bar{b}^{\;}_{\downarrow}={\cal S}^{0 2}$ are the components of the
order parameter for the spin up and down Bose-Einstein condensates. For
the magnetization we get $s^z=\frac {1}{2}({\cal S}^{1 1}-{\cal S}^{2
2})$, $s^x=\frac{1}{2}({\cal S}^{1 2}+{\cal S}^{2 1})$, and 
$s^y=\frac{1}{2i}({\cal S}^{1 2}-{\cal S}^{2 1})$. The fact that $\bar
{n}_\bj=\frac{2}{3}-{\cal S}^{0 0}$ can take any value by making
$SU(3)$ rotations of the ground states is another manifestation of the
Bose-Einstein condensation. In addition, $\bar{n}_\bj$ is the order
parameter associated to the uniform charge density wave (if the $SU(3)$
magnetization is oriented along the charge density wave direction we
obtain $\bar {n}_\bj=1$ in each site). For $N=4$ we get a model which
describes a gas of $S$=1 hard-core bosons with the bilinear-biquadratic
magnetic interaction of Eq.~((\ref{spin1})). In this case we get
coexistence of  ferromagnetism, spin-nematic, Bose-Einstein
condensation for the three different flavors, and a uniform charge
density wave. As in the previous case there is an additional phase when
$N$ is increased by one. 
 
Therefore, this family of $SU(N)$ Heisenberg Hamiltonians naturally
describes a multi-phase behavior when the original language, based upon
the generators of $SU(N)$, is translated into another based on 
generators of a lower symmetry group. This illustrates a general rule
which applies to any pair of groups $(G,{\cal G})$ such that $G$ is a
subgroup of ${\cal G}$, and the order parameter ${\bf P}$ transforms 
according to an irreducible representation of ${\cal G}$. In general,
this representation will be reducible under the operations of the
lower  symmetry group $G$, i.e., we will be able to express the given
representation as a direct sum of representations which are irreducible
under the application of $G$. We can associate an order parameter ${\bf
p_\gamma}$ to each of these representations. If we add the dimensions
of each of these ${\bf p_\gamma}$ order parameters, we will get the
dimension of ${\bf P}$. Therefore, by using the lower symmetry
language, we obtain different phases, and each of them is characterized
by one order parameter ${\bf p_\gamma}$. 


In the previous analysis we have not discussed the possible spatial
dependence of the order parameter. In other words, we assumed that it
was homogeneous over the entire lattice. However, there are many
instances in nature where the order parameter is non-uniform. The
antiferromagnetism is one of the most common examples. The previous
analysis can be easily extended to the case of non-uniform order
parameters with a well-defined wavevector $\bk$. We just need to
recognize that the non-uniform order parameters can be written as a
Fourier transform of the local order parameter ${\bf p}(\bj)$ if the
model considered has lattice translation invariance. Indeed, the
presence of a non-uniform order parameter indicates that the
translation symmetry has been spontaneously broken. Therefore the
expression for the non-uniform order parameter ${\bf {\hat p}}(\bk)$ is
\begin{equation}
{\bf {\hat p}}(\bk)=\frac{1}{N_s}\sum_\bj \exp[i \bk \cdot \bj]
\ {\bf p}(\bj) \ .
\label{nonu}
\end{equation}
Since the previous analysis is applied to the local order parameters
${\bf p}(\bj)$, Eq.~(\ref{nonu}) shows that conclusions are trivially
extended to non-uniform global order parameters. 

\subsection{Hierarchical Languages: The Quantum Phase Diagram of the
Bilinear-Biquadratic Heisenberg Model}
\label{sec6b}

We have seen that the local order parameter acquires its simplest form
when it is expressed in terms of the hierarchical language. In
addition, the generators of this language exhaust all possible local
order parameters which may result from the solution of the problem
under consideration. In other words, any local order parameter can be
written as a linear combination of generators of the hierarchical
language. The Hamiltonians considered above are special cases since
they have an $SU(N)$ invariance and therefore correspond to high 
symmetry points of an eventual phase diagram. We just considered those
cases as the simplest examples of coexistence of different phases and
unification of order parameters. We will show now that high symmetry is
not a requirement for the successful application of the present
formalism to the determination of quantum phase diagrams. To this end
we will consider the most general isotropic $SU(2)$ $S$=1 model with
nearest-neighbor interactions in a hypercubic lattice (an overal $J>0$
factor is omitted in the following)
\begin{equation} 
H_\phi =  \sqrt{2} \sum_{\langle {\bf i},{\bf j} \rangle} [
\cos \phi \ {\bf S}_{\bf i} \cdot {\bf S}_{{\bf j}} + \sin \phi
\left ( {\bf S}_{\bf i} \cdot {\bf S}_{{\bf j}} \right )^2 ] \ ,
\label{spin1x}
\end{equation}
a model already introduced in Eq.~(\ref{spin1xx}). The parameter $\phi$
sets the relative strength between the bilinear and biquadratic terms.
Eq.~(\ref{spin1}) corresponds to the particular case $\phi=5\pi/4$.
Indeed, as it is shown below, there are four isolated values of
$\phi=\{\pi/4,\pi/2,5\pi/4,3\pi/2\}$ for which $H_\phi$ is $SU(3)$
invariant. The only symmetry which is present for any value of $\phi$
is the global $SU(2)$ invariance since $H_\phi$ is a function of the
scalar products ${\bf S}_{\bf i} \cdot {\bf S}_{{\bf j}}$. 

The Hamiltonian $H_\phi$ has been the subject of several studies in the
last two decades 
\cite{Papanicolau,Solyom0,Parkinson,Barber,Klumper,Chuvukov0,
Chuvukov,Fath0,Fath,Xiang,Xian0,Xian,solyom}, nevertheless, the complete
characterization of the different phases was not, until now, completely solved. A
semiclassical treatment for $d>1$ \cite{Papanicolau} indicates that
there are four different phases: the usual ferromagnetic
($\pi/2<\phi<5\pi/4$) and antiferromagnetic ($3\pi/2<\phi<\pi/4$)
phases are separated on both sides by collinear- ($5\pi/4<\phi<3\pi/2$)
and orthogonal-nematic ($\pi/4<\phi<\pi/2$) orderings. We will show 
below that the collinear- and orthogonal-nematic phases obtained with
the semiclassical approximation are replaced by uniform- and 
staggered-nematic orderings, respectively. 

As we have seen in the previous subsection, the $SU(3)$ spins in the
fundamental representation and the $S$=1 $SU(2)$ spins are two 
equivalent languages. In addition, we have shown in the sections
\ref{sec4a} and \ref{jordanw} that the $SU(3)$ spins and the $S$=1
$SU(2)$ spins can  be respectively mapped onto $S$=1/2 hard-core
bosons.  We will use now these transformations to map the spin one
Hamiltonian $H_\phi$ onto its $SU(3)$ spin version. For pedagogical
reasons, it is convenient to use  the $S$=1/2 hard-core bosons as an
intermediate language.  

In section \ref{sec4a}, we introduced a spin-particle transformation
connecting $SU(N)$ spins and multiflavored hard-core bosons
(JW particles, in general). In particular, the fundamental
({\it quark}) representations of $SU(N)$ were mapped onto an algebra of
hard-core bosons with $N_f=N-1$ flavors (see Eq.~(\ref{fund})). For
$N=3$ the hard-core bosons have two flavors
($\alpha=\uparrow,\downarrow$) which can be associated to an internal 
spin $S$=1/2 degree of freedom. A compact way of writing the
$SU(3)$ spin in terms of hard-core bosons is
\begin{equation}
{\cal S}({\bf j})= \begin{pmatrix} \frac{2}{3} - \bar{n}_{{\bf j}}
&\bar{b}^{\;}_{{\bf j} \uparrow}& \bar{b}^{\;}_{{\bf j}
\downarrow}\\ \bar{b}^\dagger_{{\bf j} \uparrow}&\bar{n}_{{\bf j}
\uparrow} -\frac{1}{3}& \bar{b}^\dagger_{{\bf j} \uparrow}
\bar{b}^{\;}_{{\bf j} \downarrow} \\ \bar{b}^\dagger_{{\bf j}
\downarrow} &\bar{b}^\dagger_{{\bf j} \downarrow} \bar{b}^{\;}_{{\bf j}
\uparrow}&\bar{n}_{{\bf j} \downarrow}-\frac{1}{3}
\end{pmatrix} \ .
\label{spinsu3p}
\end{equation}
It is straightforward to write down each generator of the $su(3)$
algebra in terms of the Gell-Mann (traceless Hermitian) matrices
\begin{eqnarray}
\lambda_1\!&\!=\!&\!\begin{pmatrix} 0&1 & 0\\ 1&0 & 0 \\ 0& 0 & 0 \end{pmatrix} , \ 
\lambda_2\!=\!\begin{pmatrix} 0&-i & 0\\ i&0 & 0 \\ 0& 0 & 0 \end{pmatrix}, \
\lambda_3\!=\!\begin{pmatrix} 1&0 & 0\\ 0&-1 & 0 \\ 0& 0 & 0 \end{pmatrix}, 
\nonumber \\
\lambda_4\!&\!=\!&\!\begin{pmatrix} 0&0 & 1\\ 0&0 & 0 \\ 1& 0 & 0 \end{pmatrix} , \ 
\lambda_5\!=\!\begin{pmatrix} 0&0 & -i\\ 0&0 & 0 \\ i& 0 & 0 \end{pmatrix}, \
\lambda_6\!=\!\begin{pmatrix} 0&0 & 0\\ 0&0 & 1 \\ 0& 1 & 0 \end{pmatrix}
,\nonumber \\
&&\lambda_7\!=\!\begin{pmatrix} 0&0 & 0\\ 0&0 & -i \\ 0& i & 0 \end{pmatrix} , \ 
\lambda_8\!=\!\frac{1}{\sqrt{3}}\begin{pmatrix} 1&0 & 0\\ 0&1 & 0 \\ 0& 0 &
-2 \end{pmatrix} ,  \nonumber
\end{eqnarray}
with the result
\begin{eqnarray}
\bar{b}^{\dagger}_{{\bf j} \uparrow}&=& \frac{\lambda_1+i\lambda_2}{2}
\;,\;\bar{b}^{\;}_{{\bf j} \uparrow}=  \frac{\lambda_1-i\lambda_2}{2}
\ , \nonumber \\
\bar{b}^{\dagger}_{{\bf j} \downarrow}&=&\frac{\lambda_6-i\lambda_7}{2}
\;,\;\bar{b}^{\;}_{{\bf j} \downarrow}= \frac{\lambda_6+i\lambda_7}{2}
\ , \nonumber \\
\bar{b}^{\dagger}_{{\bf j} \uparrow}  \bar{b}^{\;}_{{\bf j}
\downarrow}&=& \frac{\lambda_4+i\lambda_5}{2}
\;,\;\bar{b}^{\dagger}_{{\bf j} \downarrow}\bar{b}^{\;}_{{\bf j} 
\uparrow}=\frac{\lambda_4-i\lambda_5}{2} \ , \nonumber \\
\bar{n}_{{\bf j} \uparrow}-\frac{1}{3}&=& 
\frac{\lambda_3}{2}+\frac{\lambda_8}{2\sqrt{3}} \;,\;\bar{n}_{{\bf j}
\downarrow}-\frac{1}{3}= -\frac{\lambda_8}{\sqrt{3}} \ . \nonumber
\end{eqnarray}
This constitutes the Cartan-Weyl representation of $su(3)$ and
illustrates the fact that a $S$=1 $SU(2)$ spin can be equally
represented in terms of quark fields.  

In the same way we wrote in Eq.~(\ref{spinsu3p}) the
generators of $SU(3)$ in the fundamental representation, we can write
down the corresponding expressions for the generators in the
conjugate representation
\begin{equation}
\tilde{\cal S}({\bf j})= \begin{pmatrix} \frac{2}{3} - \bar{n}_{{\bf
j}} &-\bar{b}^{\dagger}_{{\bf j} \downarrow}& -\bar{b}^{\dagger}_{{\bf
j} \uparrow}
\\ -\bar{b}^{\;}_{{\bf j} \downarrow} &\bar{n}_{{\bf j}
\downarrow} -\frac{1}{3}& \bar{b}^\dagger_{{\bf j} \uparrow}
\bar{b}^{\;}_{{\bf j} \downarrow} \\  -\bar{b}^{\;}_{{\bf j}
\uparrow}&\bar{b}^\dagger_{{\bf j} \downarrow} \bar{b}^{\;}_{{\bf j}
\uparrow}&\bar{n}_{{\bf j} \uparrow}-\frac{1}{3}
\end{pmatrix} \ .
\label{spinsu}
\end{equation}
When the $S$=1 operators are replaced by the corresponding functions
of $SU(3)$ generators in the fundamental and the conjugate
representations, it turns out that $H_{\phi}$, up to an irrelevant
constant, is a linear combination of the ferromagnetic and the
antiferromagnetic $SU(3)$ Heisenberg models
\begin{eqnarray}
H_\phi&=& \sqrt{2}\sum_{\langle {\bf i},{\bf j} \rangle}
\left [\cos \phi \ {\cal S}^{\mu \nu}({\bf i}) {\cal S}^{\nu
\mu}({\bf j})
\right . \nonumber \\
&+& \left . (\sin \phi - \cos \phi) \ {\cal S}^{\mu \nu}({\bf
i}) \tilde{\cal S}^{\nu \mu}({\bf j}) \right ] \ .
\end{eqnarray}
Repeated greek superindices are summed and the site index ${\bf i}$
runs over one of the two sublattices. This expression for $H_\phi$
illustrates the very important result that any nonlinear interaction in
the original representation is simply a bilinear term in the new
representation when mapped onto the highest rank algebra \cite{ours2}.
In particular, as mentioned above, there are certain special points in
parameter space where the Hamiltonian is highly symmetric. For example,
for $\phi = \frac{\pi}{4}$ and $\frac{5\pi}{4}$, $H_\phi$ is explicitly
invariant under uniform $SU(3)$ transformations on the spins
\cite{lai}, while for $\phi=\frac{\pi}{2}, \frac{3\pi}{2}$, $H_\phi$
is explicitly invariant under staggered conjugate rotations of the two
sublattices. These symmetries are hard to identify in
the original spin representation but are manifest in the $SU(3)$
representation.

In the following we will concentrate on the determination of the 
quantum phase diagram of $H_\phi$ for spatial dimensions $d>1$ to
avoid the strong effects of quantum fluctuations which can restore the
continuous symmetry when $d=1$. In the previous section we have
analyzed the high symmetry point $\phi=\frac{5\pi}{4}$. We found that
the ground state has a non-zero order parameter
\begin{equation}
{\cal S} = \sum_{\bf j}  {\cal S}({\bf j}) \ ,
\end{equation}
associated to a broken continuous $SU(3)$ symmetry. This order
parameter is the uniform $SU(3)$ magnetization and corresponds to the
coexistence of a ferromagnetic and a uniform spin-nematic ordering (see
Eq.~(\ref{gene})). This indicates that $\phi=\frac{5\pi}{4}$ is a
quantum phase transition point separating a ferromagnetic phase from a
uniform spin-nematic one. Let us consider now the related point
$\phi=\frac{\pi}{4}$ which differs in an overall sign from the previous
case. This sign changes the interaction from ferromagnetic to
antiferromagnetic. Therefore, for this new high symmetry point we
expect to get a ground state characterized by the staggered order
parameter
\begin{equation}
{\cal S}_{\sf ST} = \sum_{\bf j} \exp[i {\bf Q \cdot j}] \ {\cal S}({\bf j})
\ ,
\end{equation}
where ${\bf Q}$ is the antiferromagnetic wave vector. It is clear from
Eq.~(\ref{gene}), that this staggered $SU(3)$ order parameter
corresponds to the coexistence of the staggered $SU(2)$ magnetization
\begin{equation}
{\bf M}_{\sf ST} = \sum_{\bf j} \exp[i {\bf Q \cdot j}] \ {\bf S}_{\bf j} \ ,
\label{stm1}
\end{equation}
and the staggered nematic order parameter
\begin{equation}
{\bf N}_{\sf ST} = \sum_{\bf j} \exp[i {\bf Q \cdot j}] \ {\bf N}_{\bf j}
\ . 
\label{snop}
\end{equation}
${\bf N}_{\bf j}$ is the symmetric and traceless component of  the
tensor obtained from the tensorial product of two vectors ${\bf S}_{\bf
j}$. Hence, $\phi=\frac{\pi}{4}$ is a transition point separating the
usual antiferromagnetic ordering from a staggered spin-nematic phase  
characterized by the order parameter of Eq.~(\ref{snop}).

We will consider now the other two high-symmetry points, $\phi=
\frac{\pi}{2},\frac {3\pi}{2}$. For $\phi=\frac {3\pi}{2}$, the $SU(3)$
symmetry is generated by the staggered operator
\begin{equation}
{\cal S}_+= \sum_{{\bf j}\in A}  {\cal S}({\bf j}) + \sum_{{\bf j}\in B} 
{\tilde{\cal S}}({\bf j}),
\end{equation}
where $A$ and $B$ denote the two different sublattices of a hypercubic
lattice. In this case, we have a ferromagnetic interaction between
${\cal S}({\bf i})$ and ${\tilde{\cal S}}({\bf j})$, and then ${\cal
S}_+$ is the order parameter characterizing the broken $SU(3)$ symmetry
of the ground state. It is interesting to note that when the $SU(3)$
order parameter ${\cal S}_+$ is reduced with respect to the $SU(2)$
group, the two coexisting order parameters are the staggered
magnetization (see Eq.~(\ref{stm1})) and the {\it uniform} nematic
order parameter
\begin{equation}
{\bf N} = \sum_{\bf j}  {\bf N}_{\bf j} \ .
\label{snop1}
\end{equation}
In other words, if we apply an $SU(3)$ rotation generated by ${\cal
S}_+$ to the staggered magnetization we get the uniform nematic order
parameter, and vice versa, the uniform nematic order parameter is
rotated into the staggered magnetization. This can be immediately seen
by writing down the components of the local $SU(2)$ magnetization and
the nematic order parameter as a function of the local generators of
$su(3)$ in the conjugate representation ${\tilde{\cal S}}({\bf j})$
\begin{eqnarray}
S^x&=&\frac{-1}{\sqrt{2}} (\tilde{\cal S}^{0 1}+\tilde{\cal S}^{2 0}+
\tilde{\cal S}^{0 2}+\tilde{\cal S}^{1 0}) \ , \nonumber \\
S^y&=&\frac{1}{\sqrt{2}i} (\tilde{\cal S}^{0 1}+\tilde{\cal S}^{2
0}-\tilde{\cal S}^{0 2}-\tilde{\cal S}^{1 0}) \ , \nonumber \\
S^z&=&\tilde{\cal S}^{2 2}-\tilde{\cal S}^{1 1} \ , \nonumber \\
(S^x)^2-\frac{2}{3}&=&\frac{1}{2}(\tilde{\cal S}^{1 2}+\tilde{\cal
S}^{2 1}+\tilde{\cal S}^{0 0}) \ , \ (S^z)^2-\frac{2}{3}= -\tilde{\cal
S}^{0 0}  \ , \nonumber\\
\{ S^x,S^y \}&=& i (\tilde{\cal S}^{2 1}-\tilde{\cal S}^{1 2}) \ ,
\nonumber \\
\{ S^x,S^z\}&=&\frac{1}{\sqrt{2}} (\tilde{\cal S}^{0 1}-\tilde{\cal
S}^{2 0}-\tilde{\cal S}^{0 2}+\tilde{\cal S}^{1 0}) \ , \nonumber \\
\{ S^y,S^z\}&=&\frac{-1}{\sqrt{2}i} (\tilde{\cal S}^{0 1}-\tilde{\cal
S}^{2 0}+\tilde{\cal S}^{0 2}-\tilde{\cal S}^{1 0}) \ .
\label{geneconj}
\end{eqnarray}
Comparing these expressions to the ones in Eq.~(\ref{gene}), we see
that when we change from ${\cal S}({\bf j})$ to ${\tilde{\cal S}}({\bf
j})$, there is a change in sign for the three components associated to
the magnetization, while the five components corresponding to the
nematic parameter remain the same. Then, it is clear that ${\cal S}_+$
describes the coexistence of a staggered magnetization and a uniform
nematic ordering. Therefore, the conclusion is that $\phi=
\frac{3\pi}{2}$ separates an ordinary antiferromagnetic phase from the
uniform nematic ordering. 

The last high symmetry point to be considered is $\phi= \frac{\pi}{2}$.
In this case the coupling between ${\cal S}({\bf i})$ and ${\tilde{\cal
S}}({\bf j})$ turns out to be positive, i.e., antiferromagnetic, and
therefore we expect to get a broken continuous symmetry characterized
by the order parameter
\begin{equation}
{\cal S}_{-}= \sum_{{\bf j}\in A}  {\cal S}({\bf j}) - \sum_{{\bf
j}\in B} {\tilde{\cal S}}({\bf j}) \ .
\end{equation}
From the considerations above, it is clear that ${\cal S}_{-}$
describes the coexistence of ferromagnetism (uniform magnetization)
and staggered nematic order. Hence, $\phi=\frac{\pi}{2}$ is a
transition point separating these two phases. 
\begin{table}[htb]
\begin{center}
\footnotesize
\hspace*{-0.5cm}
\begin{tabular}{|c||c|c|c|}
\hline
\raisebox{0pt}[13pt][7pt]{\large $\phi$} &
\raisebox{0pt}[13pt][7pt]{\large Global $SU(3)$ OP} &
\raisebox{0pt}[13pt][7pt]{\large OP 1} &
\raisebox{0pt}[13pt][7pt]{\large OP 2}\\ 
\hline \hline 
$5\pi/4$ (FM-UN) &
$\displaystyle {\cal S} \!= \!\sum_{\bf j}  {\cal S}({\bf j})$ &
$\displaystyle {\bf M}$ & 
$\displaystyle {\bf N}$ \\
\hline 
$\pi/4$ (AF-SN) &
$\displaystyle {\cal S}_{\sf ST} \!= \!\sum_{\bf j}  
\exp[i {\bf Q} \cdot \bj] {\cal S}({\bf j})$ &
$\displaystyle {\bf M}_{\sf ST}$ & 
$\displaystyle {\bf N}_{\sf ST}$ \\
\hline 
$3\pi/2$ (AF-UN) &
$\displaystyle {\cal S}_{+}\! =\! \sum_{{\bf j}\in A}{\cal S}({\bf j}) 
+ \sum_{{\bf j}\in B}  {\tilde{\cal S}}({\bf j})$ &
$\displaystyle {\bf M}_{\sf ST}$ & 
$\displaystyle {\bf N}$ \\
\hline 
$\pi/2$ (FM-SN) &
$\displaystyle {\cal S}_{-} \!= \!\sum_{{\bf j}\in A}{\cal S}({\bf j}) 
- \sum_{{\bf j}\in B}  {\tilde{\cal S}}({\bf j})$ &
$\displaystyle {\bf M}$ & 
$\displaystyle {\bf N}_{\sf ST}$ \\
\hline
\end{tabular}
\end{center}
\caption{Order parameters describing the different phases of the
bilinear-biquadratic $S$=1 Heisenberg model for $d>1$. $\phi$ indicates
the phase boundary where the two phases in parentheses coexist.}
\label{table2}
\end{table}

\begin{figure}[htb]
\includegraphics[angle=0,width=8.2cm,scale=1.0]{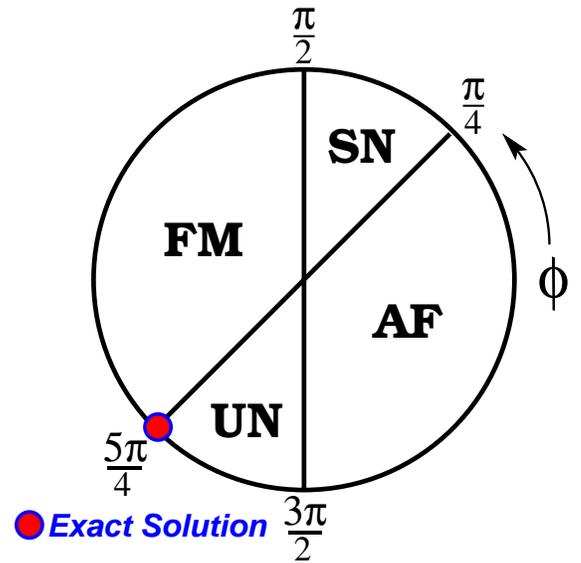}
\caption{Quantum phase diagram of the bilinear-biquadratic 
$SU(2)$ $S$=1 model for $d>1$. The phases are: AF: antiferromagnetic; 
SN: staggered-nematic, FM: ferromagnetic, and UN: uniform-nematic.}
\label{fig5}
\end{figure}

In this way, by identifying the high-symmetry points of $H_{\phi}$ we
have determined the quantum phase diagram of this model (see
Fig.~\ref{fig5}). In addition to the transition points, we have
obtained explicit expressions for the order parameters associated to
each phase for any $d>1$; these are summarized in table \ref{table2}. 
We can also predict from this analysis that the four transition points
(high-symmetry points) correspond to first order quantum phase
transitions. In each phase, the corresponding order parameter has a a
finite value and they coexist pairwise at the high-symmetry points.
However, as soon as we depart from this point in one or the other
direction in $\phi$, the $SU(3)$ symmetry is removed and one of the 
order parameters goes discontinuously to zero. In other words, the
states with pure magnetic (ferro or antiferromagnetic) and nematic
orderings belong to different representations of $SU(2)$ (the remaining
symmetry) so only one of them remains as the ground state when the
$SU(3)$ symmetry is lifted. 

In closing this section let us summarize the main steps to follow in
order to obtain and classify the local order parameters. 
\begin{description}
\item $\color{red} \bullet$
Identify the group ${\cal G}_{\sf HL} =  SU(D)$ associated to the
hierarchical language whose fundamental representation has the same
dimension $D$ as the local Hilbert space of the problem. The generators
of this language exhaust all possible local order parameters.
\vspace*{0.08in}

\item $\color{red} \bullet$
Identify the group of global symmetries of the Hamiltonian $\cal G$
which are direct products of local transformations. 
\vspace*{0.08in}

\item $\color{red} \bullet$
Given that ${\cal G} \subseteq {\cal G}_{\sf HL}$, then one can 
classify the generators of ${\cal G}_{\sf HL}$ in the fundamental
representation according to the irreps of ${\cal G}$. Each irrep leads
to a different broken symmetry order parameter.
\vspace*{0.08in}

\item $\color{red} \bullet$
Key: existence of a general set of $SU(D)$ transformations.

\end{description}

\section{Lattice Gauge Theories and the Quantum Link Models}
\label{sec7}

Wilson's lattice formulation of quantum field theory provides a 
non-perturbative regularization to Euclidean functional integrals
\cite{montvay}. The space-time lattice supplies the field theory with a
cutoff such that loop integrations in perturbation theory yields finite
results instead of divergences. 

In this way, a natural analogy emerges between quantum field theory on
a $d$-dimensional lattice and classical statistical mechanics in $d+1$,
which for a pure gauge theory (i.e., gauge theories without matter
fields) amounts to consider the classical lattice action:
\begin{eqnarray}
S[u] = - \sum_{{\bf i}, \mu \neq \nu} {\rm Tr} \left \{ u_{{\bf i}, \mu}    
u_{{\bf i}+\hat{\mu}, \nu}u^\dagger_{{\bf i}+\hat{\nu}, \mu}
u^\dagger_{{\bf i}, \nu} \right \} \ .
\label{gmag}
\end{eqnarray}
The classical complex parallel transporters $u_{{\bf i}, \mu}$ 
transform as:
\begin{equation}
u'_{{\bf i}, \mu}= \Lambda^{-1}({\bf i}) \ u_{{\bf i}, \mu} \
\Lambda({\bf i}+\hat{\mu})  
\end{equation}
under local gauge transformations $\Lambda({\bf i})$ which leave the
action $S[u]$ invariant. The elementary parallel transporters are
associated with an ordered pair of points in the lattice (links). For
instance, the link (${\bf i}, \mu$) corresponds to the straight path
from lattice site ${\bf i}$ to ${\bf i}+ \hat{\mu}$. 

Recently, Wiese and co-workers \cite{wiese} have further elaborated
work started by Horn in 1981 \cite{horn} on a new way of
non-perturbative regularization of field theories which they named
D-theory. In D-theory the Euclidean action in the standard Wilsonian
formulation of a $d$-dimensional lattice field theory is replaced by a
Hamilton operator $H$ 
\begin{eqnarray}
H = \sum_{{\bf i}, \mu \neq \nu} {\rm Tr} \left \{ U_{{\bf i}, \mu}    
U_{{\bf i}+\hat{\mu}, \nu}U^\dagger_{{\bf i}+\hat{\nu}, \mu}
U^\dagger_{{\bf i}, \nu} \right \}
\label{qgm}
\end{eqnarray}
with quantum link operators $U_{{\bf i}, \mu}$ on a $d$-dimensional
lattice, which constitute generators of an algebra acting on a Hilbert
space (e.g, a $U(1)$ quantum link model can be realized with quantum
links that satisfy an $su(2)$ algebra). $H$ commutes with the local
generators of gauge transformations ${\bf G}_{\bf i}$ ($U'_{{\bf i},
\mu}= \prod_{\bf j} \Lambda^{-1}({\bf j}) \ U_{{\bf i}, \mu} \
\prod_{\bf l} \Lambda({\bf l})$ with $\Lambda({\bf i})= e^{i
\mathbf{\alpha_\bi} \cdot {\bf G}_{\bf i}}$) and the theory is defined
through the quantum partition function ${\cal Z} = {\rm Tr} \left \{
e^{- \beta H} \right \}$, where the trace is taken on the Hilbert space. 

As emphasized in Ref. \cite{wiese}, D-theory is not a new set of field
theories but another lattice regularization and quantization of the
corresponding classical models. The main important and attractive
feature is the use of discrete quantized variables and finite Hilbert
spaces. However, it turns out to be necessary to formulate the theory
with an additional Euclidean dimension (i.e., in $d+1$). In this regard
dimensional reduction is a fundamental component to relate quantum link
models to ordinary gauge theories. The dimensional reduction hypothesis
relies on the existence of a massless phase in $d+1$, assumption that
must be verified on a case-by-case basis. 

The existence of a quantum link model connecting lattice gauge theories
to spin (or other algebraic) theories opens the possibility of formal
connections between gauge theories of high-energy physics and strongly
correlated problems of condensed matter. Indeed, since there is a
connection between the ``Gauge World'' and the ``Spin World'' and we
exhausted the connection between the ``Spin World'' and the ``Particle
World'' the consequence is that one can find isomorphisms between the
``Gauge World'' and the ``Particle World.'' In this way, for instance,
one may look for the exact equivalent of confinement (in gauge
theories) in the particle (condensed-matter) language. Even though some
of these ideas were speculated in the literature there was no formal
rigorous relation established: {\it The connection is our fundamental
theorem}.  

\subsection{$U(1)$ Gauge Magnet}
\label{sec7a}

For pedagogical purposes, it is convenient to illustrate the 
connection between gauge and condensed matter theories by considering
the simplest $U(1)$ gauge theory, usually called {\it gauge magnet}.
Gauge magnets are gauge-invariant generalizations of the Heisenberg
models. The corresponding classical model has a $U(1)$ parallel
transporter
\begin{equation}
u_{{\bf j}, \mu}= e^{i \Phi_{{\bf j}, \mu}}
\end{equation}
associated to each link $({\bf j}, \mu)$ and the action $S[u]$ is
given by Eq.~(\ref{gmag}). The dagger in this case denotes complex
conjugation. As it is shown by Chandrasekharan and Wiese
\cite{Chandrasekharan}, after the quantization process the $u_{{\bf
j}, \mu}$ fields become generators of a local $su(2)$ algebra
\begin{eqnarray}
U^{\;}_{{\bf i}, \mu}=S^{+}_{{\bf i}, \mu} \ ,
\nonumber \\
U^{\dagger}_{{\bf i}, \mu}=S^{-}_{{\bf i}, \mu} \ .
\label{gst}
\end{eqnarray}
In other words, each link variable is replaced by a spin operator. Like
in the case of quantum spin systems, this quantization can be realized
with any spin representation of $SU(2)$. The simplest case corresponds
to $S$=1/2. In this case there are two possible states for each link
variable which are denoted by the two possible eigenvalues of
$S^{z}_{{\bf i}, \mu}=\pm 1/2$. The Hamiltonian of the quantum $U(1)$
gauge magnet is obtained by replacing Eq.~(\ref{gst}) into
Eq.~(\ref{qgm})
\begin{eqnarray}
H_{\sf gm} = \sum_{{\bf i}, \mu \neq \nu}  S^{-}_{{\bf i}, \mu}    
S^{-}_{{\bf i}+\hat{\mu}, \nu} S^{+}_{{\bf i}+\hat{\nu}, \mu}
S^{+}_{{\bf i}, \nu} \ ,
\label{qgmm}
\end{eqnarray}
and the generator of the $U(1)$ gauge symmetry is
\begin{equation}
G({\bf i})=\sum_{\mu} (S^{z}_{{\bf i-\hat{\mu}},\mu}+S^{z}_{{\bf
i},\mu}) \ .
\label{gsym}
\end{equation}

\begin{figure}[htb]
\includegraphics[angle=0,width=8.2cm,scale=1.0]{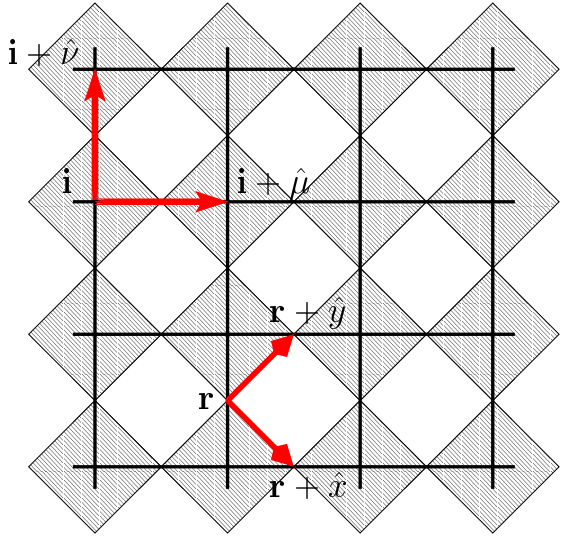}
\caption{Transformation from the direct square lattice, with lattice
sites $\bi$ and links $(\bi,\mu)$, $(\bi,\nu)$, to the dual 
checkerboard lattice, with sites $\bf r$ and links $({\bf r},x)$ and
$({\bf r},y)$.}
\label{fig6}
\end{figure}

It is natural to ask what is the equivalent particle Hamiltonian which
is obtained by applying a spin-particle transformation to
Eq.~(\ref{qgmm}). We will only consider the simplest $S$=1/2 case for
which Orland \cite{Orland} obtained the exact solution in $2+1$
dimensions. The key observation made by Orland is that this spin model
is equivalent to a gas of transversely oscillating fermionic strings.
The string tension can be obtained from the exact solution and it was
shown that the fermionic charges are confined. To obtain this result we
need to introduce a dual transformation which replaces the links by
sites: $({{\bf i}, \nu}) \rightarrow {\bf r}$, $({{\bf i},\mu})
\rightarrow {\bf r}+\hat{x}$, and $({{\bf i}+\hat{\nu}, \mu})
\rightarrow {\bf r} + \hat{y}$. This transformation is illustrated in
Fig.~\ref{fig6}, where we can see that the dual lattice is divided into
a checkerboard pattern. This means that a plaquette with vertices ${\bf
r}=(r_1,r_2)$, ${\bf r}+\hat{x}$, ${\bf r}+\hat{x}+\hat{y}$, and ${\bf
r}+\hat{y}$ is black if $r_1+r_2$ is even and white if $r_1+r_2$ is
odd. The sublattice of points ${\bf r}$ such that $r_1+r_2$ is even
will be denoted by A. 

\begin{figure}[htb]
\includegraphics[angle=0,width=8.2cm,scale=1.0]{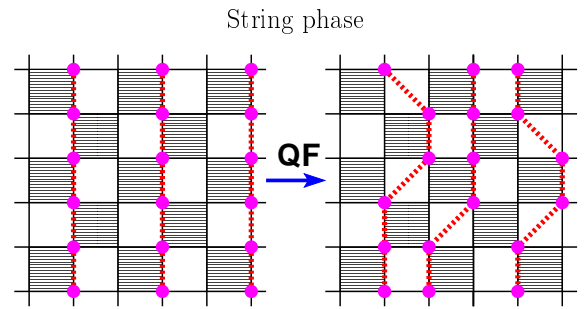}
\caption{Schematic picture of a ground state of the $S$=1/2  gauge
magnet in the particle language. Quantum fluctuations (QF) correspond
to the horizontal motion of vertical pairs in the active (black)
plaquettes.}
\label{fig7}
\end{figure}

The expression for $H_{\sf gm}$ after the dual transformation is
\begin{eqnarray}
H_{\sf gm} &=& \sum_{{\bf r} \in {\rm A}} S^{-}_{\bf r} S^{-}_{{\bf
r}+\hat{y}} S^{+}_{{\bf r}+\hat{x}+\hat{y}} S^{+}_{{\bf r}+\hat{x}}
\nonumber \\
&+&
S^{-}_{{\bf r}+\hat{x}+\hat{y}} S^{-}_{{\bf r}+\hat{x}} S^{+}_{\bf r}
S^{+}_{{\bf r}+\hat{y}} \ .
\label{qgmm2}
\end{eqnarray}
It is clear from this expression that the black plaquettes are 
active and the white ones are passive. Equation (\ref{gsym}) implies that 
the total charge in each white (passive) plaquette is conserved. 
In absence of background sources, the physical states are those
which are annihilated by the `Gauss law' operator $G({\bf i})$. 
In the particle language, this condition means that there are 
two particles on each white plaquette (see for instance the state 
illustrated in Fig.~\ref{fig7}). 

We have seen in section \ref{jordanw}, that a spin $S$=1/2 can be 
transformed into a spinless hard-core anyon through the generalized
JW mapping
\begin{eqnarray}
S^{+}_{\bf r}&=&\bar{a}_{\bf r}^{\dagger} K_{\bf r}^{\theta} 
\ , \nonumber \\
S^{-}_{\bf r}&=& (K_{\bf r}^{\theta})^\dagger\bar{a}_{\bf r}^{\;} 
\ , \nonumber \\
S^{z}_{\bf r}&=& \bar{a}_{\bf r}^{\dagger} \bar{a}_{\bf r}^{\;} - 
\frac{1}{2} \ . 
\label{satr}
\end{eqnarray}
For $\theta=0$, the anyons become hard-core bosons and for 
$\theta=\pi$ they are spinless fermions. By replacing these expressions
into Eq.~(\ref{qgmm2}) we get the following Hamiltonian representing
interacting hard-core anyons
\begin{eqnarray}
H_{\sf gm} &=& \sum_{{\bf r} \in A} \bar{a}^{\;}_{\bf r}
\bar{a}^{\;}_{{\bf r}+\hat{y}} \bar{a}^{\dagger}_{{\bf
r}+\hat{x}+\hat{y}} \bar{a}^{\dagger}_{{\bf r}+\hat{x}} \nonumber \\
&+&
\bar{a}^{\;}_{{\bf r}+\hat{x}} \bar{a}^{\;}_{{\bf r}+\hat{x}+\hat{y}}
\bar{a}^{\dagger}_{{\bf r}+\hat{y}} \bar{a}^{\dagger}_{\bf r} \ .
\label{qgmm3}
\end{eqnarray}
Notice that the transmutators $K_{\bf r}^{\theta \;}$ and $(K_{\bf
r}^{\theta})^\dagger$ do not appear in the Hamiltonian. This means that
$H_{\sf gm}$ is invariant  under transmutation of the statistics. To
physically understand the origin of this invariance we just need to
realize that the dynamics imposed  by $H_{\sf gm}$ only allows the
motion of particles along the horizontal axis within a passive
plaquette (see Fig. \ref{fig7}). Therefore, the original ordering of
the particles is preserved and the  statistics turns to be irrelevant. 

The exact ground states of this model \cite{Orland} correspond to 
parallel strings which are aligned in the $\hat{y}$ direction (see Fig.
\ref{fig7}). The model can be exactly solved because the quantum 
fluctuations of each string are described by an effective
one-dimensional $S$=1/2 $XY$ model \cite{Orland}. Notice that this
solution relates to the new paradigm in strongly correlated matter
where the appearance of local inhomogeneous (stripe-like) structures
seems to be a common feature of many different physical systems. 

There are other possible connections between the gauge magnet and 
condensed matter theories which correspond to different changes  of
language. If each link is associated to a singlet bond state (Cooper
pair), Orland \cite{Orland2} has shown that the $S$=1/2 gauge magnet is
mapped onto the Rokhsar-Kivelson model \cite{Rokhsar} (without the
diagonal term) if the new `Gauss Law': $[G({\bf i})+2]\ket{\psi}=0$ is
imposed.  In this way we see how the formal connections between 
condensed matter and lattice gauge theories can be exploited to predict
new physical behaviors in one or the other field. 

\section{Quantum Information and Computation}
\label{sec8}

A new challenge in information theory and computer science has recently
emerged as the result of applying the fundamental laws of quantum
mechanics and using the quantum effects to advantage. This new set of
ideas comprise what is known as ``Theory of Quantum Computation and
Quantum Information'' and has as a major objective to process
information in a way that exceeds the capabilities of classical
information \cite{qcbook}. The device that performs the manipulation of
information is named quantum computer and the standard unit of
information is the {\it qubit} (i.e., a two-level system). The close
relationship between information processing and the physical phenomena
leading to it is perhaps the most remarkable aspect of this new
paradigm. Since information can be represented in many different
physical forms, and easily converted from one form to another without
changing its meaning, quantum information represents a new abstract
archetype for information processing independent of the precise
implementation of the quantum computer, only requiring at least one
physical representation to be useful.

A key fundamental concept in information theory is the realization
\cite{sfer} that a model of computation is intimately connected to a
physical system through a closed operator algebra. In other words, each
physical system is associated to a certain language (e.g., spin
$S$=1/2) and thus to an algebra realizing it (e.g., Pauli algebra), and
that particular algebra may become a possible model of computation. An
immediate consequence is that an arbitrary physical system can be
simulated by another physical system (e.g., a quantum computer)
whenever there exists an isomorphic mapping between the different
operator algebras representing the systems \cite{sfer}. A very simple
example is provided in Refs. \cite{sfer} and \cite{sfer1}, where it is
shown how to simulate a 1$d$ impurity Anderson model using a quantum
computer based on an array of spins $S$=1/2. Another resource of
quantum information that can help us understand complex phenomena in
strongly correlated matter is the notion and measures of {\it
entanglement}.

It is very clear the power that our fundamental theorem has by
providing the formal connections (isomorphisms) between the different
languages of nature. Therefore, the implications for quantum
information and computation are rather obvious, namely that one can
identify quantum resources and define convenient models of computation,
or imitate an arbitrary quantum phenomena with a given quantum computer
given the appropriate dictionaries to translate nature's language to
the {\it machine language}. In this way, one can recognize three
complementary areas where the concepts of language and dictionaries
developed in the present manuscript are of particular relevance, areas
we will expand on in the following.

\begin{description}

\item \underline{Models of computation and quantum resources}

A model of computation consists of an algebra of operators, a set of
controllable Hamiltonians and measurable observables, and an initial
state of the physical system. The set of controllable Hamiltonians must
be universal in the sense that any unitary operation can be performed
with such a set. The {\it standard model} of computation is a particular
example which uses the Pauli operator algebra and the qubit as basic
unit. Another possible model is the fermion model \cite{sfer} which is
isomorphically related to the standard one through the JW
mapping and, thus, is equivalent. The main point is that the choice of
model depends upon the nature of the available physical resources and
their quantum control. For example, in a liquid NMR quantum computer
the nuclear spins of the molecules ($S$=1/2) are the units which can be
controlled and the standard model is the appropriate one. 

The control of quantum mechanical systems is hampered by quantum noise
and decoherence and, therefore, identifying static quantum resources
for information processing is a great challenge. Given a physical
information processing device there will be a language which will the
most natural one for the elementary static resources of the device. For
instance, if we could control $^4$He atoms at the quantum level it
would be natural to consider a hard-core boson model of computation.   
The importance of our fundamental theorem is that once one identifies
the best quantum resources it allows us to build the model of
computation accordingly.

\item \underline{Simulation of physical phenomena}

Physical phenomena can be simulated or {\it imitated} by a quantum
network \cite{sfer2} with the help of a quantum computer. Imitation is
realized through a quantum algorithm which consists of a quantum
network with a means to repeat blocks of instructions. A quantum
network is defined by a sequence of universal gates (unitary
operations), applied to the system for the purpose of information
processing, and measurements in a fixed temporal order. The
measurement operation is mostly needed to classically access information
about the state of the system. Every matrix which represents a
reversible operation on quantum states can be expressed as a product of
the one and two-qubit gates, and the minimum set needed to represent
any such matrices is called a universal set of gates. 

When trying to simulate a problem using quantum information processing,
an important issue is to determine how many physical resources are
needed for the solution. The main resources are {\it quantum space},
the number of qubits needed, and {\it quantum time}, the number of
quantum gates required. The accounting of algorithmic resources forms
the foundations of quantum complexity theory. One of the objectives in 
quantum information theory is to accomplish imitation efficiently, i.e,
with polynomial complexity, and the hope is that quantum imitation is
more efficient (i.e., needs less resources) than classical imitation. 
There are examples that support such hope (e.g., fermion simulations
with polynomially bounded statistical errors \cite{sfer,sfer2}),
although there is no general proof that indicates the superiority of
quantum over classical imitations, regarding efficiency. Indeed, there
is, so far, no efficient quantum algorithm that can determine the
ground state (or, in general, the spectrum) of a given Hermitian
operator \cite{sfer2}, despite occasional claims. It is known that the
ability to resolve this question leads to efficient algorithms for
NP-complete problems like the traveling salesman conundrum.

A very important observation, in connection with the notion of
efficiency, is a corollary of our fundamental theorem: Given two
languages, the generators of one of them can be written as a polynomial
function, with {\it polynomial complexity in the number of modes or
resources}, of the generators of the other and vice versa. This result
implies that the important algorithmic step of {\it translation} from
the language of the system to be imitated to the {\it machine language}
does not change the complexity of the quantum space and time. 

Certainly, a general purpose quantum computer is not the only device
that allows simulation of physical phenomena in nature (with its many
languages). Imitation can also be achieved in a conceptually different
manner using a quantum simulator. The main distinction is the lack of
universality of the latter. An example of a quantum simulator is an
optical lattice \cite{greiner} which is specifically designed to
imitate a given physical Hamiltonian and where there is limited quantum
control. The possibility of control and tunability of the interactions
of the elementary constituents offers the potential to design new
states of matter \cite{greiner,ours5}. This is of particular relevance
in strongly correlated matter where these quantum simulators furnish the 
benchmark to test theories and approximations \cite{ours5}. Again, the
importance of the languages and dictionaries developed in this
manuscript is clear and concrete.

\item \underline{Quantum information measures}

Entanglement, a word that Schr\"odinger coined to distinguish quantum
from classical mechanics, is that {\it bizarre} feature of composite
quantum systems that led to so much controversy in the past (like
quantum non-locality in the Einstein-Podolsky-Rosen gedanken
experiment). Essentially, {\it entanglement} is a quantum property
whereby a pure state of a composite quantum system may cease
to be determined by the states of its constituent subsystems. Entangled
pure states are those that have mixed subsystem states. An independent
concept is the notion of {\it separability} which refers to the
property of a quantum state of a composite system of being able to be
written as a direct product of states of its component subsystems. The
theory of entanglement, and its generalizations, is currently under
development \cite{qcbook,barnum}. In Ref.~\cite{barnum} entanglement is
viewed as an {\it observer-dependent} concept, whose properties are
determined by  the expectations of a distinguished subspace of
observables of the  system of interest (experimental access), without
reference to a preferred subsystem decomposition. The standard notion
of entanglement is recovered when these  means are limited to local
observables acting on subsystems. A tremendous effort is put in trying
to understand the properties of entanglement that can be used as a
resource in quantum information. 

In a sense, that must be defined more precisely \cite{barnum}, the
notion of entanglement is a relative of the notion of quantum
correlations and a very relevant question is whether one can construct
useful measures of entanglement to understand the emergence of complex
phenomena in strongly correlated matter. A quantum phase transition
involves a qualitative change in the correlations of the ground state 
of the system as a result of tuning parameters of its Hamiltonian. In
some cases an order parameter is associated to the transition, in
others a topological order. It is intuitively expected that this change
is also associated to a change in the nature of entanglement.
Therefore, quantifying and classifying entanglement is very important
to characterize a quantum phase transition: Can measures of
entanglement distinguish between broken and non-broken \cite{wen}
symmetry phase transitions? 
%

\end{description}

\section{Summary}
\label{sec9}

We have introduced an algebraic framework for interacting quantum
systems to study complex phenomena characterized by the coexistence and
competition of various ordered states. We argued that symmetry, and
topology, are key guiding principles behind such complex emergent
behavior. Emphasis has been made in developing a systematic
mathematical structure that allows one to attack these problems within
a single unifying approach. 

The core result of the paper, from which all other results follow, is
the proof of a fundamental theorem that permits to connect the various
operator languages used in the description of the properties of
physical systems. This theorem together with the notion of
transmutation of statistics provide the tools necessary to unifying the
quantum description of matter. To formalize this unification we needed
to rigorously define the concepts of language and dictionary: To model
a particular physical phenomena we commonly identify the main degrees
of freedom of the problem and associate to them certain operators. One
can furnish to the resulting set of operators (that we call language)
with an algebraic structure and ask whether two different languages
have something in common. The fundamental theorem tells us that two
languages can be connected whenever the dimension of their local
Hilbert spaces are equal. We expanded the notion of local Hilbert space
to embrace different Hilbert space decompositions (we saw, for
instance, how to map the Hilbert space of a bond to a site). The
resulting one-to-one language mappings we named dictionaries (a
traditional example of which is the Jordan-Wigner mapping). In the
course of the presentation we showed, through example, many different
dictionaries relating diverse operator languages. In this way we
defined universality of behavior as an equivalence relation between 
seemingly different physical phenomena which share exactly the same
underlying mathematical structure as a result of one-to-one language
mappings (for example, the spin nematic order and Bose-Einstein
condensation of flavored hard-core bosons). Out of the many languages
one can use to describe a given physical problem there is a class, we
named hierarchical language, which has the advantage that any local
operator can be expressed as a linear combination of its generators. In
this way, hierarchical languages provide the tools necessary to
classify order parameters.

A new, formally developed, notion is the idea of {\it emergent
symmetry}, i.e., the fact that new symmetries not realized in the
Hamiltonian describing the system can emerge at low energies. From a
broader perspective, an emergent symmetry is a {\it non-local} 
transformation which commutes with the system Hamiltonian and becomes
{\it local} when restricted to a low-energy subspace of the Hilbert
space. In some instances like the quasi-exactly solvable problems, to
which the 1$d$ $t$-$J_z$ model belongs \cite{ours4}, or the family of
spin Hamiltonians for which the ground state is a product of spin
singlets (e.g., the Majumdar-Ghosh model), the low-energy effective
theory can be derived in an exact way and the emergent symmetry is
exact. In other cases, the emergent symmetry is approximate, however,
it still provides a guiding principle to identify the nature of the
ground state and its low-energy excitations. In the same way that
rigidity is related to a broken symmetry phenomenon, localization (in
real or any other space) is connected to the emergence of a gauge
symmetry. Indeed, the metal-insulator transition may be interpreted as
a discontinous change in the group of emergent symmetries
characterizing the two phases.

\begin{figure}[htb] \hspace*{0.0cm}
\includegraphics[angle=-90,width=9.4cm,scale=1.0]{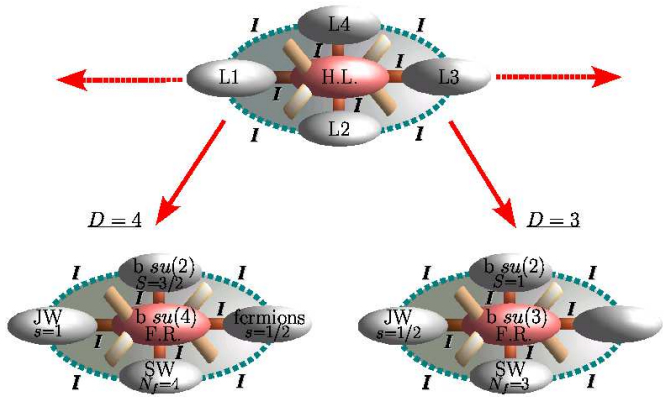}
\caption{Unified Framework. For each $D$ there is more than one diagram 
with a different hierarchical language (H.L.). $I$ indicates that there
is an isomorphic mapping between two languages, e.g., L1 and L2.}
\label{fig3}
\end{figure}

Figure \ref{fig3} summarizes the spirit and fundamental concepts that
emerge from our fundamental theorem. 

There are several reasons for our algebraic framework to constitute a
powerful method to study complex phenomena in interacting quantum
systems. Most importantly: To connect seemingly unrelated physical
phenomena (e.g., high-$T_c$ or heavy fermions and quantum spin
theories); identify the general symmetry principles behind complex
phase diagrams; unveil hidden symmetries (and associated order
parameters) to explore new states of matter; obtain exact solutions of
relevant physical models that display complex ordering at certain
points in Hamiltonian space; and find new approximations which do not
favor any of the competing interactions. The power of the present
approach is reflected in the unlimited number of potential applications
where they could be used, and ranges from condensed matter and
statistical mechanics to lattice gauge theories and quantum information
and computation. We hope and envision this formalism will find future
applications in contexts different than the ones we have developed
here.

\acknowledgments

We thank J.E. Gubernatis, L. Gurvits, W. Lee, and B.S. Shastry for
useful discussions. This work was sponsored by the US DOE under
contract W-7405-ENG-36.

\vspace*{-0cm}



\begin{thebibliography}{99}

\bibitem{anderson0}
P.W. Anderson, Science {\bf 177}, 393 (1972).

\bibitem{wen}
See, for example, X.-G. Wen, Phys. Rev. B 65, 165113 (2002). 

\bibitem{jordan}
P. Jordan and E. Wigner, Z. Phys. {\bf 47}, 631 (1928).

\bibitem{matsubara}
T. Matsubara and H. Matsuda, Prog. Theor. Phys. {\bf 16}, 569 (1956). 

\bibitem{ushveridze}
A.G. Ushveridze, {\it Quasi-Exactly Solvable Models in Quantum
Mechanics} (IOP, Bristol, 1994). 

\bibitem{ours4}
C.D. Batista and G. Ortiz, Phys. Rev. Lett. {\bf 85}, 4755 (2000).

\bibitem{Landau} L.D. Landau and E.M. Lifshitz, {\it Statistical
Physics} (Butterworth-Heinemann, Oxford, 2001).

\bibitem{andersonrev}
P.W. Anderson, {\it Basic Notions of Condensed Matter Physics}
(Addison-Wesley, Redwood City, 1992).

\bibitem{Majumdar}
C.K. Majumdar and  D.K. Gosh, J. Math. Phys. {\bf 10}, 1388 (1969);
{\bf 10}, 1399 (1969); C.K. Majumdar, J. Phys. C {\bf 3}, 911 (1970).

\bibitem{Shastry1}
B.S. Shastry and B. Sutherland, Physica B \& C {\bf 108}, 1069 (1981). 

\bibitem{Shastry2}
B.S. Shastry and B. Sutherland, Phys. Rev Lett. {\bf 47}, 964 (1981).

\bibitem{Shastry3}
B. Sutherland and B.S. Shastry, J. Stat. Phys. {\bf 33}, 477 (1983).

\bibitem{Lin}
H.Q. Lin, J.L. Shen, and H.Y. Shik, cond-mat/0109027. 

\bibitem{Papanicolau}
N. Papanicolau, Phys. Lett. A {\bf 116}, 89 (1986); Nucl. Phys. {\bf
305}, 367 (1988).

\bibitem{qcbook}
See for example M.A. Nielsen and I.L. Chuang, {\it Quantum Computation
and Information} (Cambridge, Cambridge, 2000).

\bibitem{sfer}
G. Ortiz, J.E. Gubernatis, E. Knill, and R. Laflamme, Phys. Rev. A {\bf
64}, 22319 (2001).

\bibitem{gQM}
See for example A. Bohm, {\it Quantum Mechanics: Foundations and
Applications} (Springer-Verlag, New York, 1993).

\bibitem{Noten0}
A monoid can also be defined as a semigroup $(\fmm,\Box)$ with an
element that is the unit for $\Box$, i.e., ${\bf 1}$.
 
\bibitem{Noten1}
Commutators (Anticommutators) will be represented by square (curly) 
brackets, i.e., $[A,B]=AB - BA$ ($\{A,B\}=AB + BA$). Sometimes we will 
invoke the generalized deformed commutator defined as: $[A,B]_\theta=AB
- \exp[i \theta] BA$. We may also use $[A,B]_\pm=AB \pm BA$. Canonical
commutation (anticommutation) relations for the corresponding creation 
and annihilation operators are generically defined as
\begin{eqnarray}
[b^{\;}_{\bj\alpha},b^\dagger_{\bj\beta}] = \langle\varphi_\alpha |
\varphi_\beta \rangle \ , [b^{\;}_{\bj\alpha},b^{\;}_{\bj\beta}] =0
\nonumber \\
(\{c^{\;}_{\bj\alpha},c^\dagger_{\bj\beta} \} = \langle \varphi_{\alpha} | 
\varphi_\beta \rangle \ , \{c^{\;}_{\bj\alpha},c^{\;}_{\bj\beta} \}
= 0 )\ ,\nonumber
\end{eqnarray}
where the inner product $\langle \varphi_\alpha | \varphi_\beta
\rangle$ between elements of a single-particle basis ($\alpha,\beta \in
[1,D]$) is defined on ${\cal H}_\bj$. For the sake of clarity and with
no loss of generality, we will always assume that the single-particle
basis $\{\varphi_\alpha\}$ is orthonormal, i.e., $\langle
\varphi_\alpha | \varphi_\beta \rangle=\delta_{\alpha\beta}$, but, of
course, this condition can be relaxed.

\bibitem{jacobson}
N. Jacobson, {\it Basic Algebra I} (W.H. Freeman and Company, New York,
1985), and {\it Basic Algebra II} (W.H. Freeman and Company, New York,
1989).

\bibitem{Noten6} 
For a $d$-dimensional lattice with $N_s$ sites, the operator
$S_\bj^\mu$ is defined in terms of a Kronecker product $\otimes$ as
\[
S_\bj^\mu = \one \otimes \one \otimes \cdots \otimes 
\underbrace{S^\mu}_{\bj^{th}} \otimes \cdots \otimes \one \ ,
\]
where $\one$ is the $D \times D$ unit matrix, $D=2S+1$, and $S^\mu$ is
a spin-$S$ operator. Thus $S_\bj^\mu$ admits a matrix representation
of dimension $D^{N_s} \times D^{N_s}$.

\bibitem{ours1} 
C.D. Batista and G. Ortiz, Phys. Rev. Lett. {\bf 86}, 1082 (2001). 

\bibitem{commzhang}
The $SO(5)$ theory of Zhang \cite{zhang} constitutes a particular
example of a unified (high-symmetry) order parameter used to describe
the coexistence of more than one phase (in this case superconductivity 
and antiferromagnetism). However, contrary to the philosophy of this
paper, this description relies on the assumption of an $SO(5)$ symmetry
of the low-energy effective theory, assumption whose validity is
unclear. According to Zhang \cite{zhang}, this high-symmetry effective
theory can be obtained by applying the renormalization group.  

\bibitem{zhang}
S-C. Zhang, Science {\bf 275}, 1089 (1997). 

\bibitem{oursanyon}
B. Abdullaev, C.D. Batista, and G. Ortiz (unpublished).

\bibitem{cornwell}
J.F. Cornwell, {\it Group Theory in Physics} (Academic Press, London, 1989).

\bibitem{Noten5}
If the system has long-range interactions the dimension of the lattice
becomes irrelevant, since any $d$-dimensional system can be described
in a lower dimensional lattice at the cost of including long-range
interactions. 

\bibitem{rpmbt}
G. Ortiz and C.D. Batista in {\it Recent Progress in Many-Body
Theories}, Eds. Raymond F Bishop, Tobias Brandes, Klaus A Gernoth,
Niels R Walet, and Yang Xian (World Scientific, Singapore, 2002), p.
425; Int. J. of Modern Phys. B 17, 5413 (2003).

\bibitem{eduardo}
E. Fradkin, Phys. Rev. Lett. {\bf 63}, 322 (1989); D. Eliezer and G.W.
Semenoff, Phys. Lett. B {\bf 286}, 118 (1992).

\bibitem{wang2}
Y.R. Wang, Phys. Rev. {\bf B 43}, 3786 (1991).

\bibitem{huerta}
L. Huerta and J. Zanelli, Phys. Rev. Lett. {\bf 71}, 3622 (1993).

\bibitem{wang}
S. Wang, Phys. Rev. E {\bf 51}, 1004 (1995).

\bibitem{ordering}
Suppose that we want to establish a particular ordering for
two-flavored particles ($\alpha=\uparrow,\downarrow$) in a finite
$N_s=N_x\times N_y$ 2$d$ lattice. A possible ordering from $(\bj,\alpha)$
to an ordered set of integers I could be
\[
{\rm I}=j_1+(j_2-1)N_x +(\frac{1}{2}-\sigma)N_x N_y \ ,
\]
where$\sigma=\frac{1}{2}(-\frac{1}{2})$ for
$\alpha=\uparrow(\downarrow)$.

\bibitem{green1}
H.S. Green, Phys. Rev. {\bf 90}, 270 (1953).
 
\bibitem{cmt24}
C.D. Batista and G. Ortiz, in {\it Condensed Matter Theories}, vol.
16, edited by Susana Hernandez and W. John Clark (Nova Science
Publishers, Inc., Huntington,  New York, (2001)) pp. 1.

\bibitem{huang}
K. Huang, {\it Statistical Mechanics} (Wiley, New York, 1963), Chap. 16.

\bibitem{Noten4p} 
These isomorphisms represent useful dictionaries translating a spin
language $S$ into itinerant quantum particle languages with {\it
effective spin} $s=S-1/2$.

\bibitem{gentile}
G. Gentile, Nuovo Cim. {\bf 17}, 493 (1940).

\bibitem{avinash}
See for example A. Khare, {\it Fractional Statistics and Quantum
Theory} (World Scientific, Singapore, 1997).

\bibitem{fract}
F.D.M. Haldane, Phys. Rev. Lett. {\bf 67}, 937 (1991).

\bibitem{Noten3}
Fermions correspond to the case $p=2$, while for bosons $p$ is an
arbitrary integer number larger than 2.

\bibitem{greenberg}
O.W. Greenberg, Phys. Rev. Lett. {\bf 13}, 598 (1964).

\bibitem{green2}
H.S. Green, Prog. Theor. Phys. {\bf 47}, 1400 (1972).

\bibitem{liepara}
S. Kamefuchi and Y. Takahashi, Nucl. Phys. {\bf 36}, 177 (1962); C. Ryan
and E.C.G. Sudarshan, {\it ibid} {\bf 47}, 207 (1963); K. Kademova, 
Nucl. Phys. B {\bf 15}, 350 (1970); A.J. Bracken and H.S. Green, Nuovo
Cim. {\bf 9}, 349 (1972). 

\bibitem{boundary}
Care should be exercised regarding the boundary conditions of the model. 
Keeping this warning in mind we will omit any reference to boundary
conditions in the following.

\bibitem{ours2} 
C.D. Batista, G. Ortiz, and J.E. Gubernatis, Phys. Rev. B {\bf 65},
180402(R) (2002).

\bibitem{Baxter}
R.J. Baxter, {\it Exactly Solved Models in Statistical Mechanics}
(Academic Press, London, 1990). 

\bibitem{lai}
C.K. Lai, J. Math. Phys. {\bf 15}, 1675 (1974); B. Sutherland, Phys.
Rev. B {\bf 12}, 3795 (1975). 

\bibitem{tatba}
L.A. Takhtajan, Phys. Lett. A {\bf 87}, 479 (1982); H.M. Babujian, Phys.
Lett. A {\bf 90}, 479 (1982); Nucl. Phys. B {\bf 215}, 317 (1983).

\bibitem{aklt}
I. Affleck, T. Kennedy, E.H. Lieb, and H. Tasaki, Phys. Rev. Lett. {\bf
59} 799 (1987).

\bibitem{klum}
A. Kl\"umper, Europhys. Lett. {\bf 9}, 815 (1989).

\bibitem{Arrachea}
L. Arrachea and A.A. Aligia, Phys. Rev. Lett. {\bf 73} 2240 (1994).

\bibitem{bill}
B. Sutherland, Phys. Rev. B {\bf 62}, 11499 (2000).

\bibitem{Nam}
D. C. Mattis and S. B. Nam, J. Math. Phys. {\bf 13}, 1185 (1972).

\bibitem{bos}
C.D. Batista, G. Ortiz, and B.S. Shastry (unpublished).

\bibitem{lieb}
E.H. Lieb and F.Y. Wu, Phys. Rev. Lett. {\bf 20}, 1443 (1968).

\bibitem{Auerbach}
A. Auerbach, {\it Interacting Electrons and Quantum Magnetism}
(Springer-Verlag, New York, 1994). 

\bibitem{anderson}
P.W. Anderson, Phys. Rev. {\bf 110}, 827 (1958); {\bf 112}, 1900 (1958). 

\bibitem{Noten4}
Other representations with different particle statistics are possible; 
for example, we could have used a {\it hard-core} fermion representation
\cite{ours1}. In such a case, the resulting Hamiltonian would be
similar in form to Eq.~(\ref{hamiltJb}) with a kinetic energy term
modified to include a non-local gauge field.

\bibitem{ours5}
G. Ortiz and C.D. Batista, Phys. Rev. B 67, 134301 (2003).

\bibitem{phmartin}
P.A. Martin, Nuovo Cimento {\bf 68}B, 302 (1982).

\bibitem{kumar}
B. Kumar, Phys. Rev. B {\bf 66}, 024406 (2002).

\bibitem{batshas}
C.D. Batista and B.S. Shastry, Phys. Rev. Lett. {\bf 91}, 116401 (2003).

\bibitem{Nozieres}
Ph. Nozi\`eres, {\it Theory of Interacting Fermi Systems}
(Addison-Wesley, Reading, 1997).

\bibitem{Ha}
Z.N.C. Ha, {\it Quantum Many-Body Systems in One Dimension} (World
Scientific, Singapore, 1996).

\bibitem{kekule}
After Friedrich August Kekul\'e, a German chemist, who solely on
intuition proposed the known resonant structure of benzene in 1872,
before quantum mechanics was born. It was Linus Pauling who formally
developed the resonance theory in 1931.

\bibitem{Solyom0}
J. S\'olyom, Phys. Rev. B {\bf 36}, 8642 (1987).

\bibitem{Parkinson}
J. B. Parkinson, J. Phys. C {\bf 21}, 3793 (1988).

\bibitem{Barber}
M. N. Barber and M. T. Batchelor, Phys. Rev. B {\bf 40}, 4621 (1989).

\bibitem{Klumper}
A. Klumper, Europhys Lett. {\bf 9}, 815 (1989); J. Phys. A {\bf 23},
809 (1990).

\bibitem{Chuvukov0}
A.V. Chubukov, J. Phys. Condens. Matter {\bf 2}, 1593 (1990).

\bibitem{Chuvukov}
A.V. Chubukov, Phys. Rev. B {\bf 43}, 3337 (1991).

\bibitem{Fath0}
G. F\'ath and J. S\'olyom, Phys. Rev. B {\bf 44}, 11836 (1991).

\bibitem{Fath}
G. F\'ath and J. S\'olyom, Phys. Rev. B {\bf 47}, 872 (1993).

\bibitem{Xiang}
T. Xiang and G. A. Gehring, Phys. Rev. B {\bf 48}, 303 (1993).

\bibitem{Xian0}
Y. Xian, J. Phys. Condens. Matter {\bf 5}, 7489 (1993).

\bibitem{Xian}
Y. Xian, J. Phys. Lett. A {\bf 183}, 437 (1993).

\bibitem{solyom}
G. F\'ath and J. S\'olyom, Phys. Rev. B {\bf 51}, 3620 (1995).


\bibitem{montvay}
See, for example, I. Montvay and G. M\"unster, {\it Quantum Fields on a
Lattice} (Cambridge, Cambridge, 1994).

\bibitem{wiese}
U.-J. Wiese, Nucl. Phys. B {\bf 73}, 146 (1999) and references therein. 

\bibitem{horn}
D. Horn, Phys. Lett. B {\bf 100}, 149 (1981).

\bibitem{Chandrasekharan}
S. Chandrasekharan and U.-J. Wiese, hep-lat/9609042. 

\bibitem{Orland}
P. Orland, Nucl. Phys. B {\bf 372}, 635 (1992).

\bibitem{Orland2} 
P. Orland, Phys. Rev. B {\bf 49}, 3423 (1994).

\bibitem{Rokhsar}
D. S. Rokhsar and S. A. Kivelson, Phys. Rev Lett. {\bf 61}, 2376 (1988).

\bibitem{sfer1}
G. Ortiz, E. Knill, and J.E. Gubernatis, Nuc. Phys. B {\bf 106}, 151
(2002).

\bibitem{sfer2}
R. Somma, G. Ortiz, J.E. Gubernatis, R. Laflamme, and E. Knill, Phys.
Rev. A {\bf 65}, 42323 (2002); R. Somma, G. Ortiz, E. Knill, and J.E.
Gubernatis, Int. J. of Quant. Inf. 1, 189 (2003).

\bibitem{greiner}
M. Greiner {\it et al.}, Nature {\bf 415}, 39 (2002).

\bibitem{barnum}
H. Barnum, E. Knill, G. Ortiz, and L. Viola, Phys. Rev. A 68, 032308
(2003). See also quantum-ph/0305023. 









%
%
%
%
%

\end{thebibliography}
\end{document}